\documentclass[%
 reprint,
 amsmath,amssymb,
 aps,
prb,
]{revtex4-2}

\usepackage{graphicx}
\usepackage{dcolumn}
\usepackage{bm}

\begin{document}

\title{Optical properties of InGaN quantum wells:\\ accurately modeling the effects of disorder
}

\author{Aurelien David}
 \email{aureliendavid@google.com}
\affiliation{%
 Google / Raxium, 1600 Amphitheater Pkwy, Mountain View CA 94043, USA
}%

\date{\today}
             
\begin{abstract}
A model including random alloy disorder is used to account for the outstanding optical properties of InGaN quantum wells. The model provides excellent agreement to experimental observations on various structures. This study clarifies the prevalent role played by disorder in optical features such as the luminescence lineshape, the Stokes shift, and the radiative rate. Finally, the relationship between disorder and the peculiar properties of long-wavelength InGaN emitters is investigated.
\end{abstract}

\maketitle

\section{\label{sec:intro}Introduction and context}

The physics of III-nitride quantum wells (QW) are of high interest, both due to their applications in light emitting diodes (LEDs) and because they display unusual optical behaviors. Two important properties differentiate them from other III-V QWs: the strong polarization fields and the existence of InGaN compositional disorder.

The polarization fields, on the order of a MV/cm, lead to a giant quantum-confined Stark effect (QCSE) \cite{Takeuchi97}. This red-shifts the emission wavelength and reduces the electron-hole wavefunction overlap, slowing down the recombination dynamics. The effect of fields has been studied by many works, and is generally well quantified by a standard one-dimensional Schrodinger equation with the proper field magnitude.

The nature and consequences of material disorder have also attracted considerable attention, but their understanding is less complete. Regarding the nature of disorder, initial observations of strong indium clustering \cite{Gerthsen00} were disproved by careful studies \cite{Galtrey08}, which showed instead that the In and Ga atoms closely follow a random alloy (RA) distribution. The RA disorder nonetheless does produce regions with relatively high [In], leading to wavefunction localization effects. This induces the appearance of an Urbach tail in the absorption spectrum of InGaN QWs -- this tail is a general optical signature of disorder effects \cite{John86}.

The consequences of this disorder remain a topic of debate. Early on, it was proposed that it could fully localize carriers in nm-scale In clusters, leading to a high internal quantum efficiency (IQE) despite the large dislocation density that characterizes III-nitride heteroepitaxy; this theory has been challenged more recently. In addition, various other effects of disorder have been suggested: it could suppress the radiative rate and cause the green gap; enhance Auger-Meitner scattering; lead to excitonic localization and cause a giant Stokes shift; etc. However, many of these proposals remain qualitative, and lack a direct link between models and experiments.

A more quantitative understanding of the effects of disorder remains desirable. In this Article, we introduce an optical model that properly accounts for disorder effects. We show that it enables highly accurate predictions of the outstanding optical properties of InGaN QWs, and we clarify the underlying physics. 

The article is organized as follows. It begins with a quick review of past work on the topic. The key elements of the model are introduced. The model is then applied to three types of structures to explain their optical properties: a single-QW test structure; a series of multiple-QW LEDs of varying wavelength; and a red-emitting LED.

\section{Past work}

\subsection{Modeling in III-nitrides}

We first review a few key works that modeled the effects of RA disorder, and point out some possible limitations and remaining open questions.

Prompted by observations of RA distributions in atom probe tomography measurements~\cite{Galtrey08}, Ref.~\cite{Watson-Parris11} provided an early investigation of RA disorder effects, based on a two-band model. The authors found that low-energy hole states are strongly localized, whereas electron states are weakly localized or delocalized. This key finding has been validated by numerous other teams since then. Ref.~\cite{Watson-Parris11} also proposed a link between the RA disorder and the luminescence width (see however our comment below).

Ref.~\cite{Aufdermaur15} modeled RA effects using an Empirical Tight Binding (ETB) approach. It was found that RA led to a moderate reduction in the radiative rate (due to in-plane separation of electron-hole wavefunctions), which could be a contributing factor to the green gap. However, this analysis was based on calculating only a few quantum states, so that we believe the conclusion to be imprecise. As will be shown in this work, computing many more quantum states is essential to accurately predict optical properties, although this is challenging in the ETB framework due to the high computational burden.

Refs.~\cite{Schulz15,Tanner18} also used an ETB model to compute a few low-energy electron and hole states, and derived similar conclusions to Ref.~\cite{Watson-Parris11} regarding their localization properties, in particular finding strongly-localized hole states. In Refs.~\cite{Tanner20b,McMahon20}, the same team extended their calculations to a larger number of states to compute the low-temperature luminescence and Urbach tails for QWs of varying composition; they found good semi-quantitative agreements with experimental values of the Stokes shift and luminescence width. 

To our understanding, these previous works that computed luminescence spectra (i.e. Refs.~\cite{Watson-Parris11,Tanner20b,McMahon20}) did not provide a model for carrier distributions, but instead assumed all calculated states are populated. Hence, luminescence properties (e.g. peak energy, spectral width) are dependent on the number of calculated states, which makes agreement with experimental data somewhat coincidental.

Ref.~\cite{Piccardo17} used a new computation method (the Localization Landscape theory \cite{Filoche12}) to compute approximate quantum states and derive Urbach tails, which agreed well with their extensive experimental characterization. The number of statistical configurations used for averaging, however, remained small (10 per sample), which is insufficient to resolve the Urbach tail unless empirical broadening is included. Refs.~\cite{Yang14,Li17} also used the Landscape framework to compute the current-voltage characteristic of InGaN LEDs. They concluded that RA creates low-energy percolation paths between quantum wells, which help explain the remarkably low experimental voltage of InGaN LEDs~\cite{David16b}.

Ref.~\cite{Pant22} used a two-band effective mass approximation to compute the effect of carrier density on the wavelength shift and spectral width of a blue LED, and found good agreement with experimental data. The spectrum lineshape itself was not predicted, presumably due to the use of a large inhomogeneous broadening parameter (50~meV). 

In summary, the reviewed work established some crucial effects of RA disorder. However, a thorough computation of optical properties such as absorption and luminescence that quantitatively matches experiments remains lacking, as does an exploration of a sufficient number of states with a carrier distribution model.

\subsection{Broader context of localization physics}

Carrier localization in disordered materials is a vast and evolving topic of research, which we won't attempt to summarize. Here, we merely clarify a few semantic points.

Strictly speaking, in a two-dimensional disordered system of infinite size, \textit{all states} are localized -- as shown in Anderson's seminal work on strong localization \cite{Anderson79}. Accordingly, the mobility edge (the energy boundary between localized and delocalized carriers) does not formally exist. However, the Anderson localization length increases exponentially with the carrier energy, and can therefore be very large (exceeding microns or even millimeters). 

In the present work, we use the term \textit{localized} to refer to low-energy states confined 'trivially' within a single fluctuation of the InGaN potential on a few-nanometers scale, and the term \textit{delocalized} to describe states whose wavefunctions spreads across a large number of alloy fluctuations. This is consistent with the vocabulary used in the III-nitrides field, albeit at odds with conventions in other communities. Likewise, we call \textit{mobility edge} the energy at which states transition between these two natures. The reader should remain aware that our so-called delocalized states actually remain localized, in the Anderson sense, on a much-larger length scale.



\section{\label{sec:model}Model overview}

The model follows a conventional, textbook-like approach to computing optical properties, with RA disorder as the key non-standard ingredient. The successive computation steps are: the quantum states for electrons and holes; the corresponding densities of states, and the resulting absorption spectrum; the carrier distributions functions; and the luminescence spectrum. We give a short discussion below; more details are in Appendix I.

The simulation region, which includes the QW and surrounding barriers, is discretized on a spatial grid. The distribution of In and Ga atoms inside the QW is drawn randomly to mimic a RA, and smoothed out by a Gaussian filter of appropriate standard deviation $\sigma$; this is similar to multiple previous works based on a continuum approach \cite{Watson-Parris11,Piccardo17,Jones17}. The value of $\sigma$ is not prescribed, and can be seen as a fitting parameter that controls the strength of localization effects \cite{Divito20,Banon22}. In this work, we find that $\sigma=0.4$~nm results in excellent agreement with multiple experimental data sets. This value is consistent with our past work \cite{David19b} but differs from that used in other works \cite{Piccardo17,Li17}. The effect of $\sigma$ is further discussed in Appendix I-E.

The Hamiltonian is a simple two-band effective-mass Hamiltonian. The band structure parameters employ typical values found in Ref.~\cite{Vurgaftman03}; however, we use the hole mass `far from the band edge' \cite{Chuang96} (rather than at the $\Gamma$ point) to describe localized wavefunctions, as has been proposed in other works \cite{Watson-Parris11}. This choice is important: it leads to a heavier hole mass and a stronger localization. More considerations on this choice and the details of the valence band structure are discussed in Appendix I-B.

The polarization fields are computed following the approach of Ref.~\cite{Christmas05}. The junction field is obtained analytically, with the applied voltage as an input. The screening potential caused by confined carriers at high density is computed through a self-consistent Poisson-Schrodinger loop (with a one-dimensional approximation for simplicity, see Appendix I-A).

The Hamiltonian does not include the many-body effects caused by the Coulomb interaction. In past work, we have studied these effects in InGaN QWs, with the following conclusions. First, the excitonic population is low at intermediate carrier densities and room temperature \cite{David22}, especially in thick-enough QWs, as the polarization fields suppress the binding energy. Excitons can thus reasonably be neglected in the structures considered in this Article, where such conditions are met. Second, separate from excitons, the Coulomb interaction causes an enhancement of the absolute value of the radiative rate (without much effect on the spectral lineshape) \cite{David19b}; this effect remains weak for thick-enough QWs and sufficient carrier density, and would not significantly affect the conclusions presented in this Article. Neglecting the Coulomb interaction is thus acceptable (more discussion is provided in Appendix I-D), and leads to a significant reduction in computational cost, enabling the computation of many states.

Once the potential is established, the Schrodinger equation is discretized with a standard finite difference scheme, yielding a sparse-matrix eigenvalue problem whose solutions are the particle energies and wavefunctions. The computational domain is 20$\times$20~nm in the in-plane directions; the vertical domain size depends on the structure. For each InGaN configuration, a sufficient number of electron and hole states (typically a few hundred) are computed to find excited states up to hundreds of meVs from the ground state.  

This process is repeated a sufficient number of times to achieve statistical averaging over InGaN configurations, and in particular to resolve the Urbach tail down to four decades from the mobility edge. The necessary number of configurations can be estimated as follows. Hole confinement regions due to RA fluctuations have lateral dimensions of about 4--5~nm. Given the simulation domain size, one configuration thus yields about 20 localized states. The likelihood of finding a localized state with an energy $N$ decades below the mobility edge is roughly $10^{-N}$. Therefore, to barely resolve a density of states (DOS) down to N decades requires averaging over $C$ configurations such that $20 C=10^N$. For instance, barely resolving four decades of DOS require $C=500$ configurations. In practice, we use 1,000--5,000 configurations to reduce numerical noise (see Appendix I-C for a discussion of convergence versus $C$).

The DOS for electrons and holes ($\rho_e$, $\rho_h$) are computed as histograms of the quantum levels, with an appropriate energy step ($dE \sim 5-10$ meV):

\begin{equation}
\rho_e(E) dE = \frac{2}{V} \sum_{E<E_e<E+dE}{1}
\end{equation}

No numerical broadening is applied in the DOS calculation, to ensure its width is dominated by disorder. The DOS
histogram is built from all the levels of all InGaN configurations ($>$100,000 states), which ensures smooth tails even at low energy.

\begin{figure}
\includegraphics[width=7.5cm]{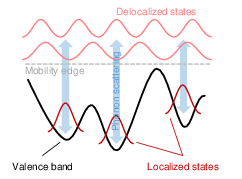}
\caption{\label{fig_thermalization0} Sketch of proposed thermalization process: phonon scattering between localized and delocalized states is efficient (because their wavefunctions always overlap), which ensures indirect thermalization between localized states. See Appendix II for a detailed discussion.
}
\end{figure}

The electron and hole distributions are assumed to follow thermalized Fermi-Dirac distributions $f_e, f_h$. Given a carrier density ($n_e=n_h=n$), the quasi-Fermi levels are obtained (by requiring $\int{\rho_e f_e dE}=n$), and states are filled according to $f_e$. Although our assumption of carrier thermalization may appear obvious, this is not necessarily the case in the presence of disorder. We discuss and justify this assumption in detail in Appendix II. In short, we propose that carrier-carrier scattering between localized and delocalized states leads to overall thermalization for all states (Fig.~\ref{fig_thermalization0}). As will be shown, thermalized populations lead to close agreement with experimental data, validating this assumption \textit{a posteriori}. 

The joint optical density of states (JDOS) $\rho_J$ is obtained, similarly to $\rho_e$ and $\rho_h$, as a histogram of all transitions between each electron level and each hole level, weighed by the squared wavefunction overlap:

\begin{equation}
\label{eq_JDOS}
\rho_J(E) dE = \frac{2}{V} \sum_{E<E_e+E_h<E+dE}{\left<\psi_e | \psi_h \right>^2}
\end{equation}

The overlap term accounts for the oscillator strength of each individual electron-hole transition. This ingredient distinguishes the present model from effective band-tail models (where disorder is accounted for with empirical tail states \cite{Romer21}), and is enabled by the explicit calculation of state wavefunctions.

Eq.~\ref{eq_JDOS} gives the ``bare'' JDOS, i.e. in the absence of carrier populations ($n=0$). We further define the ``loaded'' JDOS at carrier density $n$ as:

\begin{equation}
\rho_{Jn}(E) dE = \frac{2}{V} \sum_{E<E_e+E_h<E+dE}{\left<\psi_e | \psi_h \right>^2} (f_h(E_h)-f_e(E_e))
\end{equation}

This quantity accounts for the change in $\rho_J$ as the levels become occupied by carriers. 

The bare absorption spectrum $\alpha$ (at $n=0$) is

\begin{equation}
\label{eq:alpha}
\alpha(E) = \frac{\pi e^2 \mu^2 \hbar}{\epsilon_0 c n_o m_0^2 E} \rho_J 
\end{equation}

where $\mu$ is the momentum matrix element, $n_o$ is the optical index of GaN, and the other symbols have standard physical meanings. Likewise, the loaded absorption spectrum $\alpha_n$ at carrier density $n$ is given by Eq.~\ref{eq:alpha} with $\rho_J \rightarrow\rho_{Jn}$. At high $n$, $\alpha_n$ can become negative, corresponding to transparency.

The luminescence spectrum $L$ is obtained from $\alpha_n$ and the quasi-Fermi level splitting $\Delta E_F$:

\begin{equation}
\label{eq_L}
L(E)=\frac{E^2 n_o^2}{\pi^2 c^2 \hbar^3}\frac{\alpha_n}{exp[(E-\Delta E_F)/kT]-1}
\end{equation}

The raw luminescence spectrum thus obtained is then slightly smoothed (by a Gaussian filter of standard deviation 10~meV); this removes residual numerical noise, without having an effect on the width or overall shape. Finally, empirical LO-phonon tails are added to the spectra, as will be discussed further in Section~\ref{sec:pin_lum}.

The radiative coefficient is

\begin{equation}
B=\frac{1}{n^2}\int{L(E)dE}
\end{equation}

In summary, some of the key features of the model are: (i) the use of use a two-band effective mass Hamiltonian, including RA disorder and neglecting Coulomb interaction; (ii) the computation of a large number of quantum states and configurations for proper convergence; (iii) the assumption of carrier thermalization.

After this overview, the following sections discuss predictions of the model and comparisons to experiments.


\newpage
\section{\label{sec:pin}Blue single quantum well}

\begin{figure}
\includegraphics[width=8cm]{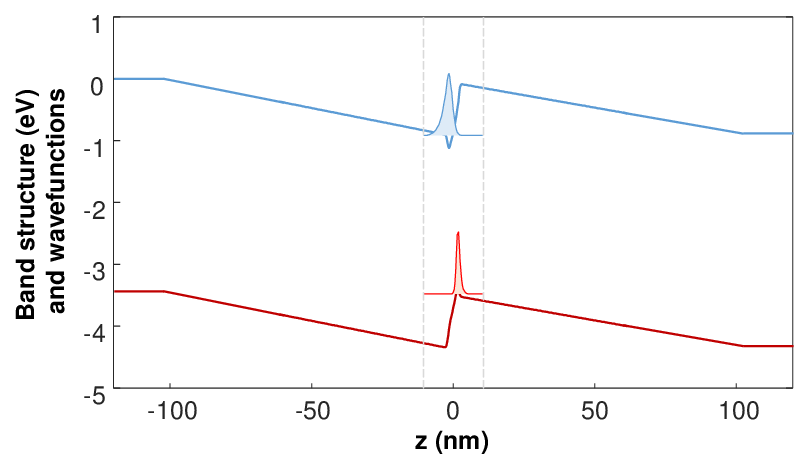}
\caption{\label{fig_pin_BS} Band structure of the single QW sample. Shaded areas: ground state wavefunctions (integrated along $(x,y)$). Dashed lines: boundaries of the simulation domain along $z$.
}
\end{figure}

Blue QWs typically have an In content $\sim$~15\% and a thickness $\sim 3$~nm. This section deals with a test structure with such a QW, on which we have previously reported extensive experimental results~\cite{David17a,David20,David21}. The sample, grown on a high-quality bulk GaN substrate by metal-organic chemical vapor deposition, is a $p-i-n$ diode with a 200 nm thick undoped region having a single QW ([In]~=~15\%, thickness~3.9~nm) at its center. This structure was designed to ensure $n=p$ in the QW (the Fermi level is at midgap in the QW in the absence of excitation, avoiding the accidental modulation doping otherwise common in PL test structures) and to preclude carrier sweep-out by the junction field (which often affects PL measurements if the depletion width is too narrow). Fig.~\ref{fig_pin_BS} shows the band structure of the single QW.

\subsection{Density of states / Urbach tail}

\begin{figure}
\includegraphics[width=8cm]{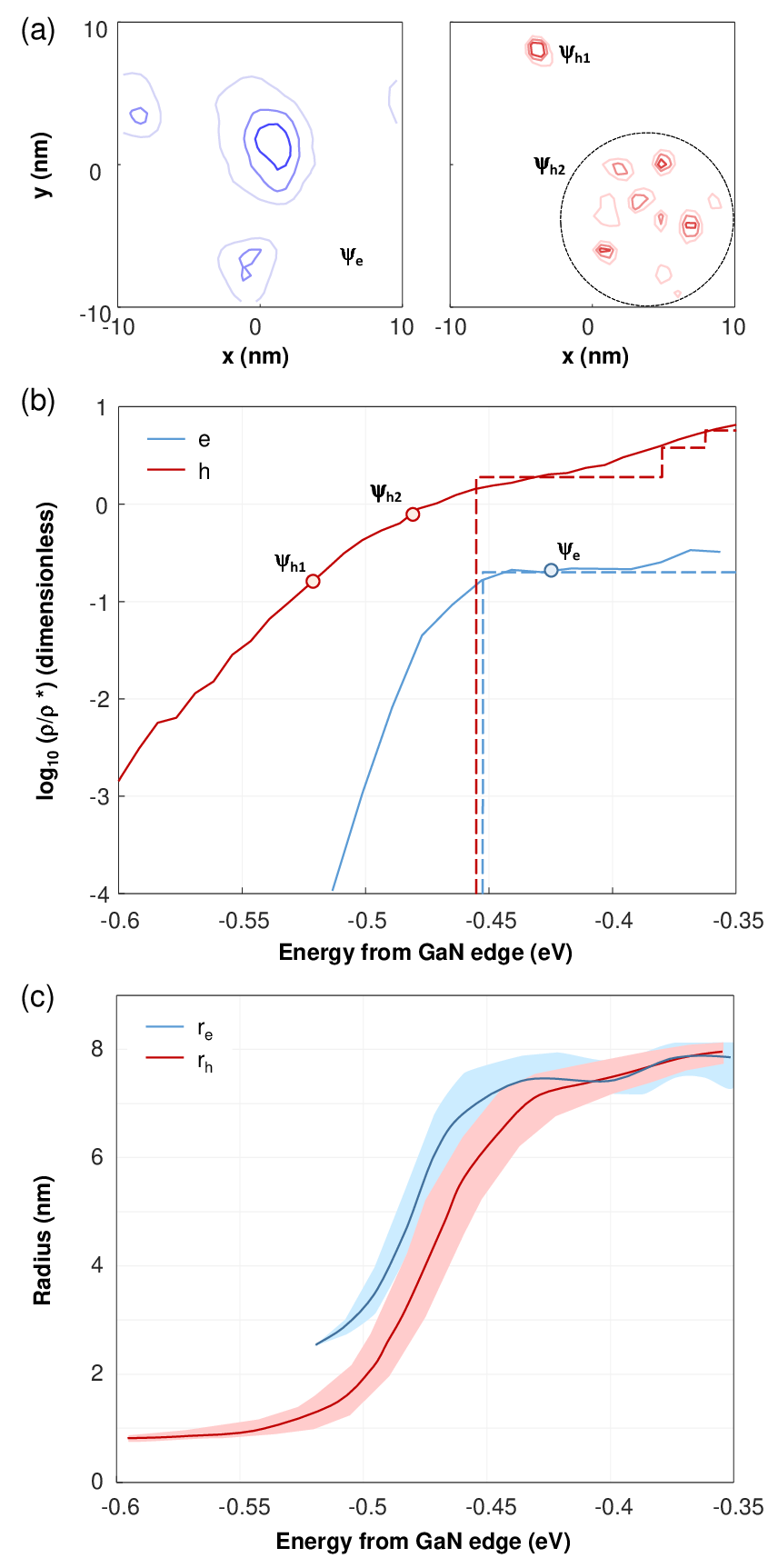}
\caption{\label{fig_pin_rho} Model results for a blue SQW. (a) $e$ and $h$ wavefunctions. (b) Modeled $e/h$ densities of states (solid lines: full model; dashed lines: without disorder). The dots illustrate the energies of the wavefunctions in (a). (c) Statistics of radius of wavefunctions. Solid lines: median values; shaded area: 25\%-75\% quartiles.
}
\end{figure}

Fig.~\ref{fig_pin_rho} shows modeling results for electrons ($e$) and holes ($h$) for this structure. Fig.~\ref{fig_pin_rho}(a) shows a few examples of wavefunctions. Electron wavefunctions are mostly delocalized, with $\psi_e$ showing several lobes across the simulation domain. The character of hole wavefunctions depends on their energy; the low-energy state $\psi_{h1}$ is strongly localized on a scale of a few nm, whereas at higher-energy, $\psi_{h2}$ is delocalized. 

Fig.~\ref{fig_pin_rho}(b) shows the calculated densities of states, and compares them to the no-disorder equivalents. Here and throughout this Article, values of $\rho$ are shown in units of $\rho^*=m_0/\pi \hbar^2 t$, with $t$ the QW thickness. This normalization factor is chosen because, in a square quantum well, the textbook value of $\rho_{e,h}$ is $m_{e,h} \times \rho^*$ for each quantum level. The energies of electrons and holes are referenced to conduction and valence band edges of GaN at the center of the QW ($z=0$), and are therefore negative. The energies of the wavefunctions of (a) are shown on (b).

$\rho_e$ has a weak Urbach tail ($E_u$~=~6~meV, obtained by fitting the exponential decay of $\rho_e$ a low energy, as illustrated in Appendix I-G). This is indicative of a small number of localized states. The wavefunctions in this tail are loosely localized, with a radius extending over several nm. At higher energy, $\rho_e$ corresponds to delocalized states and is mostly flat. 

$\rho_h$ shows much stronger localization effects. The Urbach tail is pronounced ($E_u$~=~17~meV). The low-energy part of the tail corresponds to strongly localized states, with a radius of about 1~nm. The density of these strongly-localized states is significant. We observe a mobility edge for holes at --0.5~eV -- corresponding to the knee in $\rho_h$, and to the threshold between delocalized and localized states (as will be shown in Fig.~\ref{fig_pin_rho}(c)). The often-used concept of ``band edge'' becomes ambiguous in the presence of such pronounced disorder effects. We propose that the mobility edge can be thought of as an effective band edge -- indeed, it separates localized from delocalized states and, as we will see, it is well correlated to the luminescence energy. Appendix I-G discusses how the mobility edge value is extracted.

Fig.~\ref{fig_pin_rho}(b) also show the values of $\rho$ in the absence of disorder (i.e. solving a one-dimensional Schrodinger equation along $z$, with the QW potential averaged in the ($xy$) plane). These display the usual step-like shape corresponding to discrete energy levels. Disorder has two effects: it smears out the step-like behavior, and gives rise to Urbach tails.

Fig.~\ref{fig_pin_rho}(c) show the in-plane radius of the wavefunctions (defined as $\sqrt{\left< \psi | r^2 | \psi_h \right>}$, with $r$ the distance from the wavefunction maximum). Holes have a small radius $\sim$~1~nm at low energy in the Urbach tail, corresponding to localization in a single minimum of the potential. At higher energy, the radius remains finite but grows exponentially with energy (saturating quickly at a value of 8~nm, limited by the simulation domain size): this is a manifestation of Anderson localization of high-energy states \cite{Anderson79}. For practical purposes, we will consider these states as delocalized. The energy where the radius reaches around 2~nm ($E\sim -0.5$~eV) corresponds to hole states that spread over more than one localization domain, and therefore indicates the mobility edge -- this also corresponds to the knee in the hole DOS. Regarding electrons, even low-energy states are loosely confined with a radius of a few nm.

\begin{figure}
\includegraphics[width=8cm]{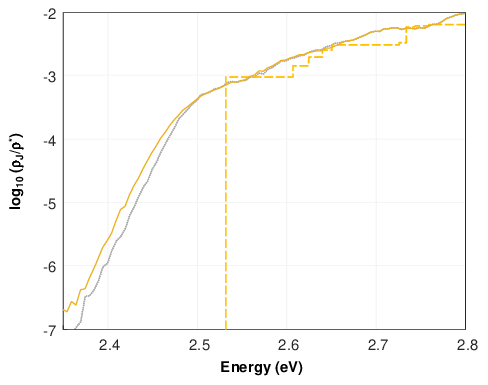}
\caption{\label{fig_pin_rhoo} Modeled JDOS. Solid line: full model. Dotted line: disorder for holes but not for electrons. Dashed line: No disorder.
}
\end{figure}

Fig.~\ref{fig_pin_rhoo} shows the joint density of states $\rho_J$. It displays a pronounced Urbach tail, similar to that of the holes ($E_u$~=~17~meV). This is expected, as $\rho_J$ is roughly the convolution product of $\rho_e$ and $\rho_h$, and its exponential tail is dominated by the strong tail of $\rho_h$. To illustrate this, $\rho_J$ is also shown in a calculation when disorder is included for holes but not electrons: this barely affects the Urbach tail (by less than 1 meV). Finally, $\rho_J$ is also shown in the absence of disorder, displaying the same step-like behavior as $\rho_{e,h}$. The absolute value of $\rho_J$ is low (a fraction of $\rho^*$) even above the Urbach tail. This is because the field-induced separation of electron and hole wavefunctions along $z$ suppresses the oscillator strength -- a pronounced effect in this sample with a rather thick QW.

In analogy with $\rho_h$, we can define an optical mobility edge from the knee of $\rho_o$, at an energy $E=2.5$~eV. Below this edge, optical transitions are mostly due to localized hole states.

\subsection{\label{sec:pin_FD}Carrier occupation}

\begin{figure}
\includegraphics[width=8cm]{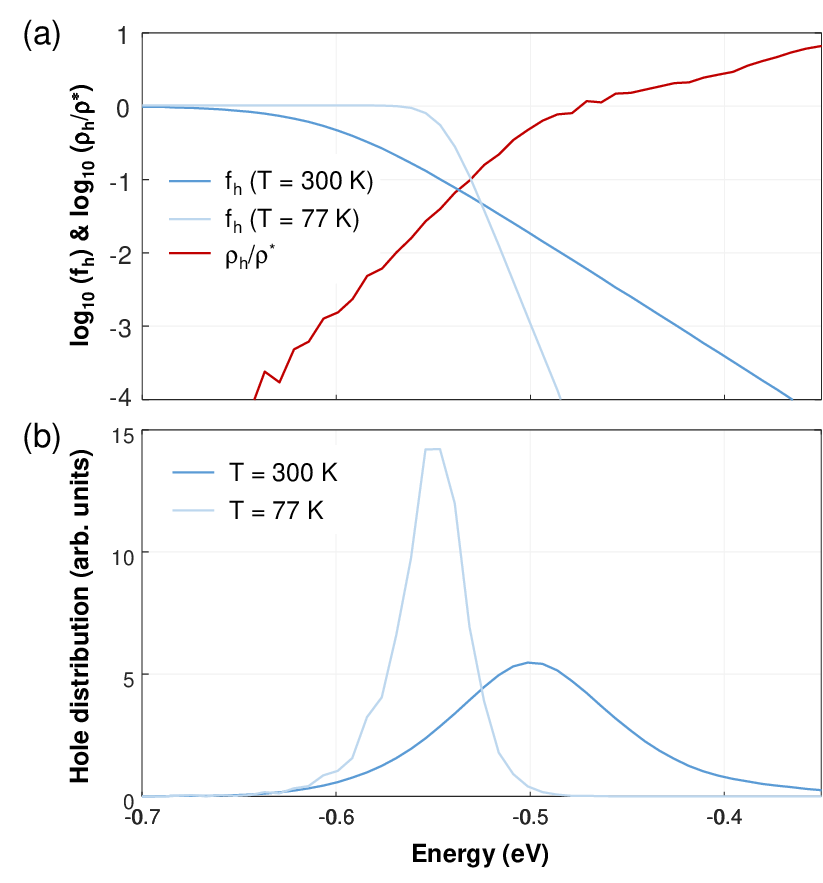}
\caption{\label{fig_pin_FD} Modeled carrier distribution. (a) Fermi-Dirac distribution $f_h$ at $T=$~77~K and~300~K, and hole DOS $\rho_h$. (b) Resulting hole distribution $f_h \times \rho_h$.
}
\end{figure}

We focus on the behavior of the hole population, as its strong localization effects dominate the optical properties. Fig.~\ref{fig_pin_FD}(a) shows Fermi-Dirac distributions $f_h$ at two temperatures ($T=$~77~K,~300~K), at $n=10^{18}$~cm$^{-3}$. Fig.~\ref{fig_pin_FD}(b) shows the resulting energy-resolved hole distribution $f_h \times \rho_h$. At room temperature, the Boltzmann tail of $f_h$ overlaps with the delocalized carrier region -- meaning that a sizable fraction of the carriers (tens of \%) are delocalized. At 77~K on the other hand, the hole population remains in the Urbach tail, indicating that populated states are mostly localized.

\begin{figure}
\includegraphics[width=8cm]{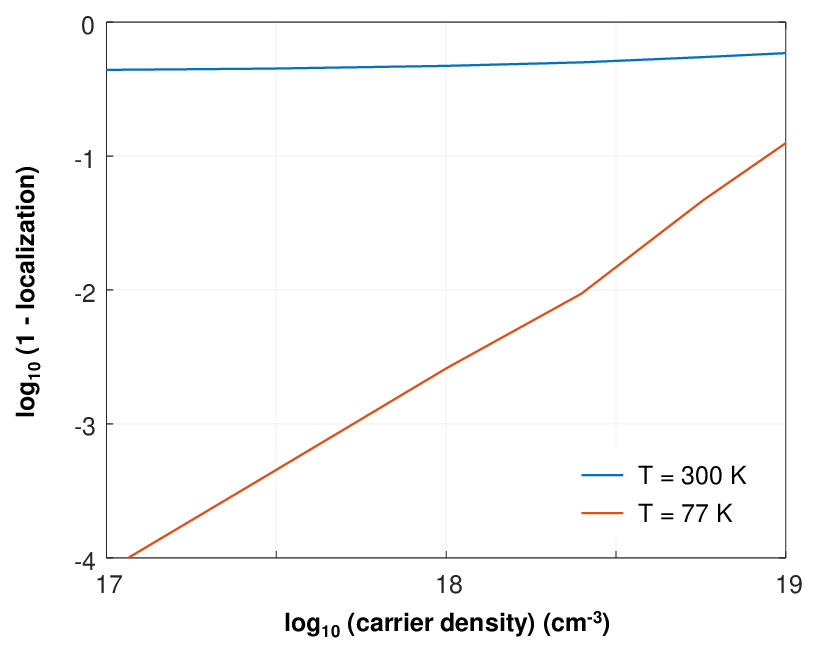}
\caption{\label{fig_pin_loc_deg} Modeled localization degree vs carrier density $n$. The plotted quantity is the delocalized fraction $(1-d)$. At 300$K$, $1-d$ is nearly constant around 50\%, indicating a significant delocalized population. At 77~K, $1-d$ is low and grows exponentially with $n$.
}
\end{figure}

To quantify this further, we define the degree of localization $d$ as the fraction of the hole population that occupies localized states (using the mobility edge of $\rho_h$ to classify states as localized or delocalized). Fig.~\ref{fig_pin_loc_deg} shows how $d$ varies with $T$ and $n$. At room temperature, $d \sim 50\%$, indicating that thermalized holes populate both localized and delocalized states, regardless of $n$. At 77~K on the other hand, the delocalized population $(1-d)$ is low and grows exponentially with $n$: the localized states are fully populated ($f_h = 1$), and only the high-energy tail of $f_h$ reaches the delocalized states.

It has been argued in various works that the remarkable efficiency of InGaN LEDs might be caused by strong localization, whereby all carrier are localized on a scale of a few nm, making them impervious to non-radiative recombinations at dislocations and other defects \cite{Nakamura98,Odonnell99,Chichibu06,Hammersley12}. However, explicit experimental evidence of this strong localization picture is lacking. In contrast, the present model finds that at room temperature, about half of the holes in a blue QW are delocalized. This finding is compatible with widespread experimental evidence. This includes the recent observation of long-range carrier diffusion in InGaN QWs \cite{David21,Becht23,Slawinska25}, and reports of a localized-to-delocalized transition at intermediate cryogenic temperatures \cite{Schomig04,Sauty22}. We believe the high IQE of InGaN LEDs is caused by other factors, discussed e.g. in Ref.~\cite{David20}.

\subsection{\label{sec:pin_lum}Luminescence}

InGaN emitters display a broad linewidth, even in high-quality samples. At room temperature, blue LEDs typically have a FWHM of about 110~meV, or 20~nm. This FWHM is obviously much larger than the thermal broadening $kT$ -- in contrast with conventional III-Vs, where thermal broadening is often the dominant factor at room temperature. The temperature dependence of the luminescence lineshape is also non-trivial. Even at cryogenic temperatures (4~K to 77~K), the FWHM remains as large as 50~meV. To date, no model has provided a quantitative prediction of this complex lineshape. As we will now show, these properties stem from the RA disorder.

We compute luminescence as described in Section~\ref{sec:model}. In addition, III-nitride luminescence spectra are known to exhibit LO phonon replicas, which are most obvious at low temperature. We account for LO phonon replicas empirically, by replicating the zero-phonon-line spectrum, to obtain the final luminescence spectrum:

\begin{equation}
L'(E) = \sum_p{S_p \times L(E+pE_{ph})}
\end{equation}

$E_{ph}=90$~meV is the phonon energy. $S_p$ is the \textit{p}-th Huang-Rhys parameter. According to Huang-Rhys theory, one expects $S_p=S_1^p/p!$. However, in III-nitrides, it has been reported that the higher-order replicas do not follow this scaling \cite{Paskov02,Tan06}. We believe localization plays an important role in this effect; however, is investigation is left for future work. In this Article, for simplicity, we use empirical values of $S_p$ to fit the spectra. When considering luminescence spectra on a linear scale, the first replica is most visible: we use a value $S_1=0.2$, which gives good agreement with experimental results and is consistent with multiple literature reports \cite{Kalliakos02,Graham05}.

\begin{figure}
\includegraphics[width=8cm]{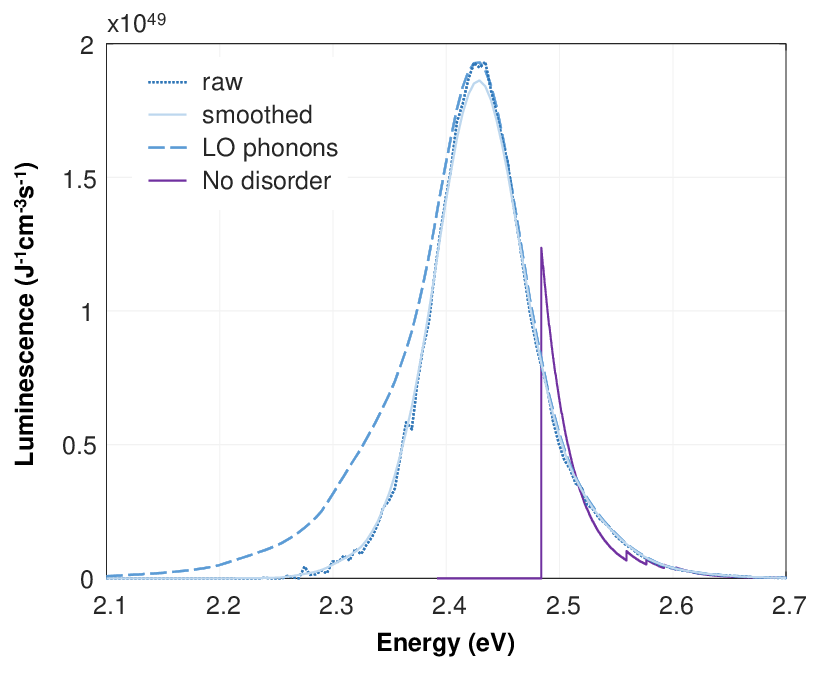}
\caption{\label{fig_pin_lum_1} Modeled luminescence spectrum at various steps: raw value; after smoothing; after addition of LO phonon replicas. The no-disorder spectrum, with peaky features, is also shown.
}
\end{figure}

Fig.~\ref{fig_pin_lum_1} illustrates the stages of the luminescence calculation, and shows how the addition of LO phonons broadens the low-energy tail. For comparison, a luminescence spectrum computed without disorder is also shown: in this case, the steplike features of $\rho_J$ (see Fig.~\ref{fig_pin_rhoo}) gives rise to a series of peaks of width $\sim kT$. Clearly, $L$ is substantially broader in the presence of disorder. When using a disorder-free luminescence model, it is customary to convolve the spectra with an empirical broadening function (e.g. a Lorentzian) to obtain a realistic spectral width \cite{Coldren95b}. Here, in contrast, the spectral shape stems naturally from the disordered states, and no additional broadening is required.

\begin{figure}
\includegraphics[width=8cm]{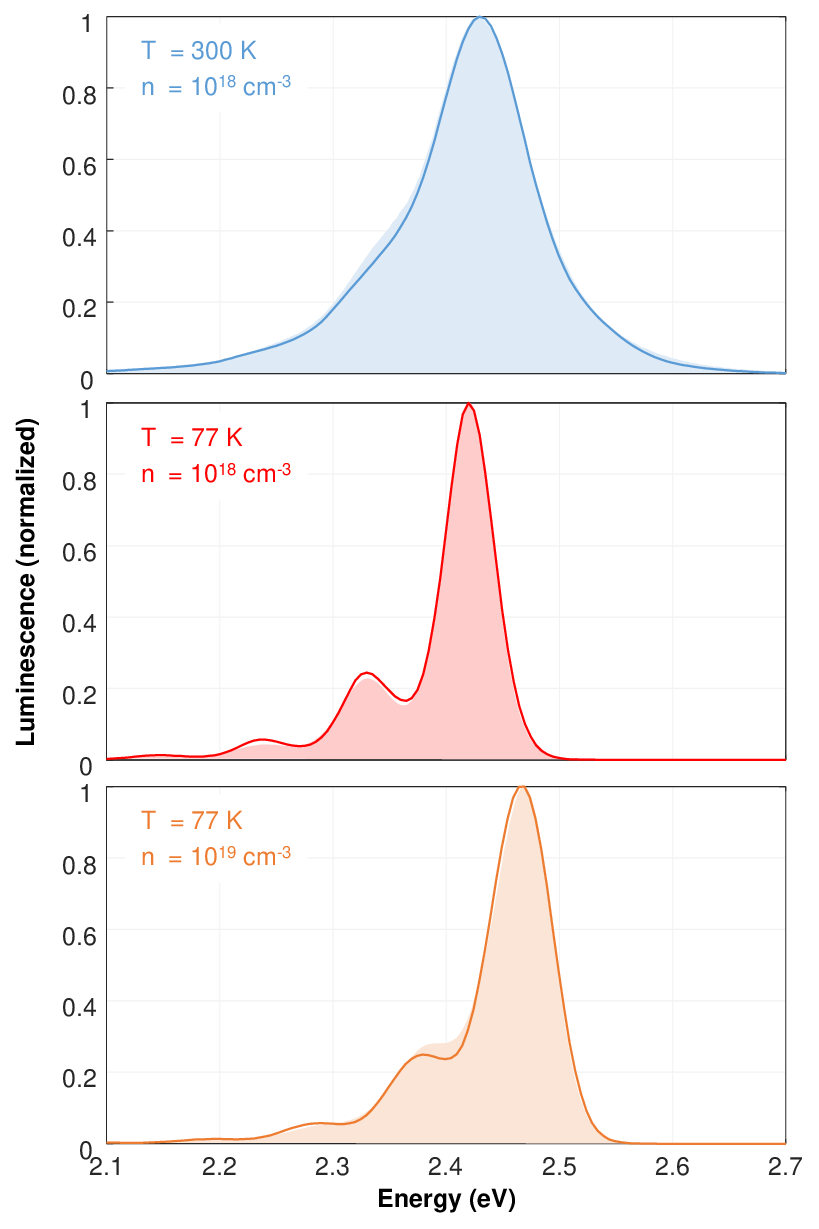}
\caption{\label{fig_pin_lum_2} Luminescence spectra at various temperatures $T$ and carrier densities $n$. Shaded areas: experimental data; solid lines: model.
}
\end{figure}

We now turn to a comparison between the model and experimental data. Fig.~\ref{fig_pin_lum_2} shows PL spectra at 300~K and 77~K. At 77~K, two carrier densities are considered ($n=10^{18}$~cm$^{-3}$ and $10^{19}$~cm$^{-3}$). The experimental carrier densities are extracted from optical carrier lifetime measurements \cite{David17a}. For the modeled spectra, no parameters are adjusted in the simulation; the epitaxial structure follows the nominal thicknesses, and uses the QW composition ([In]=15\%) determined from carefully-calibrated X-ray diffraction measurements.

The model is excellent agreement with the experiment. It accurately predicts the the overall spectral shape, including its width and low and high energy tails. This close agreement validates the accuracy of the model, and confirms that the luminescence lineshape of InGaN is dominated by RA disorder. 

The peak wavelength deserves a few comments. The theoretical spectrum at $300$~K is obtained by modeling the nominal QW design, then rigidly shifted by --50~mV to better align with the experimental data; this small adjustment corresponds to an In variation of $\sim 1\%$, within the experimental accuracy of the QW's composition. 
At low temperature, we apply a bandgap shift of +50~meV from 300K to 77K (consistent with literature \cite{Vurgaftman03}). 
At $n=10^{19}$~cm$^{-3}$, the bandgap is further shifted by --150~meV to account for bandgap renormalization \cite{Pant22,Nagai04}. 
Notwithstanding these small rigid shifts, the lineshape is accurately predicted in all cases, without tweaking any model parameters to improve the fits.

As already mentioned, the low-energy phonon tail is fitted empirically. The same Huang-Rhys parameter $S=0.2$ is used across all calculations. The model accurately predicts the contrast of the phonon-related oscillations, which stems from the overlap of the low-energy tail of the zero-phonon-line and of the high-energy tail of the phonon replicas. 

\begin{figure}
\includegraphics[width=8cm]{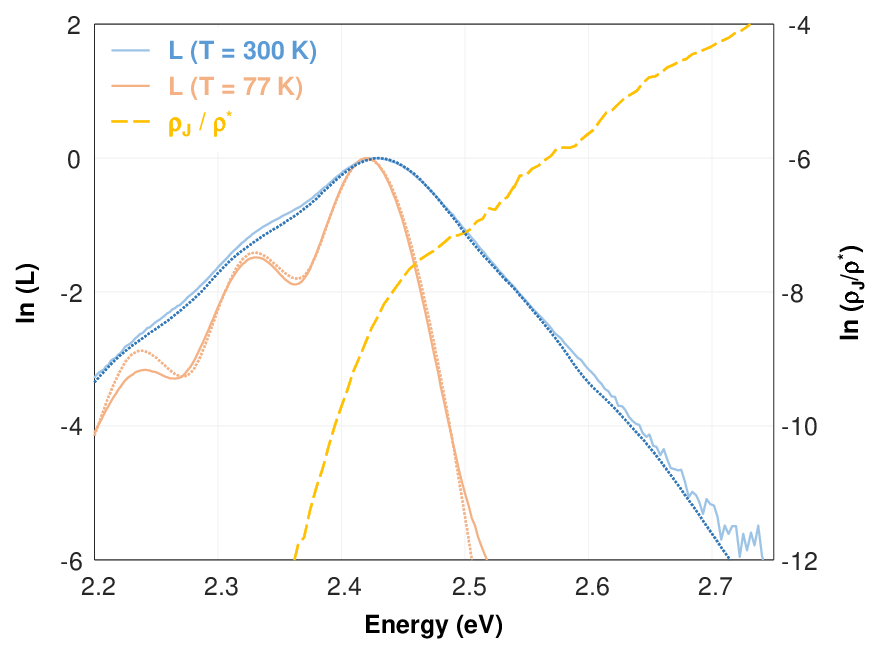}
\caption{\label{fig_pin_Tc} Optical properties of the SQW on a natural logarithmic scale. Solid lines: experimental luminescence at $n=10^{18}$~cm$^{-3}$. Dotted lines: modeled luminescence. These agree both in the low and high-energy tails. Dashed line: modeled JDOS at 300~K (the same rigid shift of --50~meV is applied to the modeled luminescence and JDOS at 300K, to align with the experiment).
}
\end{figure}

Fig.~\ref{fig_pin_Tc} shows the optical properties on a logarithmic scale, displaying excellent agreement between theory and experiment in the low and high-energy tails. This confirms that at high energy, carriers obey a thermal distribution, as will be detailed in the following section. At low energy and low temperature, the zero phonon line of the model matches the experimental spectrum until an energy of 2.35~eV (after which the thermal tail of the first LO phonon replicas dominates over the zero-phonon line). This energy corresponds to a JDOS that is reduced by three decades form the mobility edge. This confirms that the hole distribution does follow a Fermi-Dirac distribution at low energy, at least down to three decades in the Urbach tail. 

The main discrepancy between model and experiment is the detailed spectral shape of the phonon replicas. The model rigidly replicates the zero-phonon line. In contrast, the experimental replicas become progressively broader (this is best seen for the second replica at 77~K). We attribute this to disorder: localized states have a larger electron-hole charge separation than delocalized states and hence a large $S$ factor \cite{Kalliakos02}, which our model ignores. A more accurate treatment, considering an $S$ value for each electron-hole transition, is left as a topic of future work.

Overall, the agreement between model and experiment validates our hypothesis of a thermalized carrier distribution from low-to-high energy. At low energy, the agreement confirms localized carriers deep in the Urbach tail are thermalized. At high energy, the observation of a thermal tail, which can only originate from delocalized states, confirms the existence of a substantial delocalized population.

Our conclusions preclude less-standard carrier distributions, such as strongly-localized distributions into low-energy states: these would affect the luminescence lineshape in a way that is incompatible with experimental data. For instance, it has been proposed that carriers may remain localized in high-energy localized states at low temperature, which could explain the so-called S-shape of emission versus temperature \cite{Cho98}. We do not find evidence of such a behavior at 77~K (though this mechanism remains plausible at lower temperature). It has also been proposed that the spectral width of InGaN LEDs cannot be explained solely by RA disorder, and that additional In clustering has to be assumed for modeling to match experimental data \cite{DiVito19}. We reach the opposite conclusion: at least in our high-quality sample, RA exactly predicts the spectral width. Furthermore, we find no evidence of a steady-state hot carrier population, as will be discussed shortly.

\subsection{Localization and magnitude of luminescence}
\label{sec_disorder_B}

In this section, we scrutinize the impact of disorder on the strength of optical transitions. This is an oft-argued topic. Some works have proposed that the in-plane separation between electrons and localized holes leads to a reduction in the oscillator strength which is detrimental to the IQE~\cite{Aufdermaur16,Pant23}. Other works have arrived at the opposite conclusion \cite{Jones17}. As we will see, disorder actually affects the radiative rate in several ways, with a non-trivial interplay.

First, we explore how localization affects the coupling between hole and electron wavefunctions. To assess this, we define the oscillator strength of a given hole state (with index $p$) to all electron states (with indices $q$):

\begin{equation}
\label{eq_Oh}
O_p = \sum_{q} {\left<\psi_{h,p} | \psi_{e,q} \right>^2}
\end{equation}

$O_p$ has a maximum possible value of 1, e.g. if there were perfect wavefunction overlap between an electron and a hole. Of course, in InGaN QWs, $O_p$ is strongly suppressed by the electron-hole separation along $z$ caused by the polarization fields, and is further affected by in-plane localization. Note that Eq.~\ref{eq_Oh} does not take into account the thermal occupation of states -- it is merely a quantification of overlap effects. Therefore, no conclusions can be drawn immediately on luminescence.
\footnote{More precisely, the sum is computed over all electron states $q$ that are in the ground state along the $z$ direction, i.e. whose wavefunction has only one lobe along $z$. If higher-order states along $z$ were included in the sum, $O_p$ could exceed unity.}

\begin{figure}
\includegraphics[width=8cm]{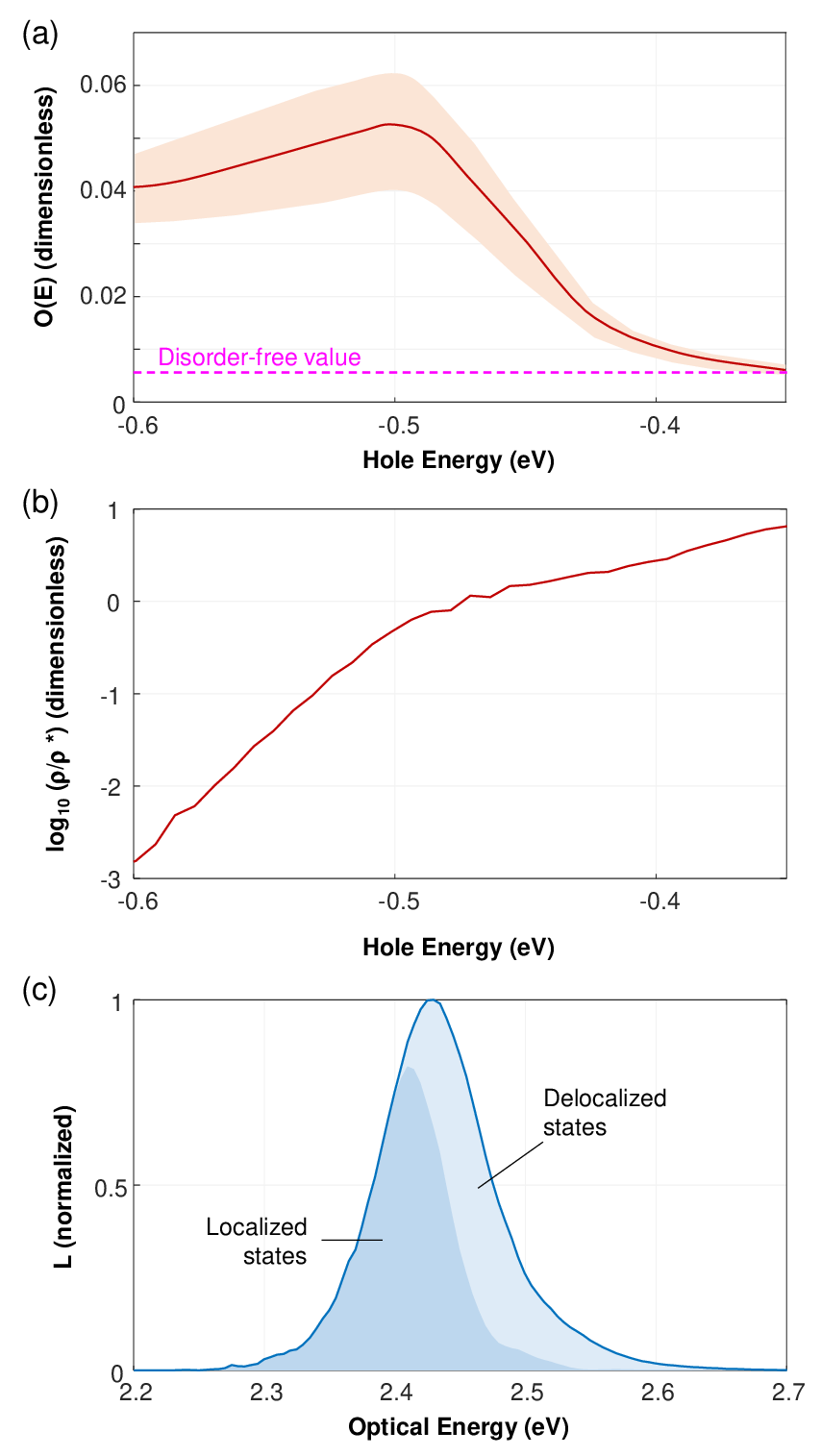}
\caption{\label{fig_pin_opt_strength} Oscillator strength and localization. (a) Statistics of the hole oscillator strength $O(E)$. Solid line: median value; shaded area: 25\%-75\% quartiles. Dashed line: value of $O_{1D}$ in a disorder-free one-dimensional model. $O(E)$ is equal to $O_{1D}$ for high-energy delocalized states, and increases for localized states, reaching a maximum at the mobility edge. (b) Hole DOS (same as Fig.~\ref{fig_pin_rho}(b)). (c) Luminescence spectrum $L$. Dark / light shades: contribution from localized / delocalized hole states, respectively.
}
\end{figure}

We bin $O_p$ as a function of the hole energy, and compute the resulting statistics $O(E)$, which characterize the oscillator strength of the hole population at a given energy $E$. $O(E)$ is show on Fig.~\ref{fig_pin_opt_strength}(a). The hole DOS is shown again on Fig.~\ref{fig_pin_opt_strength}(b), to clarify the nature of hole states versus their energy. At high energy (far enough above the mobility edge), $O$ converges to a value equal to the oscillator strength $O_{1D}$ in the absence of disorder (e.g. $\left<\psi_{h} | \psi_{e} \right>^2$ computed in a conventional one-dimensional model). This result is expected: for delocalized hole states, disorder has little influence on the oscillator strength. 
At lower energy, $O$ increases, reaching a maximum value just below the mobility edge before decreasing slowly in the Urbach tail. This shows that localized hole states have a higher total coupling strength to electrons.

Fig.~\ref{fig_pin_opt_strength}(c) again shows the luminescence spectrum $L$ at 300~K (without phonon tails), and illustrates the contribution of localized hole states. As expected, the low-energy side of $L$ comes from localized hole states. Overall, 60\% of $L$ stems from localized states -- as previously mentioned, these states make up 50\% of the hole population, but their higher oscillator strength leads to a slightly higher contribution to luminescence.

However, the effect of localization on the oscillator strength should not be taken to indicate that disorder increases the radiative rate. Indeed, the luminescence spectrum is proportional to the product of the JDOS (which is affected by the oscillator strength) and of the occupation factor $1/(exp[(E-\Delta E_F)/kT]-1)$ (see Eq.~\ref{eq_L}). Because of the occupation factor, the JDOS near $\Delta E_F$ has a strong influence on the radiative rate. Disorder smears out the JDOS, reducing its value at energies near $\Delta E_F$ (see Fig.~\ref{fig_JDOS_vs_sigma}
in Appendix I-F). Ultimately, this leads to a \textit{reduction} in the total luminescence rate, which can supersede the increase in oscillator strength. This effect depends on the magnitude of the disorder and on the carrier temperature. 

To explore this effect in more detail, we expand our model in order to manipulate the magnitude of disorder in the QW. To do so, the broadening parameter $\sigma$ is varied. A high  value of $\sigma$ mostly averages out RA fluctuations, whereas a small small values exacerbates their effect. Technical details are discussed in Appendix I-F.

For various values of $\sigma$, we compute the Urbach energy $E_U$ which quantifies the magnitude of disorder, and the $B$ coefficient. This calculation takes into account the effect of RA disorder on the oscillator strength and on the occupation factor. The reader is reminded that the present model ignores the Coulomb interaction, and may therefore have limited quantitative accuracy to predict the value of $B$ (in particular, its \textit{absolute} value, which is enhanced by Coulomb interaction). However, we believe the qualitative trends discussed below to be relevant.

\begin{figure}
\includegraphics[width=8cm]{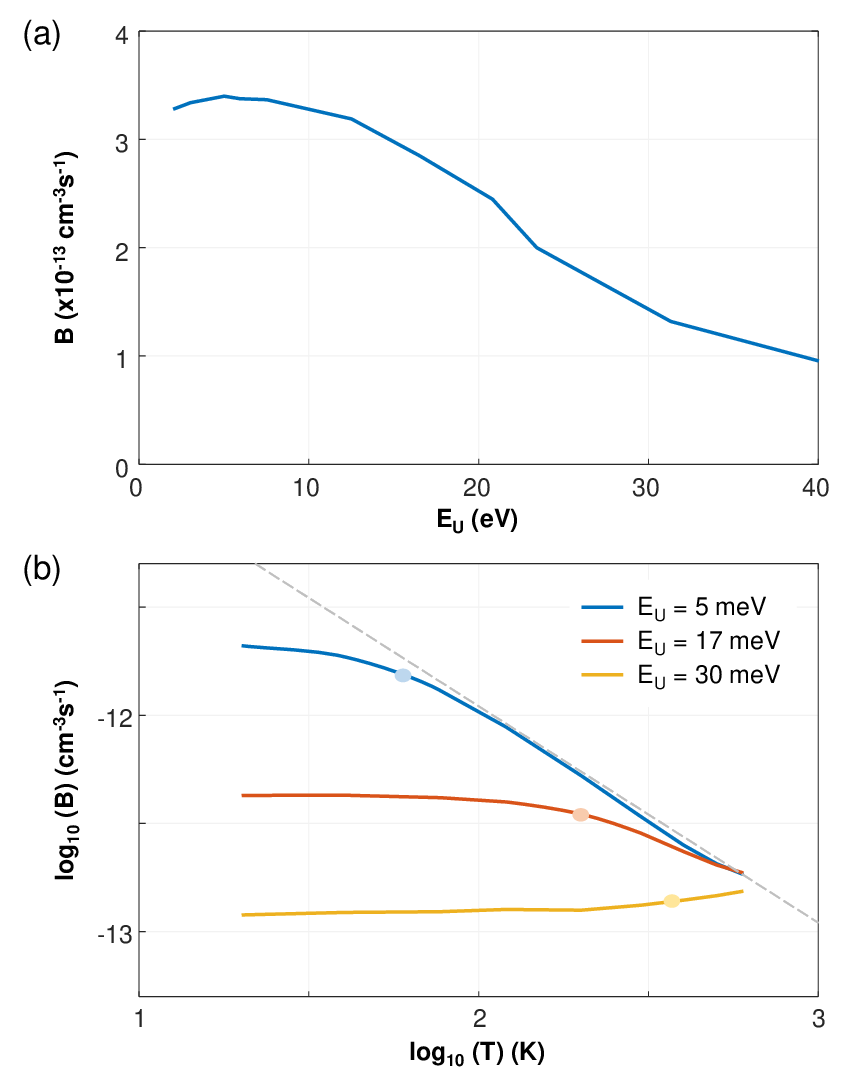}
\caption{\label{fig_Eu_B} Effect of disorder on the radiative coefficient $B$. (a) $B$ as a function of the Urbach energy $E_U$, at $T=300$~K. Disorder decreases $B$. (b) $B$ as a function of temperature $T$, for various values of $E_U$. Dashed line: textbook behavior in a QW: $B \sim 1/T$. The dots denote the temperatures at which $kT = E_U$.
}
\end{figure}

The results are shown in Fig.~\ref{fig_Eu_B}(a). For very small values of $E_U$, $B$ reaches a plateau that characterizes the disorder-free limit. As $E_U$ increases, the dominant effect of disorder is to reduce $B$ due to smearing of the JDOS. 

It is also instructive to consider the temperature-dependence of $B$, show in Fig.~\ref{fig_Eu_B}(b). Two regimes can be discerned. $B$ is temperature-independent (the behavior of a zero-dimensional system) when the thermal energy $kT$ is smaller than the localization energy $E_U$; indeed, in this regime, most of the luminescence stems from localized states whose occupation factor is degenerate, regardless of the temperature. Conversely, when $kT$ is larger than $E_U$, $B$ is proportional to $1/T$ (the behavior of a 2-dimensional system) due to the contribution of delocalized states. For $E_U=17$~meV (the value characterizing the InGaN QW under investigation), this transition occurs around $T=200$~K. Various experimental studies have found that $B$ decreases with increasing temperature, though the reported magnitude of this effect varies \cite{Galler12,Tian14,David19_HC}.

To close this section, we caution that the effect of RA disorder on $B$ should not be equated with its effect on the LED's IQE (a common misconception). Indeed, the IQE results from the competition between radiative and non-radiative effects. Non-radiative recombinations are also modulated by wavefunction overlaps \cite{David20} and by occupation factors, and are therefore likely to be affected by disorder in a way similar to radiative recombinations. For instance, Ref.~\cite{David19_BC} explored wavefunction-overlap effects for the radiative and Auger-Meitner processes, and concluded that $RA$ disorder did not affect their \textit{relative magnitude}. A quantitative model of the effect of disorder on various recombination types would be necessary to assess the net effect on IQE.

\subsection{Carrier temperature}

The carrier temperature $T_c$ of an electron-hole plasma is customarily extracted from the exponential tail of luminescence spectrum. In III-nitrides, some works have reported a high $T_c$ that exceeds the lattice temperature by 100~K or more. This has been interpreted as a sign of incomplete carrier thermalization, due to a bottleneck in relaxation processes \cite{Kudo02,Shen07,Keppens10,Han20}.

In such works, the extraction of $T_c$ is based on the assumption that the JDOS is constant or slowly-varying (e.g. $\sim \sqrt{E}$) in the high-energy luminescence tail, so that the luminescence is simply proportional to the Boltzmann occupation $e^{-E/kTc}$. This would be valid in a square and ordered QW. In InGaN however, the JDOS shows a slow exponential increase, even at high energy, i.e. $\rho_J\sim~e^{aE}$. This is due to the large number of transitions enabled by the QCSE, and their smearing by disorder. As a consequence, the luminescence tail actually decays like $e^{-E(1/kT_c-a)}$, leading to an overestimated \textit{apparent} carrier temperature if a naive exponential fit is used.

Fig.~\ref{fig_pin_Tc} illustrates this. The modeled $\rho_J$ grows exponentially at high energy, and therefore affects the exponential decay of the luminescence tails. A direct Boltzmann fit of the experimental tails, ignoring the contribution from $\rho_J$, would lead to artificially-hot apparent temperatures: respectively $T_c=110$~K and 500~K. In contrast, to best fit the tails with the luminescence model, we used carrier temperatures $T_c=77$~K and 320~K. The latter temperature is slightly above the crystal temperature, but rather than interpreting this as a slight hot-carrier effect, we attribute it to the model’s imperfect prediction of $\rho_J$ just above the valence band edge (i.e. the use of a single-band effective mass model ignores the fine structure of the valence band in this energy range), which can easily affect $T_c$ by a few tens of K. 

Although the data of Fig.~\ref{fig_pin_Tc} is at an intermediate carrier density, the experimental thermal tail (and in fact, the overall lineshape) remains the same down to $10^{17}$~cm$^{-3}$, indicating that thermalization occurs even at low density.

In summary, a proper extraction of the carrier temperature requires an accurate knowledge of the JDOS. This results in realistic carrier temperatures, nearly equal to the crystal temperatures. We believe sporadic literature reports of steady-state hot carriers are caused by an improper extraction of $T_c$. Indeed, the thermalization dynamics are known to occur on a picosecond time scale, as shown by ultrafast spectroscopy \cite{Sun97,Binder13b} (note that such time-resolved measurements can confirm the thermalization dynamics; however, when it comes to extracting the carrier temperature, they suffer from the same challenge as steady-state measurements: an accurate knowledge of the high-energy JDOS is necessary).

\subsection{\label{sec:pin_Stokes}Stokes Shift}

\begin{figure}
\includegraphics[width=8cm]{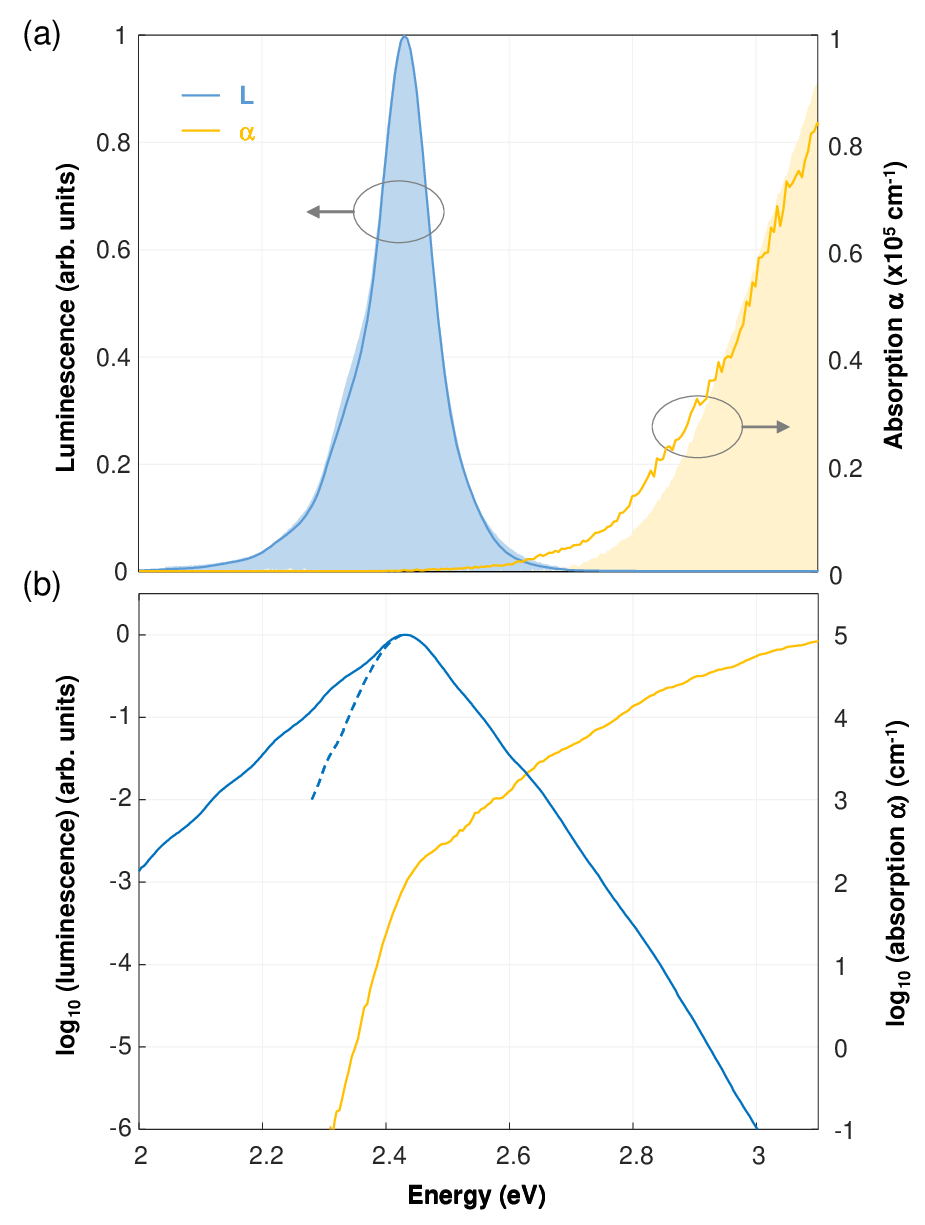}
\caption{\label{fig_pin_Stokes} Stokes shift. (a) Comparison of the experimental luminescence and absorption (shaded areas) and modeled counterparts (solid lines). The absorption is in absolute units. (b) Modeled luminescence (solid/dashed line: with/without LO phonon tails) and absorption on the log scale, showing that the states responsible for luminescence are on the cusp of the mobility edge, far away from the 'bulk-like' absorption edge.
}
\end{figure}

III-Nitrides stand out for the large magnitude of their Stokes shift (SS), i.e.~the energy shift between the absorption band edge and the luminescence spectrum. While it is generally accepted that this is related to disorder, there is no consensus on a quantitative understanding of the Stokes shift, or the detailed microscopic mechanisms that underlie it.

Early on, various qualitative models were introduced. For instance, Refs.~\cite{Odonnell99,Martin99} observed that the SS varies linearly with the emission energy, reaching very high values (300~meV for a blue LED), and proposed that ``the localization of excitons increases linearly with decreasing emission energy in InGaN structures''. Likewise, Ref.~\cite{Chichibu99} reported a SS $>200$~meV for blue structures, and attributed this to in-plane localization in disk-like regions. Meanwhile, Ref.~\cite{Shan98} proposed that electron are becoming strongly localized at room temperature. 

While these interpretations were plausible at the time they were published, they now seem to be in conflict with the improved understanding of disorder-related effects that has emerged in the last decade. For instance, Ref.~\cite{Piccardo17} showed experimentally that the Urbach tail (which characterizes the ‘strength’ of the disordered potential) does not vary much for violet-to-green LEDs, with values remaining in the range 20-30~meV (see Section~\ref{sec:MQW}). Besides, the theoretical binding energy of excitons in InGaN QWs is only 20-40~meV \cite{David22}, and exciton binding is the weakest in thick QWs, where the SS is most pronounced. In summary, the localization potential and excitonic binding have characteristic energies an order of magnitude smaller than the SS, and their composition-dependence does not match that of the SS -- all findings at odds with the aforementioned  models for the SS.

Actually, as we will show, reports of a large SS are largely a matter of semantics -- namely, of the way the numerical value of the SS is defined.

Fig.~\ref{fig_pin_Stokes}(a) illustrates the experimental Stokes shift for the SQW sample. The luminescence spectrum is the same as in other figures ($T=300$~K, $n=10^{18}$~cm$^{-3}$). The QW absorption $\alpha$ is measured by placing the sample in an integrating sphere and illuminating it with a collimated white-light beam. This measurement is accurate down to an absorption signal of ~0.1\%, i.e. $\alpha \sim 10^3$~cm$^{-1}$ (therefore, it does not give access to the Urbach tail, for which photocurrent measurements would be necessary). Note that the absolute absorption is shown here. Often in literature, $\alpha$ is show in arbitrary units, which is ill-suited to study the SS quantitatively.

The modeled luminescence and absorption shown on Fig.~\ref{fig_pin_Stokes}(a) display an excellent agreement with the experimental data. The absolute magnitude of $\alpha$ is well matched, and the energy shift between $\alpha$ and $L$ is accurately predicted. The details of $\alpha$ around 2.6-2.8~eV are imperfectly predicted; we attribute this to the inaccurate description of the valence band structure near the band edge by a two-band model. Note, in order to calculate $\alpha$ up to high energy, we computed 6,000 electron states and 12,000 hole states -- about 30 times more levels than for other calculations.

\begin{figure}
\includegraphics[width=8cm]{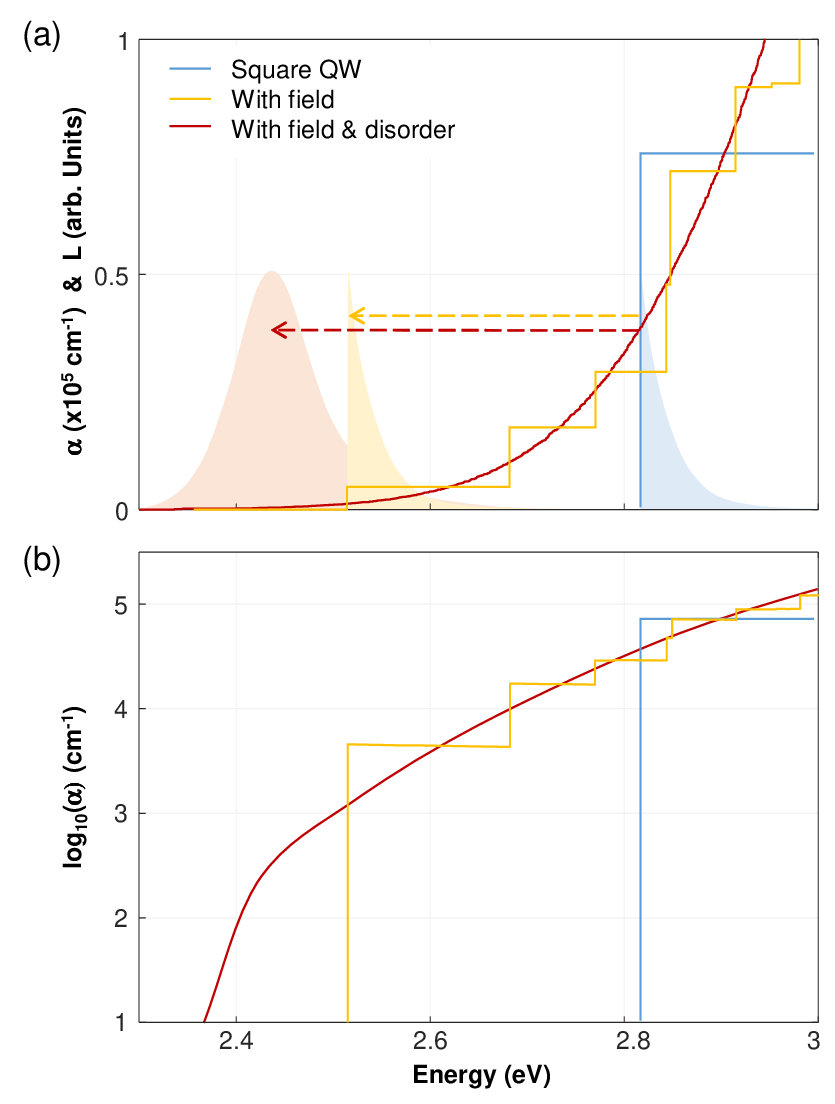}
\caption{\label{fig_pin_Stokes_toy} Qualitative sketch of the origin of the SS. (a) $\alpha$ (solid lines) and $L$ (shades) in linear scale. (b) $\alpha$ in log scale. For a square QW, the onset of $\alpha$ and $L$ is at the same energy and the SS is zero. In the presence of fields, $\alpha$ displays many steps, and $L$ arises at the lowest-energy transition. With disorder, an Urbach tail appears and $L$ shifts to lower energy. The dashed arrows show the SS measured from $\alpha \sim 5\times 10^4$~cm$^{-1}$.
}
\end{figure}

We now clarify how the SS arises from field-induced QCSE and disorder. Fig.~\ref{fig_pin_Stokes_toy} qualitatively illustrates $\alpha$ and $L$ in three cases: in a square flat-band QW, when polarization fields are present, and when disorder is further included. For a square QW, $\alpha$ is a step function (the transition between the first quantized $e$ and $h$ levels), the onset of $\alpha$ and $L$ coincides, and the SS is zero. In the presence of fields, the QW become triangular; the resulting QCSE shifts states to lower energies, and multiple electron-hole transitions become allowed with a reduced oscillator strength, leading to an irregular steplike $\alpha$. The onset of $L$ is again at the first $e-h$ transition, whose energy can be hundreds of meV below the point where $\alpha$ reaches a bulk-like value ($\sim 10^5$~cm$^{-1}$). Finally, disorder broadens the step-like $\alpha$ and creates an Urbach tail which further shifts $L$ to low energy.


In summary, the SS in III-nitrides is simply caused by a combination of the QCSE (a contribution already pointed out in Ref.~\cite{Berkowicz00}) and of disorder, which create enough JDOS at low energy to give rise to luminescence. There is no need to invoke more complex processes, such as exciton localization. 

Despite this straightforward explanation, many works focus on the surprisingly-high \textit{numerical value} of the SS. As we will now argue, such a numerical value is somewhat ill-defined, as it relies on extracting an arbitrary ``effective bandgap'' from $\alpha$ (typically, by fitting it with a sigmoid function or a Tauc fit).

Indeed, both Figs.~\ref{fig_pin_Stokes}(b) and \ref{fig_pin_Stokes_toy}(b) show the absorption on a logarithmic scale, and reveal that in the presence of disorder, $L$ stems from states that are on the cusp of the mobility edge, corresponding to $\alpha \sim$~100~cm$^{-1}$. 

In contrast, the effective bandgap derived from a sigmoid fit $\alpha \sim 1 /(1+ e^{(E-E_e)/\Delta} )$ characterizes $\alpha$ around 10$^5$~cm$^{-1}$, and yields $E_e = 3$~eV and $\Delta=100$~meV. Commonly, $E_e$ is interpreted as an effective bandgap, and $\Delta$ as an Urbach energy. However, such labels are misleading. $E_e$ characterizes an energy where $\alpha$ is very high, corresponding to a bulk-like JDOS much higher than is needed to lead to luminescence; therefore $E_e$ is not an especially relevant quantity to understand luminescence. Likewise, $\Delta$ characterizes the broadening of $\alpha$ near $E_e$. This broadening is tied to the QCSE (i.e. to the existence of multiple transitions with low oscillator strength that cause a staircase-like $\rho_J$). It is not related to alloy disorder, and does not characterize the disorder-induced Urbach tail at low energy. $\Delta$ typically reaches values as high as 100-150~meV for blue-green structures \cite{Martin99}, whereas the Urbach energy has a much lower value of 20-30~meV in the same spectral range \cite{Piccardo17}. In short, the fitting parameters of a sigmoid fit characterize $\alpha$ in an high-energy / high-JDOS region that is not relevant for luminescence.

We have also considered Tauc fits as an alternative to sigmoid fits, as suggested in Ref.~\cite{Weisbuch21}. However, such fits suffer from similar flaws, and the added ambiguity of the choice of the Tauc exponent. 

In summary, the large reported value of the SS arises from the fact that luminescence is centered around the Urbach tail, whereas the bandgap energy is commonly derived from fitting standard absorption measurements that capture the ‘bulk-like’ absorption at much higher density of states. Arguably, the relevant feature of $\alpha$ for the onset of luminescence is the optical mobility edge (at $2.5$~eV in the present sample), which plays the role of an effective band edge for optical transitions. Defining the SS as the energy difference between the mobility edge and the luminescence peak would indeed result in small SS values (a few tens of meV), and would be more physically meaningful than the conventional definition.



\section{\label{sec:MQW}Multiple-QW LEDs in the violet-green range}

\begin{figure}
\includegraphics[width=8cm]{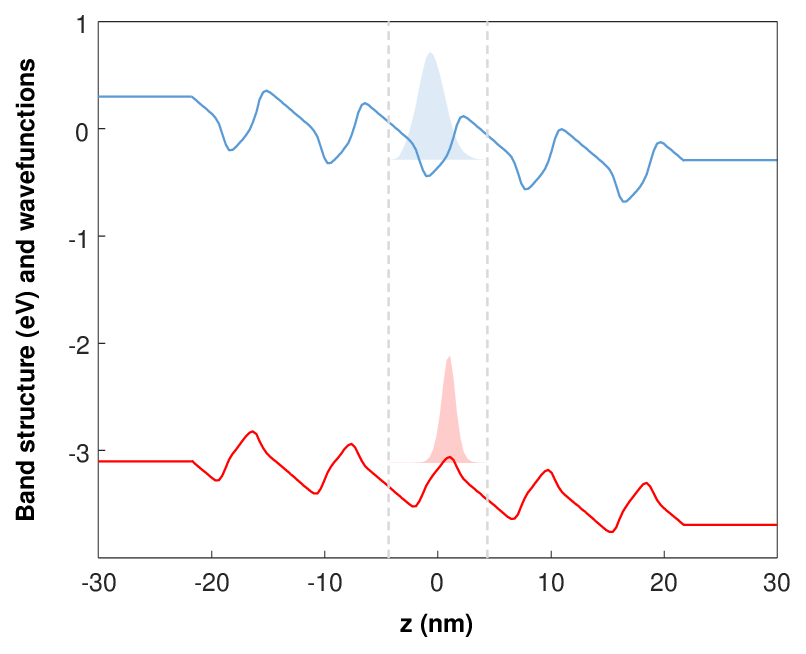}
\caption{\label{fig_MQW_bandstruct} Band structure of the MQW LED (for [In]=15\%). Shaded areas: ground state wavefunctions (integrated along $(x,y)$) for the center QW. Dashed lines: boundaries of the simulation domain along $z$. 
}
\end{figure}

Section~\ref{sec:pin} dealt with a single-QW sample. This Section considers more typical multiple-QW LED structures, with varying In content, to reassess some of the trends discussed in the previous Section. The active region includes five QWs (thickness 2.7~nm) separated by GaN barriers (thickness 6~nm). These values are representative of commercial designs. For simplicity, we assume an abrupt junction that drops only across the active region -- this assumption barely influence the results. Fig.~\ref{fig_MQW_bandstruct} shows the resulting LED band structure. As it is identical for all the QWs, we model only one QW. Of course, our model does not deal with transport effects (an important issue, where disorder also plays a role~\cite{Yang14}). Rather, we assume an applied voltage (2.8~V) and assume charge neutrality ($n_e=n_h$) in the QW to study its optical properties.

\begin{figure}
\includegraphics[width=8cm]{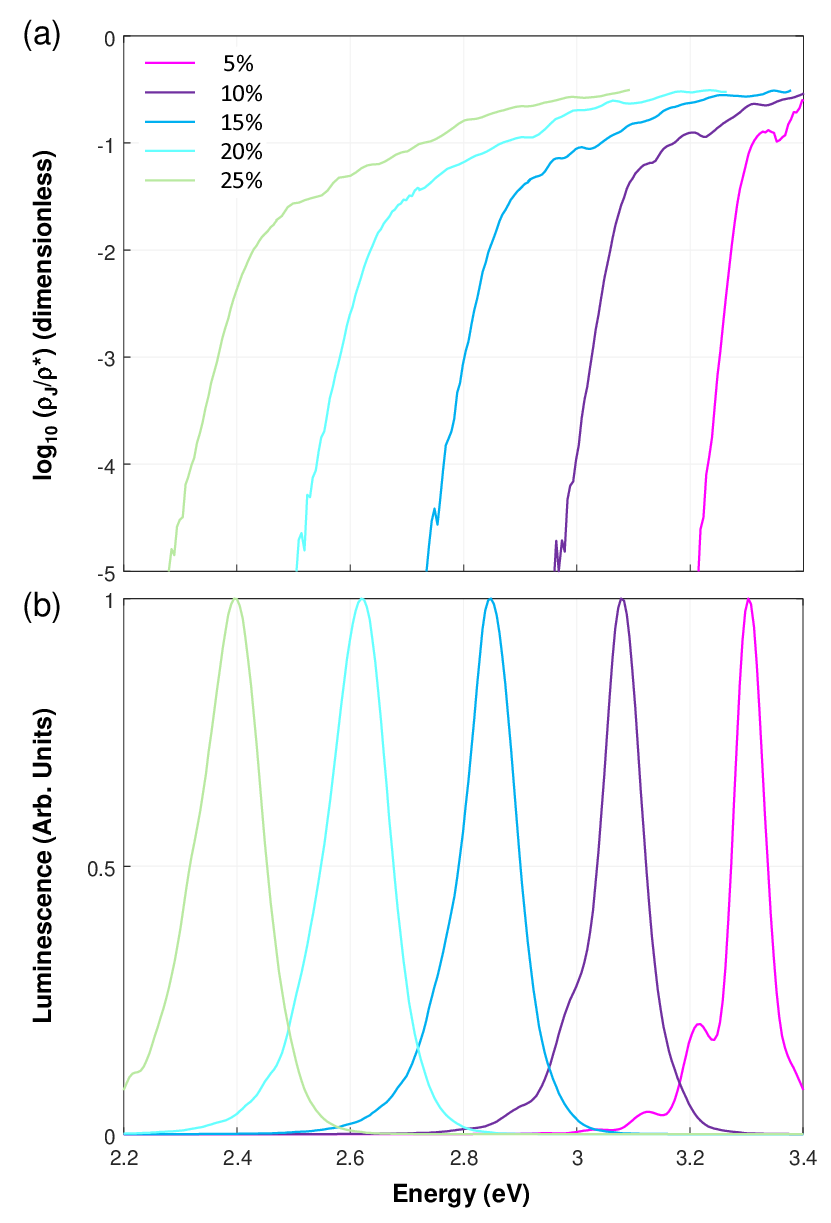}
\caption{\label{fig_MQW_rho_lum} Modeled optical properties of MQW LEDs versus [In]. (a) JDOS. (b) Luminescence spectra. 
}
\end{figure}

Fig.~\ref{fig_MQW_rho_lum}(a-b) shows $\rho_J$ and $L$ for QW compositions [In]~=~5--25\%, spanning the near-UV to green wavelength range. The Urbach tail of $\rho_J$ slowly becomes stronger as [In] increases. The luminescence spectra peak near the cusp of $\rho_J$, and become broader as [In] increases.

\begin{figure}
\includegraphics[width=8cm]{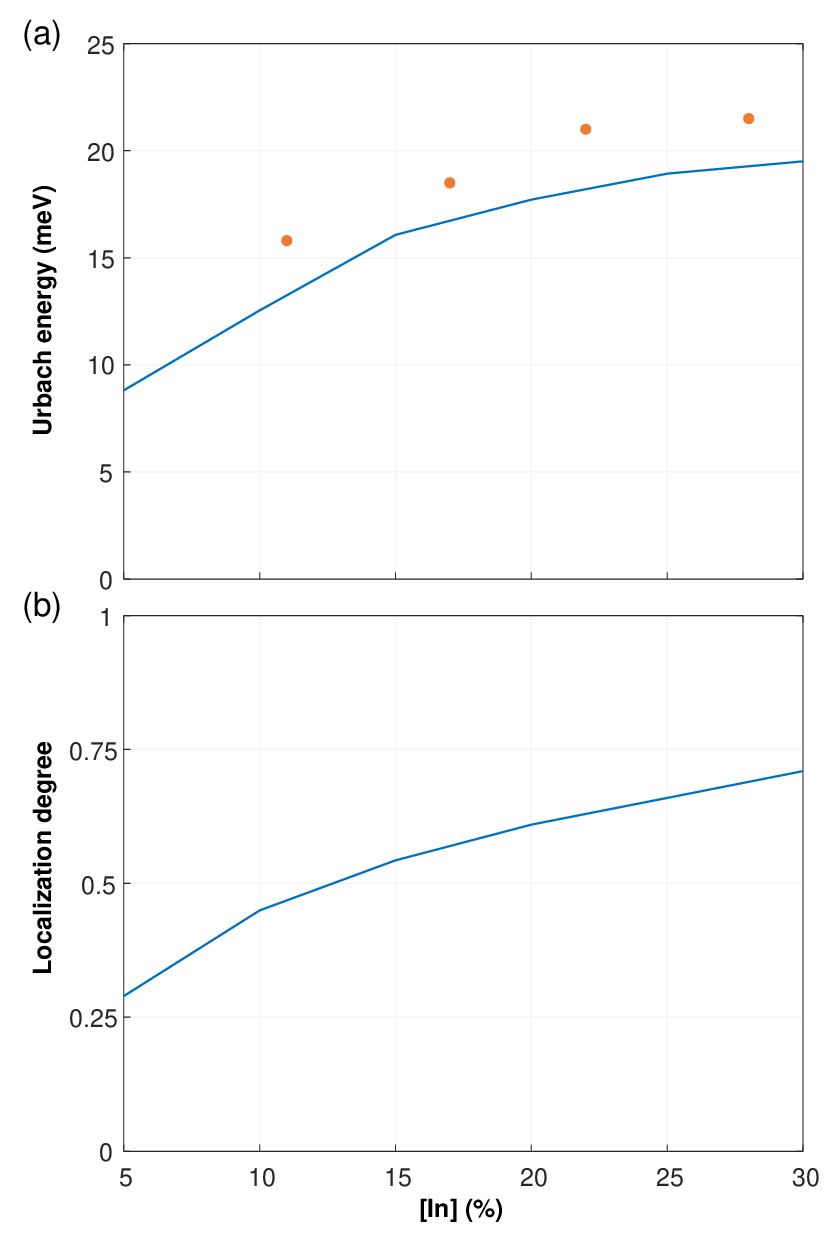}
\caption{\label{fig_MQW_Urbach} Modeled localization properties of MQW LEDs versus [In].(a) Urbach energy. Solid line: numerical model. Dots: experimental data from \cite{Piccardo17}. (b) Localization degree at 300~K.
}
\end{figure}

Fig.~\ref{fig_MQW_Urbach}(a) shows $E_U$ versus [In], and compares the model predictions to the experimental data of Ref.~\cite{Piccardo17}. The overall trend is in good agreement with the experiment, with the numerical model under-estimating $E_U$ by about 2~meVs. The most notable result is that $E_U$ varies only weakly with [In], despite a significant change in the depth of the confinement potential caused by InGaN fluctuation. This effect was already explained in \cite{Piccardo17}: each In-rich region only supports one localized quantum state, regardless of [In], leading to a weak variation in $E_u$ caused only by the confinement strength of the quantum level.

Accordingly, the degree of localization $d$ at 300~K (shown in Fig.~\ref{fig_MQW_Urbach}(b)) only depends weakly on the wavelength: $d$ increases steadily with [In], but a significant fraction of the carriers (tens of \%) remain  delocalized for all In compositions. We conclude that the existence of a macroscopic delocalized population is a general property of InGaN LEDs at room temperature. 

It has sometimes been proposed that localization effects are much more pronounced in longer-wavelength emitters, which could be an explanation for the green gap (e.g. if localization enhances non-radiative recombinations) \cite{Aufdermaur16,Nippert16b,Karpov18}. In contrast, we find that the degree of localization only increases moderately from blue to green LEDs (namely, from 50\% to 70\%), suggesting that localization effects are not a key element of the green gap. Our conclusions concur with those of another theoretical study~\cite{Tanner20b}.

\begin{figure}
\includegraphics[width=8cm]{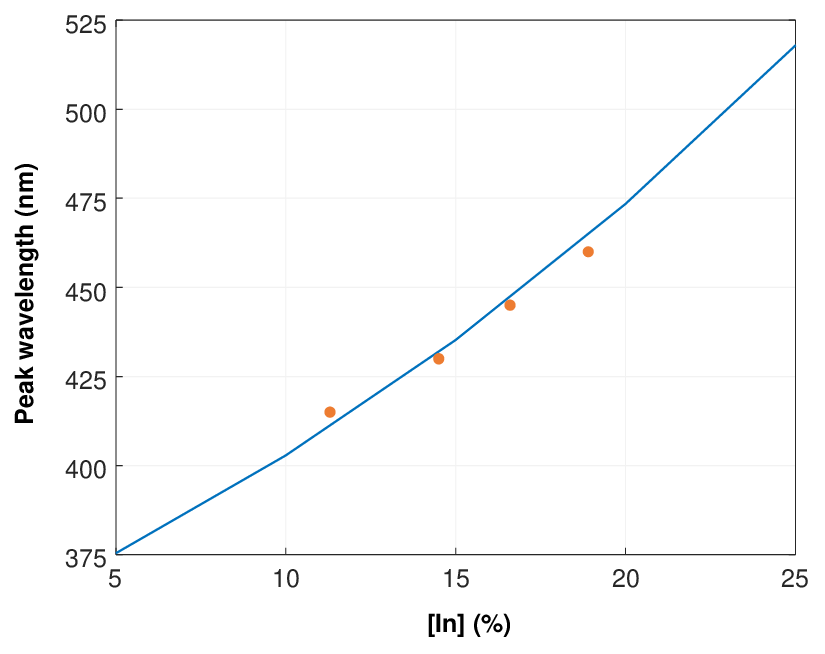}
\caption{\label{fig_MQW_In_wl} Peak wavelength vs [In]. Dots: experimental data from \cite{David10b}.
}
\end{figure}

Fig.~\ref{fig_MQW_In_wl} shows the relationship between [In] and the peak emission wavelength $\lambda$. For comparison to experiments, we use data from Ref.~\cite{David10b}, which considered LEDs with a similar MQW design, and provides X-ray diffraction estimates of the QW composition. The agreement between model and experiment is good. We note, however, that a simple model without disorder could also reproduce the slope of $\lambda$ versus [In], where disorder does not play a key role.


\begin{figure}
\includegraphics[width=8cm]{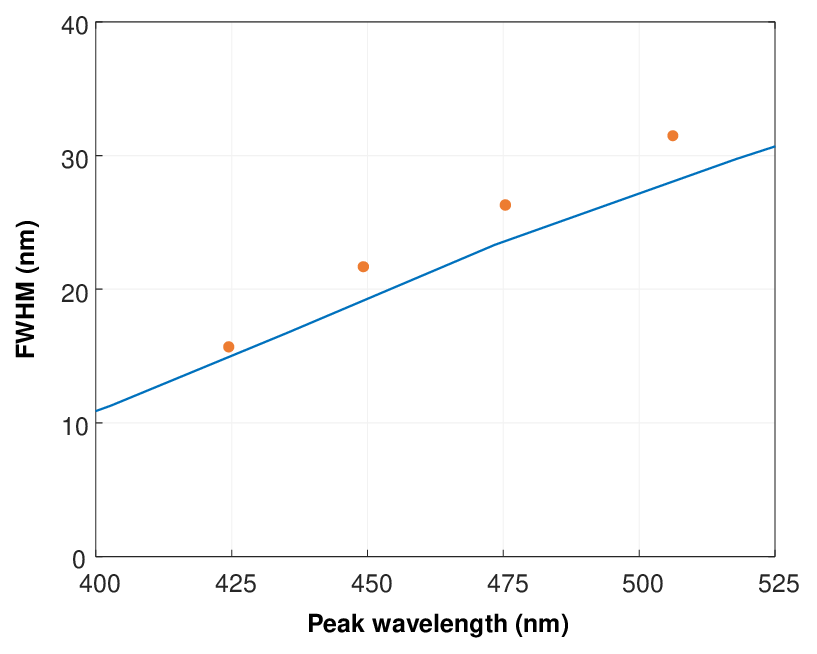}
\caption{\label{fig_MQW_FWHM} Modeled FWHM of the luminescence vs [In]. Dots: experimental data of commercial LEDs from \cite{LumiledsLuxeonC}.
}
\end{figure}

Fig.~\ref{fig_MQW_FWHM} shows the relationship between the emission wavelength and the FWHM. In contrast to $\lambda$, the FWHM cannot be obtained with a no-disorder model. For comparison with experimental data, the FWHM of commercial Lumileds LEDs was obtained from Ref.~\cite{LumiledsLuxeonC}. The model shows rather good agreement, the experimental FWHM being slightly broader. This could stem from the existence of additional inhomogeneous broadening in real-world long-wavelength LEDs. In addition, we kept the same Huang-Rhys parameter at all wavelengths for simplicity (S=0.2), although at long wavelength, S is know to increase, which should slightly broaden the FWHM. Regardless, the key result is that the majority of the spectral width is caused by disorder.

\begin{figure}
\includegraphics[width=8cm]{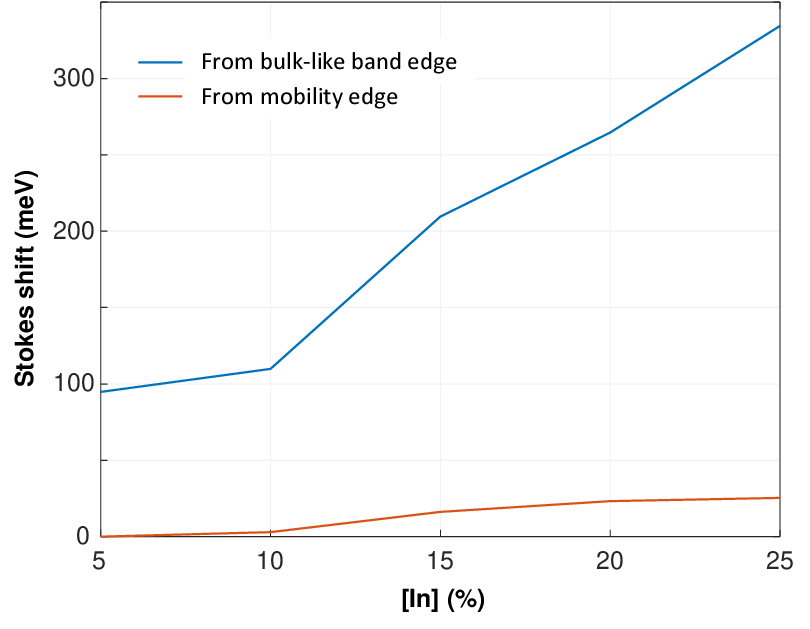}
\caption{\label{fig_MQW_Stokes} Stokes shift vs [In]. The conventional definition yields a high value, whereas referring the shift from the mobility edge leads to small values.
}
\end{figure}

Finally, Fig.~\ref{fig_MQW_Stokes} shows the Stokes shift versus [In]. As discussed in Section~\ref{sec:pin_Stokes}, the numerical value of the SS highly depends on the definition of the bandgap energy. We compare the common definition (using a sigmoid fit of the bulk-like absorption), and an alternative definition (using the mobility edge, i.e. the cusp of $\rho_J$). In the former case, the SS reaches hundreds of meV, as reported in literature and sometimes wrongly attributed to deeply localized excitons at In clusters. In contrast, the alternative definition leads to modest values, commensurate with the Urbach energy. Again, we conclude that the SS simply stems from RA disorder and does not exhibit an abnormal behavior.

As a closing remark, we note that the wavelength of an LED can be controlled through both the QW composition and its thickness. Above, we only considered composition tuning. We have checked that, for a given composition, variations in thickness have no effect on the Urbach tail. Therefore, given a desired emission wavelength, it is possible to engineer the value of the Urbach tail (i.e. the strength of disorder effects) by jointly tuning the QW thickness and composition.



\section{\label{sec:long}Red LEDs}

This section focuses on red InGaN LEDs, which have attracted significant interest in the last few years, in particular due to their potential use in display applications. Even though it is challenging to develop high-efficiency long-wavelength InGaN emitters, there has been remarkable progress, with peak EQEs above 10\% demonstrated e.g. by Nanchang University~\cite{Zhang20} and Lumileds~\cite{Armitage24}.

In this Section, we first show how the model reproduces some of the unexpected experimental properties of these LEDs. We then investigate the underlying physics.

\subsection{Comparison to experiment}

Red InGaN LEDs stand out for their peculiar optical properties. For illustration, we focus on the data published in Ref.~\cite{Iida22}, which demonstrates efficient single-QW red LEDs and provides experimental details including the epitaxial structure.

\begin{figure}
\includegraphics[width=8cm]{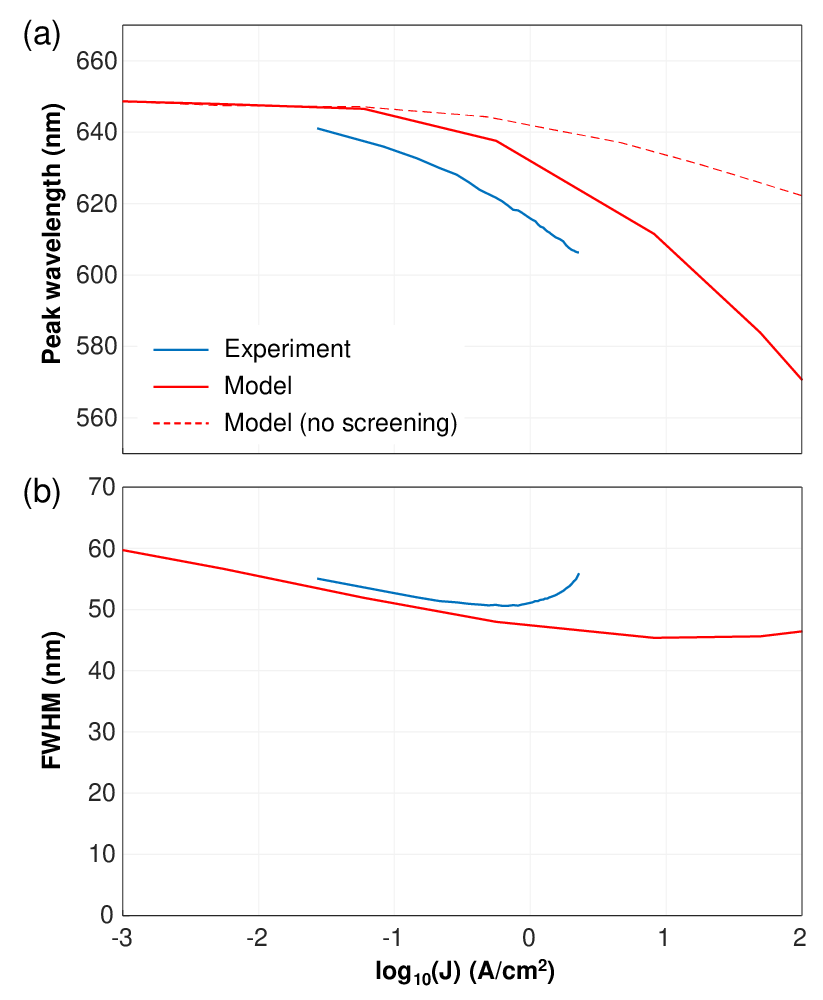}
\caption{\label{fig_red_Iida} Optical properties of red InGaN LED: experimental data (from \cite{Iida22}) and model predictions. (a) Wavelength versus current density. Dashed line: model in the absence of field screening, showing the contribution from band-filling effects only. (b) FWHM versus current density. In the experimental data, the broadening above 1~A/cm$^2$ is caused by carrier heating, which is not considered in the model.}
\end{figure}

Multiple works have reported a large wavelength shift with current, which already occurs at current densities below 1~A/cm$^2$ (well below the onset of polarization field screening in conventional InGaN LEDs); the origin for this shift has not been explained quantitatively. Fig.~\ref{fig_red_Iida}(a) shows the wavelength shift reported in Ref.~\cite{Iida22}. Shifts of similar magnitude are reported in other works \cite{Hwang14,Zhang20}. Note that, for comparison to the model, the contribution of non-radiative recombinations to $J$ was removed before plotting the experimental data of Ref.~\cite{Iida22}.~\footnote{Specifically, the plotted current density is the radiative current density $J_{rad}$, calculated from $J_rad=J \times IQE$.}

In addition, red InGaN spectra are characterized by a wide FWHM (50~nm or more), and by a remarkable FWHM \textit{increase} at low current. In a textbook luminescence model, carriers obey Boltzmann statistics at low current; hence the relative level occupation, and the luminescence lineshape, are not carrier-dependent. Therefore, the experimental FWHM increase is a non-trivial behavior. Like the wavelength shift, this trend appears widespread in red InGaN LEDs \cite{Funato06,Hwang14,Muyeed21}. In fact, it can also be observed, to a weaker extent, in some green InGaN LEDs \cite{Funato06,Lv18}. Fig.~\ref{fig_red_Iida}(b) shows the FWHM data from Ref.~\cite{Iida22} (there is also a FWHM increase at high current -- this is merely due to heating and is not surprising). Ref.~\cite{Iida22} attributed this trend to the filling of localized states. Our model confirms this interpretation, as will now be shown.

\begin{figure}
\includegraphics[width=8cm]{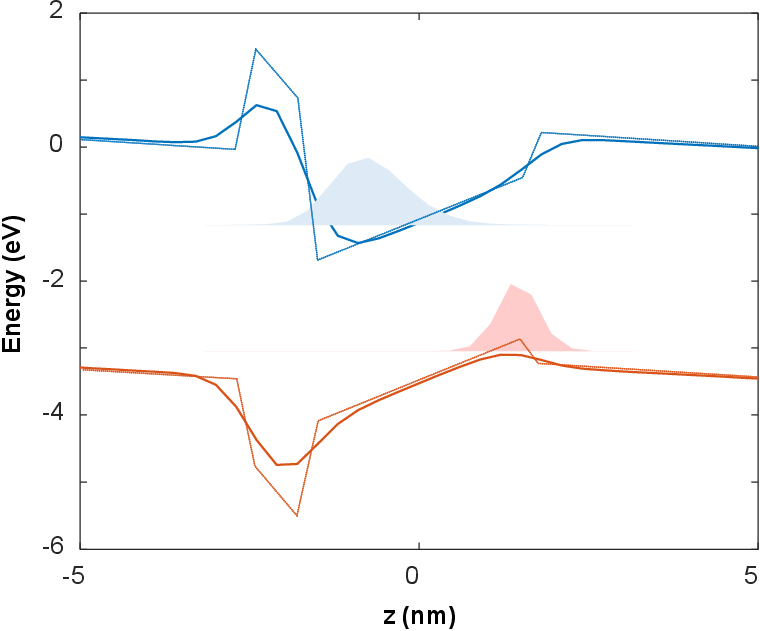}
\caption{\label{fig_red_bandstruct} Modeled band diagram. The dark/light lines show the band structure with/without alloy disorder, respectively. The shades are examples of ground state wavefunctions. All quantities are averaged over $(x, y)$.
}
\end{figure}

\begin{figure}
\includegraphics[width=8cm]{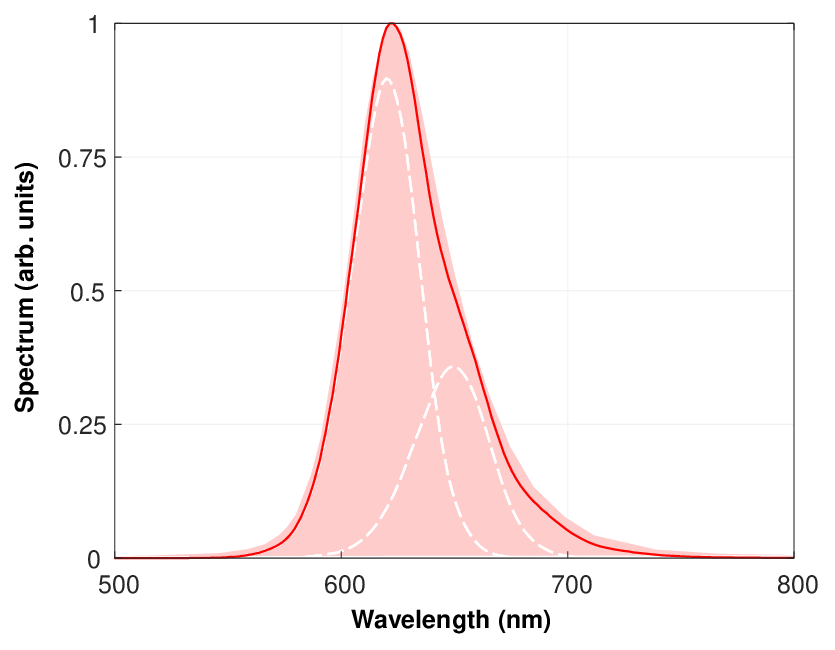}
\caption{\label{fig_red_spectrum} Emission spectrum. Shaded area: experimental data from Ref.~\cite{Iida22}. Solid line: model. Dashed lines: contributions of the zero-phonon-line and the first LO phonon replica to the modeled spectrum.
}
\end{figure}

Red InGaN LEDs are more challenging to fit than standard blue-green LEDs. This is at least in part due to a lack of details in the epitaxy of published structures (for instance, Ref.~\cite{Iida22} does not mention [In] in their red-emitting QWs; we use [In]=30\%, similar to values found in their other works \cite{Hsiao23}). Besides, band structure parameters are less accurate at high [In], as many of these parameters stem from experimental data on shorter-wavelength structures.  

To achieve good quantitative agreement with the data of Ref.~\cite{Iida22}, we slightly modified the value of the InGaN bowing parameter from from 1.4~eV (a typical value suggested in \cite{Vurgaftman03}) to 1~eV. This smaller value remains in the accuracy range of published experimental data. We assume a voltage drop of 1.5V (corresponding to an applied voltage of 2V) across a 30~nm thick depletion region (in line with the thickness of the undoped layers). Fig.~\ref{fig_red_bandstruct} shows the QW band structure.

Fig.~\ref{fig_red_Iida} shows the modeled wavelength and FWHM. They are in very good agreement with the experimental data. In particular, the model predicts the onset of wavelength shift at low current density, and the broadening of the FWHM at low current density.

Fig.~\ref{fig_red_spectrum} compares the experimental and theoretical spectrum at a current density of 10~A/cm$^2$. They are in very good agreement. Note that, for a good match of the low-energy tail, we adopted a larger Huang-Rhys parameter than for blue-green LEDs ($S=0.4$ instead of 0.2). This value is in very good agreement with previous experimental investigations of the value of S, which increases at longer wavelength \cite{Kalliakos02,Graham05}.

\subsection{Underlying physics}

Beyond achieving a fit to experimental data, the model provides insight into the mechanisms that cause these optical properties. 

The wavelength shift has two origins. The largest contribution is the screening of the QCSE, with an additional contribution from band filling. Fig.~\ref{fig_red_Iida} shows the breakdown between these two contributions. The shift due to QCSE screening is pronounced. This stems from the strong field value (4~MV/cm before screening) and the use of a rather thick QW (3.6~nm). 

\begin{figure}
\includegraphics[width=8cm]{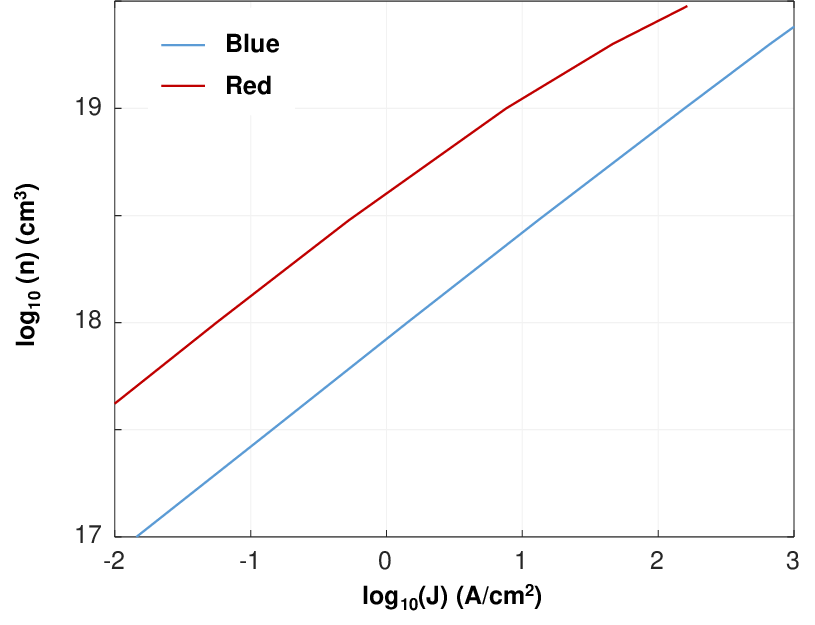}
\caption{\label{fig_red_nJ} Relationship between current density $J$ and carrier density $n$ for a standard blue LED (with a 2.7~nm thick QW) and the red structure. The slower recombination rate in the red structure leads to a higher $n$ at the same $J$.
}
\end{figure}

Interestingly, the onset of QCSE screening occurs at low current density (below 1~A/cm$^2$). This is because the $B$ coefficient is small, due to the large electron-hole separation in this epitaxial design. In turn, this leads to a slow carrier lifetime and to a high carrier density at a given current density. Indeed, measurements of low $B$ coefficient were reported in Refs.~\cite{Muyeed21,Muyeed23}; most recently, Ref.~\cite{Kawakami25} reported very slow radiative lifetimes of 1--10~$\mu$s for the material of Ref.~\cite{Iida22}. 

Fig.~\ref{fig_red_nJ} illustrates this, by comparing the $J-n$ relationship for a blue LED and a red LED. The slower lifetime of the red LED shifts this curve by a decade, leading to a carrier density $\sim 3 \times 10^{18}$ (sufficient to induce some field screening) at $J=1$~A/cm$^2$.

The broad FWHM has two origins: a pronounced broadening by alloy disorder, and an additional contribution from phonon replicas. Fig.~\ref{fig_red_spectrum} shows the breakdown of the modeled spectrum. The direct emission (zero-phonon line) has a width of 36 nm. The first LO replica further broadens the spectrum to 45 nm; this large contribution stems from the high value of the Huang-Rhys parameter S=0.4. 

\begin{figure}
\includegraphics[width=8cm]{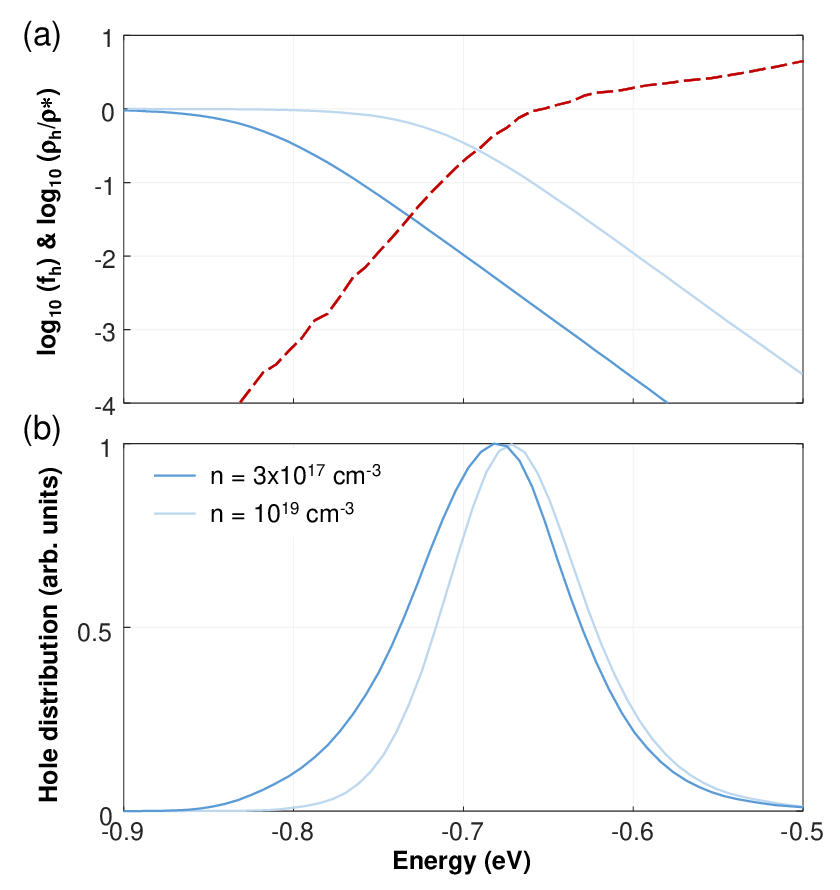}
\caption{\label{fig_red_hole_stat} Hole statistics in the red structure, at low and high carrier densities. (a) Fermi-Dirac distribution $f_h$ (solid lines) and hole density of states $\rho_h$ (dashed line). (b) Hole population $\rho_h \times f_h$. For easier visual comparison of the hole distributions, the bandgap shift caused by carrier screening at high density has been subtracted out (thus lining up $\rho_h$ at both values of $n$).
}
\end{figure}

Finally, the broadening of the FWHM at low current is indeed tied to the filling of localized states. Fig.~\ref{fig_red_hole_stat} shows the JDOS, the Fermi-Dirac distributions, and the hole distributions at various carrier densities. At low density ($n=3 \times 10^{17}$~cm$^{-3}$), the carrier distribution is Boltzmann-like for all states, whereas at higher density ($n=10^{19}$~cm$^{-3}$), the states in the Urbach tail are degenerate, leading to a narrowing of the low-energy tail of the hole distribution, and hence of the PL lineshape. Therefore, the FWHM narrowing from low to intermediate current is a manifestation of the localized tail and its increasing degeneracy.

In summary, the model appropriately accounts for key properties of red InGaN LEDs: pronounced wavelength shift with current, broad spectrum, and non-trivial FHWM trends. All these properties are intrinsic and tied to the RA, rather than being manifestations of a poor material quality or extrinsic disorder. 


\section{Discussion and conclusions}

A two-band model including the effect of random alloy disorder was introduced. Its success in fitting optical properties at various temperatures and densities, without tuning parameters, is compelling evidence that it captures the underlying physics. It reveals the following insights.

Overall, carriers in InGaN QWs behave neither like free carriers in a conventional III-V QWs, nor like a fully localized system. Rather, they exhibit a hybrid nature. Holes occupy a mix of delocalized and localized states. At room temperature, both are equally populated. Even at low temperature (77~K), experimental data shows that carriers are thermalized, despite the majority of holes occupying localized states. It is proposed that overall carrier thermalization occurs through phonon scattering between localized and delocalized states.

This hybrid aspect is imparted to the optical properties. The absorption spectrum displays a well-known Urbach tail caused by localized states, and a bulk-like region. The luminescence spectrum stems from transitions involving both localized and delocalized states. The former induce a low-energy spectral tail, and dominate the spectral broadening. 
The latter manifest as a high-energy thermal tail, confirming that the carrier populations are thermalized with the crystal.

Hence, two important quantities characterize localization in InGaN QWs. The Urbach energy indicates the strength of localization; it depends weakly on the QW composition. The Urbach energy competes with the thermal energy in determining how pronounced localization effects are. The mobility edge marks the boundary between localized and delocalized states. It plays the role of an effective band edge, with luminescence occurring around the mobility edge. The surprisingly-high reported value of the Stokes shift is mostly a matter of definition: the Stokes shift measured from the mobility edge is in fact quite small.

As the emission wavelength gets longer, these trends remain qualitatively valid and become more pronounced. Red LEDs exhibit the strongest disorder effects. This explains their large luminescence width and pronounced blue-shift with current density. Therefore, these unusual optical properties of red InGaN LEDs are found to be intrinsic, rather than caused by poor material quality.

In perspective, disorder induces pronounced effects, despite the absence of full localization; these must be considered for a proper understanding of optical properties.  This work has focused on visible InGaN quantum wells, but the underlying physics are expected to remain valid for other related materials, such as AlGaN emitters in the ultraviolet range. More broadly, our finding are likely relevant to many disordered semiconductors; indeed, regardless of the microscopic origin of disorder, its effects can be analyzed in terms of mobility edge and Urbach tail. Therefore, the present work is intended to establish and validate a framework to understand disorder effects -- not only to describe existing LEDs, but also to guide the development and understanding of future materials.



\section*{Appendix I - Model details}

\subsection{Schrodinger equation}

The model employs a two-band Hamiltonian $H=-\hbar^2\Delta/2m+V$, with $m$ the effective mass and $V$ the potential. Space is discretized on a 3-dimensional Cartesian grid, and $H$ is cast as a sparse matrix with a conventional finite-difference scheme.

The potential is $V=V_{alloy}+V_{field}$, where $V_{alloy}$ is the local band gap in the presence of alloy disorder, and $V_{field}$ is the potential from electrostatic fields.

To obtain $V_{alloy}$ for a given InGaN configuration, the InGaN map is first generated on a fine cubic grid with lattice spacing 0.3~nm (the in-plane cation spacing in InGaN). In each alloy region, a random distribution of In and Ga atoms having the desired average composition is drawn. This fine map is then smoothed out with a Gaussian filter of standard deviation $\sigma=0.4$~nm (see the following section for a discussion of this value). The map is re-sampled on the final computation grid with lattice spacing $dx=dy=0.6$~nm, $dz=0.3$~nm. The fine value along $z$ ensures heterostructures are well resolved, and the coarser value in the $(x,y)$ directions does not affect the results. $V_{alloy}$ is derived from this map with a standard bandgap bowing formula.

$V_{field}$ is the sum of three contributions $V_J$, $V_P$, $V_S$. The junction potential $V_J$ is calculated analytically by requiring that the $pn$ junction's potential drop occur over the depletion width (both of which are inputs to the calculation). 

The polarization field $V_P$ is computed following Ref.~\cite{Christmas05}. Namely, the electrical polarization $Pz$ in the $z$ direction is computed, the corresponding charge is obtained as $\rho=-Div(P_z)$, and the potential is obtained by inverting the Laplace equation $\Delta V_P = -\rho/\epsilon_r$. We note that we have not yet implemented the revised polarization field constants introduced in Ref.~\cite{Dreyer16}: these might improve the model's accuracy for structures with the highest In contents.

The screening potential $V_S$, which accounts for field screening at high carrier density $n$, is obtained from a self-consistent one-dimensional (1D) Schrodinger-Poisson solution, whose input 1D potential is the average of the 3D potential $V_{alloy}+V_J+V_P$ along $(x,y)$; the resulting one-dimensional screening potential along $z$ is then applied uniformly across the $(x,y)$ directions to obtain $V_S$. This simplified 1D approach neglects the impact of localization on screening; we have verified that this has negligible impact, because at the high densities where screening becomes meaningful, many states are populated and the carrier density is nearly uniform in-plane.

Returning to $H$, periodic boundary conditions are applied in the $(x,y)$ directions; this is slightly preferable over Dirichlet conditions, as it avoids artificial confinement effects, especially for wavefunctions near the domain boundaries. Dirichlet boundary conditions ($\psi=0$) are used at boundaries along $z$, which is appropriate for bound wavefunctions (this is slightly inaccurate for high-energy delocalized states, but bears no practical consequence). 

The energies and wavefunctions of $H$ are then obtained as solutions of the Schrodinger equation by a sparse eigenvalue solver. The energies of electrons and hole states are counted from the GaN band edges, are are therefore negative.

The following numerical values of parameters are used: effective masses $m_e=0.2$ and $m_h=1.9$ (this value is appropriate for localized wavepackets built with high-$k$ components, as discussed in the following Appendix), bowing parameter between GaN and InGaN $b=1.4$~eV (except for red LEDs, as stated in Section~\ref{sec:long}), conduction band offset 0.6, momentum matrix element $\mu$ with $2\mu^2/m_0=20$~eV, high-frequency dielectric constants $\epsilon_r=10.4$ (neglecting differences in $\epsilon_r$ between GaN and InGaN), optical index $n_o=2.5$.


\subsection{The valence band structure}

The valence band structure of wurtzite GaN displays pronounced non-parabolicity, band mixing and anisotropy effects \cite{Chuang96}. To take all these into account would require a multiband theory with spatial disorder -- a complex task that goes beyond the scope of the present work. We offer some qualitative comments below. Overall however, the use of a single effective mass must be acknowledged as a source of uncertainty which deserves future investigation.

In the effective mass approximation, one must decide what mass to use. The mass at the $\Gamma$ point may seem like a natural choice. However, heavy holes quickly acquire a heavier in-plane mass as the wavevector $k_x$ increases. For values $1 / k_x \sim 1$~nm, which are relevant to build real-space wavepackets localized on a 1~nm scale, the mass converges towards the value 'far from the band edge'. This value spans 1.6--2 $m_0$ across literature \cite{Chuang96,Vurgaftman03}. In this article, we use a value of 1.9 $m_0$. It should be pointed out that, if the conventional mass at $\Gamma$ were used, hole localization would be much weaker.

Conveniently, this high-k effective mass is also correct to describe the heavy hole DOS at high energy, where the carrier temperature is fitted. The heavy hole DOS is expected to be least accurate in a narrow energy region, for delocalized holes just above the mobility edge.

As we ignore the light hole band, the hole density of states for delocalized states is slightly under-estimated. We note however that the exponential slope of the JDOS above the mobility edge is caused by the multiple transitions allowed by the QCSE; we expect that this slope would remain mostly unaffected by including an additional hole band, and therefore we expect that this would not have a large effect on our estimation of the carrier temperature.

As a possible improvement, one could consider the use of an anisotropic hole mass, by assigning a different mass along the $z$ direction. This would mostly affect the hole wavefunction decay along $z$ and hence the absolute value of the electron-hole overlap, but would weakly affect localization trends; indeed, the confinement potentials in the in-plane and $z$ directions are strong and therefore nearly separable.

\subsection{Numerical implementation aspects}

In principle, one should compute the same large number $N$ of wavefunctions (e.g. $N=100-1,000$) for each InGaN configuration, in order to obtain both localized and delocalized states for this configuration. However, the delocalized states at high energy barely see the local details of the RA disorder, as they are rapidly-varying wavefunctions across the whole simulation domain. Because of this, delocalized states are very similar for different InGaN configurations.

This provides an opportunity to reduce the computation burden. If $C$ configurations are desired, we first compute all $N$ eigenstates for a small number C' of configurations (for instance, $C=1,000$ and $C'=10-20$). For all the ensuing configurations, we only compute a small number $N'$ of eigenstates (e.g. $N'=10$) of lowest energy, and re-use the higher-energy eigenstates from one of the first $C'$ configurations. Thus, the calculation is somewhat slow for the first $C'$ configurations and much faster for the other configurations. We have carefully checked that this approximation had no noticeable impact on numerical results.

With this approach, modeling a single quantum well for $N=400$ levels and $C=1,000$ configurations takes about one hour on a modern workstation.

\begin{figure}
\includegraphics[width=8.5cm]{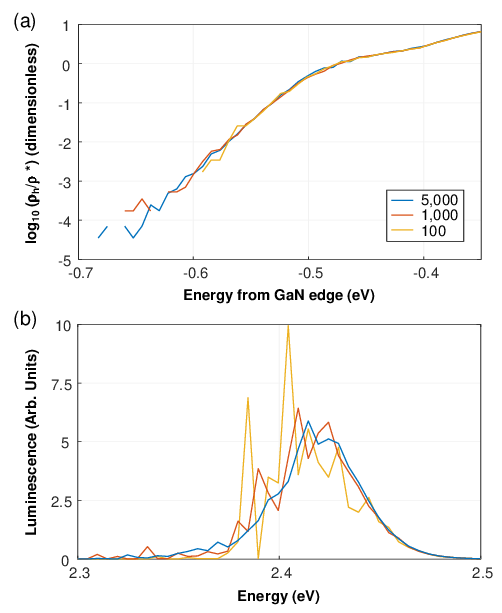}
\caption{\label{fig_convergence} Convergence study of the single QW structure versus the number of configurations $C$. (a) Hole DOS. As $C$ increases, the Urbach tail is resolved to lower energy. (b) Raw luminescence spectrum at 77~K, before any smoothing is applied. In this case, only $C=5,000$ ensures a reasonably smooth spectrum.
}
\end{figure}

We have found that a number of configurations $C=1,000-5,000$ was appropriate to obtain good convergence. More specifically, this ensures smooth luminescence spectra at low temperature (77~K), where the carriers occupy deeply localized levels and the Urbach tail must be well-resolved. This is illustrated in Fig.~\ref{fig_convergence}

\subsection{A comment on Coulomb interaction}

Coulomb interaction can affect optical properties in two distinct ways.

First, free carriers may form an exciton population. In Ref.~\cite{David22}, we computed the binding energy $E_B$ of excitons in InGaN quantum wells. At low carrier density, we found $E_B\sim 20$~meV for the quantum wells of moderate thickness considered in this Article. However, carrier screening reduces this binding energy, for instance to 5~meV or less at $n=10^{18}$~cm$^{-3}$. 

Applying the Saha equation \cite{Liu16,Snoke08,David22} then predicts that only a few \% of carriers should form excitons at room temperature, so that neglecting excitons is a safe approximation. At 77~$K$, this fraction might rise as high as 20\%, making our approximation somewhat less accurate.

Second, even if excitons do not form, the Coulomb interaction enhances light-matter coupling and increases the JDOS and the radiative rate \cite{Koch06}. We explored this effect, theoretically and experimentally, in Ref.~\cite{David19b}. It is pronounced in thin QWs at low carrier density, but weaker in thick QWs at moderate density (e.g. $t>3$~nm, $n>10^{18}$~cm$^{-3}$) -- the regime of interest of the present study. In addition, at least in thick QWs, the relative JDOS enhancement tends to occur equally at all energies (due to the broadening of optical features by QCSE and disorder); therefore, its main effect is to enhance the absolute value of the radiative rate, without introducing significant spectral signatures, at least in thick QWs.

In thinner QWs, this Coulomb enhancement of the JDOS becomes more pronounced. We have not examined whether this affects the shape of the Urbach tail -- in which case, inclusion of this effect would be necessary for accurate spectra predictions in thin QWs. We provide further comments on this in the next section.

\subsection{Impact of the smoothing parameter $\sigma$}

\begin{figure}
\includegraphics[width=8.5cm]{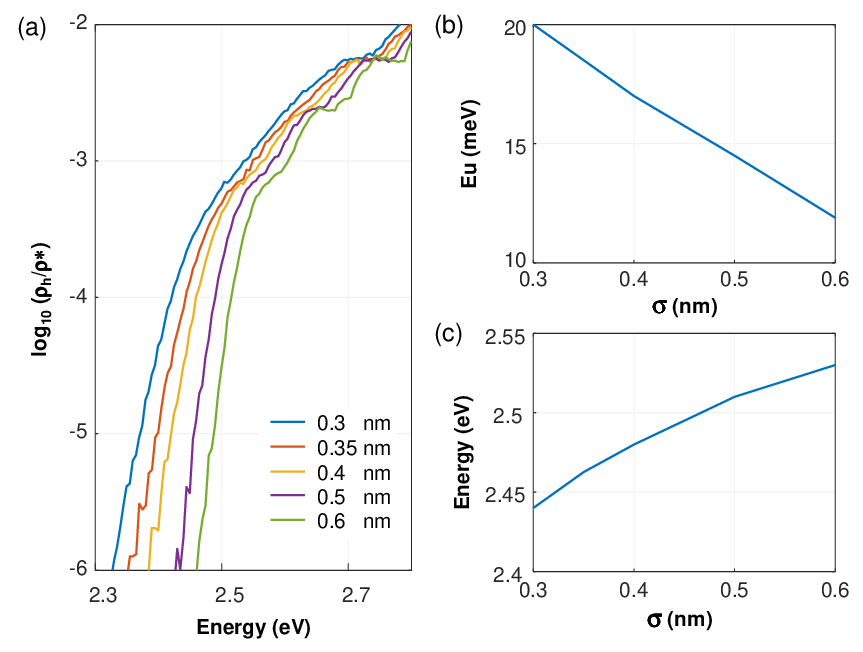}
\caption{\label{fig_sigma} Effect of the smoothing parameter $\sigma$ on: (a) the joint density of states, (b) the Urbach energy and (c) the mobility edge energy.
}
\end{figure}

The smoothing parameter $\sigma$ -- the standard deviation of the Gaussian that smooths the InGaN composition map -- is the only free parameter of the model. Its value is not prescribed, though it should intuitively be commensurate with the lattice spacing (otherwise, disorder effects vanish). Other publications have used a value $\sigma=0.6$~nm, whereas our work (here and in Ref.~\cite{David19b}) uses $\sigma=0.4$~nm. Here, we illustrate the effect of this parameter.

We consider the same single-QW structure as in Section~\ref{sec:pin}, and consider the range $\sigma=0.3-0.6$~nm. Fig.~\ref{fig_sigma}(a) shows the resulting JDOS. As $\sigma$ increases, the effects of disorder become less pronounced, the Urbach tail becomes shallower, and the transition energies are blue-shifted. Accordingly, the Urbach energy and the mobility edge (Fig.~\ref{fig_sigma}(b-c)) respectively increase and decrease. At high values of $\sigma$, the step-like features of the JDOS become more apparent. Note that the shift of the mobility edge is in large part because large values of $\sigma$ smooth the band structure along $z$, reducing the depth of the QW confinement.

For values $\sigma \sim 0.3-0.4$~nm, $E_u$ is in the range $17-20$~meV, similar to published experimental data for this composition. Indeed, we have found that values of $\sigma$ in this range gave good fits to our experimental measurements. On the other hand the effects of disorder would be too weak for $\sigma=0.6$~nm, and would not provide good agreements with experimental data.

As just discussed, the Coulomb interaction affects the shape of the JDOS, and may thus affect the Urbach tail -- especially in thin QWs, where Coulomb interaction is more pronounced. In principle, one would then not expect that the present model should predict the Urbach tail properly. On the other hand, one may then think of $\sigma$ as a knob to effectively emulate Coulomb effects at a simplistic level (by matching the shape of the Coulomb-affected JDOS, if not its amplitude). We leave further investigation of this question as potential future work.

\subsection{Modulating $\sigma$ to vary the strength of RA disorder}

As discussed in Section~\ref{sec_disorder_B}, the value of $\sigma$ can be varied to tune disorder effects.

In previous works, several authors have proposed to evaluate the impact of RA disorder by using the virtual crystal approximation as a 'disorder-free' reference. However, this approximation does not only lead to a constant composition inside the QW, but also to abrupt hetero-interfaces at the boundaries of the QW -- whereas in a disordered model, the profile along $z$ is smeared out. Both fluctuations across the QW and the band profile around the heterobarriers can affect $B$, and the virtual crystal approximation convolves both effects.

Therefore, to better isolate the contribution of carrier localization, the profile of the band structure along the $z$ direction should remain unchanged while RA disorder is varied. To achieve this, we use an anisotropic broadening parameter $\sigma$ to smooth out atomic-level RA fluctuations. Along $z$, $\sigma_z$ is maintained at a constant value (0.4~nm, like elsewhere in this work); this ensures that the QW's band profile along $z$ is unchanged. Along $(x, y)$, the value of $\sigma_{xy}$ is varied to modulate the magnitude of RA disorder. Hence, the 'disorder-free' reference is obtained for large values of $\sigma_{xy}$ which completely smooth out RA fluctuations, while retaining a constant band profile along $z$.

\begin{figure}
\includegraphics[width=8cm]{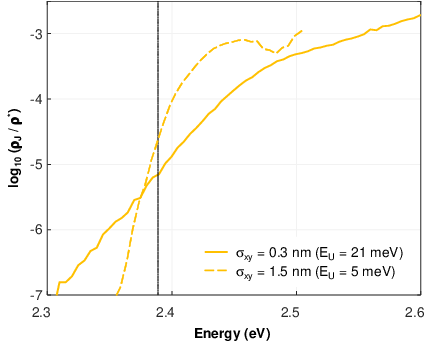}
\caption{\label{fig_JDOS_vs_sigma} Effect of the broadening $\sigma_{xy}$ on the JDOS. The dotted vertical line denotes the Fermi level splitting $\Delta E_F$.
}
\end{figure}

Fig.~\ref{fig_JDOS_vs_sigma} illustrates how $\sigma_{xy}$ affects the JDOS. Two values are considered: $\sigma_{xy}=0.3$~nm (strong RA effect) and 15~nm (weak RA effect). The JDOS for the latter case was shifted to align the quasi-Fermi level energies $\Delta E_F$. Due to the occupation factor, the JDOS value around $\Delta E_F$ has a strong influence on the total PL magnitude and the value of $B$. Less disorder leads to a higher JDOS in this energy range, and a higher $B$, as shown in Section~\ref{sec_disorder_B}.

\subsection{Fit of the Urbach energy and mobility edge}

\begin{figure}
\includegraphics[width=8cm]{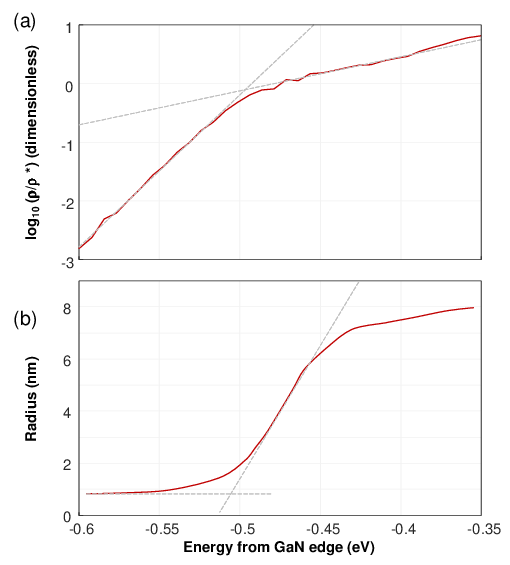}
\caption{\label{fig_mobility_fit} Extraction of the hole mobility edge for the single-QW structure (a) from the hole density of states and (b) from the hole radius. In (a), the fit of the low-energy tail also yields the Urbach energy $E_U$.
}
\end{figure}

The Urbach energy is extracted by fitting the exponential decay of the density of states in the low-energy tail. This is illustrated in Fig.~\ref{fig_mobility_fit}~(a). 

Likewise, the hole mobility edge can be quantified by graphical extraction of the inflection point of the density of states. Two linear fits of $\rho_h$ are performed in the Urbach tail and in the delocalized region, and the intersection energy is a measure of the mobility edge. This is illustrated in Fig.~\ref{fig_mobility_fit}~(a). This construction can also be applied to $\rho_J$ to extract the optical mobility edge.

Alternatively, the hole mobility edge can be extracted from the energy at which the hole radius departs from its localized value. Here, linear fits are made to the median hole radius curve in the low-energy plateau and in the transition region, and the intersection energy identifies the mobility edge (Fig.~\ref{fig_mobility_fit}~(b)). We have also considered the inverse participation ratio as an alternative measure of localization, and have found it to be well correlated with the hole radius. 

We have found these two approaches to give very similar mobility edge values (within $\sim$ 10 meV). 

\section*{Appendix II - Carrier thermalization}

Throughout this Article, we assume carriers obey a thermalized Fermi-Dirac distribution. In principle, this is not obvious in the presence of disorder, especially if hopping processes were to dominate the carrier distribution -- which should be the case at very low temperature \cite{Miller60,Mott68}. Indeed, strongly-localized hole states located in separate regions have vanishing wavefunction overlap, and thus cannot exchange energy or tunnel from one site to another. In this hopping regime, it is possible that some carriers remain captured in higher-energy localized states and cannot relax to lower energies. This has been invoked as the process underlying the S-shape of InGaN wavelength at cryogenic temperature \cite{Cho98}.

At sufficiently high temperature, on the other hand, conventional band transport should dominate. Indeed, hole states at sufficiently high energy are delocalized, and phonon-mediated inelastic hole-hole scattering between localized and delocalized states will lead to overall thermalization. To be more specific, we argue that if all localized states can exchange with the continuum of delocalized states by scattering, this leads to overall thermalization, even if there is no direct scattering between localized states.

\begin{figure}
\includegraphics[width=7.5cm]{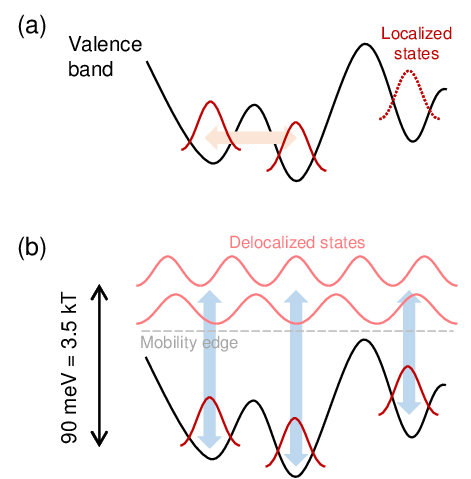}
\caption{\label{fig_thermalization} Sketch of thermalization processes in the hopping and band transport regimes. (a) At very low temperature, hole states distribute by hopping between low-energy states (lateral arrow). Localized states may have very low hopping rate, such that some states remain off-equilibrium for a long time (dotted wavefunction). (b) At higher temperature, the phonon population provides efficient scattering with the continuum of delocalized states (vertical arrows), which ensures indirect thermalization between localized states. For instance, at room temperature, LO phonons ensure thermalization of states 90~meV below the mobility edge.
}
\end{figure}

Fig.~\ref{fig_thermalization} sketches these two regimes. However, which regime prevails in specific conditions deserves some discussion. Below, we focus our attention on carrier-phonon scattering. Indeed, carrier-carrier scattering is only relevant at high carrier density (whereas we assume --and experimentally observe-- thermalization at low densities down to $10^{17}$ cm$^{-3}$), and impurity scattering is generally elastic.

At room temperature, it appears clear that band transport should be valid due to phonon scattering. Indeed, absorption and emission of LO phonon is very fast (ps time scale or faster), and provides significant energy exchange ($E_{LO}=90$~meV) which can ionize even deeply localized states towards delocalized states. This scattering time is much faster than the recombination time ($\sim 10-100$~ns), providing many scattering events before carriers recombine. 

As the temperature is reduced, the phonon population decreases, and so does the scattering probability from localized to delocalized states, which requires phonon absorption. This provides a possible bottleneck for thermalization. Properly describing this regime would require a detailed model of carrier-phonon scattering, including both optical and acoustic phonons, which is well outside the scope of the present discussion. Instead, we consider a naive toy model to extract some orders of magnitude.

Let us consider a situation where localized states at an energy $E$ (below the mobility edge) are initially in a non-thermalized state, and evaluate how fast they may scatter with delocalized states through phonon absorption. At 77~K, the hole mobility in GaN is on the order of 1000 cm$^2$V$^{-1}$s$^{-1}$ \cite{Hamaguchi21}, corresponding to a phonon scattering time of about 1~ps for phonon emission. The absorption rate of a phonon of energy $E$ is suppressed by a factor $exp(-E/kT)$, assuming the phonon population itself has thermalized. Therefore, states located at an energy 50--60~meV below the mobility edge may still scatter towards delocalized states in $\sim$ 10 ns. This provides fast thermalization compared to the recombination time. Thus, a non-thermal distribution is not expected to endure for localized states at this energy, which is somewhat deep in the Urbach tail (1.5 decade below the mobility edge). 

In this simple framework, we cannot justify thermalization for states at even deeper energy at 77~K. However, the experimental evidence of Section~\ref{sec:pin} indicates that thermalization extends to even-lower energy (at least 3 decades into the Urbach tail). 

At very low temperature (e.g. 4~K), this basic model does not provide a process for thermalization -- and indeed, there is experimental evidence of non-thermal behavior at temperatures of a few tens of K, as previously mentioned. An accurate model describing this transition would be desirable.



\begin{thebibliography}{85}%
\makeatletter
\providecommand \@ifxundefined [1]{%
 \@ifx{#1\undefined}
}%
\providecommand \@ifnum [1]{%
 \ifnum #1\expandafter \@firstoftwo
 \else \expandafter \@secondoftwo
 \fi
}%
\providecommand \@ifx [1]{%
 \ifx #1\expandafter \@firstoftwo
 \else \expandafter \@secondoftwo
 \fi
}%
\providecommand \natexlab [1]{#1}%
\providecommand \enquote  [1]{``#1''}%
\providecommand \bibnamefont  [1]{#1}%
\providecommand \bibfnamefont [1]{#1}%
\providecommand \citenamefont [1]{#1}%
\providecommand \href@noop [0]{\@secondoftwo}%
\providecommand \href [0]{\begingroup \@sanitize@url \@href}%
\providecommand \@href[1]{\@@startlink{#1}\@@href}%
\providecommand \@@href[1]{\endgroup#1\@@endlink}%
\providecommand \@sanitize@url [0]{\catcode `\\12\catcode `\$12\catcode
  `\&12\catcode `\#12\catcode `\^12\catcode `\_12\catcode `\%12\relax}%
\providecommand \@@startlink[1]{}%
\providecommand \@@endlink[0]{}%
\providecommand \url  [0]{\begingroup\@sanitize@url \@url }%
\providecommand \@url [1]{\endgroup\@href {#1}{\urlprefix }}%
\providecommand \urlprefix  [0]{URL }%
\providecommand \Eprint [0]{\href }%
\providecommand \doibase [0]{https://doi.org/}%
\providecommand \selectlanguage [0]{\@gobble}%
\providecommand \bibinfo  [0]{\@secondoftwo}%
\providecommand \bibfield  [0]{\@secondoftwo}%
\providecommand \translation [1]{[#1]}%
\providecommand \BibitemOpen [0]{}%
\providecommand \bibitemStop [0]{}%
\providecommand \bibitemNoStop [0]{.\EOS\space}%
\providecommand \EOS [0]{\spacefactor3000\relax}%
\providecommand \BibitemShut  [1]{\csname bibitem#1\endcsname}%
\let\auto@bib@innerbib\@empty
\bibitem [{\citenamefont {Takeuchi}\ \emph {et~al.}(1997)\citenamefont
  {Takeuchi}, \citenamefont {Sota}, \citenamefont {Katsuragawa}, \citenamefont
  {Komori}, \citenamefont {Takeuchi}, \citenamefont {Amano},\ and\
  \citenamefont {Akasaki}}]{Takeuchi97}%
  \BibitemOpen
  \bibfield  {author} {\bibinfo {author} {\bibfnamefont {T.}~\bibnamefont
  {Takeuchi}}, \bibinfo {author} {\bibfnamefont {S.}~\bibnamefont {Sota}},
  \bibinfo {author} {\bibfnamefont {M.}~\bibnamefont {Katsuragawa}}, \bibinfo
  {author} {\bibfnamefont {M.}~\bibnamefont {Komori}}, \bibinfo {author}
  {\bibfnamefont {H.}~\bibnamefont {Takeuchi}}, \bibinfo {author}
  {\bibfnamefont {H.~A.~H.}\ \bibnamefont {Amano}},\ and\ \bibinfo {author}
  {\bibfnamefont {I.~A.~I.}\ \bibnamefont {Akasaki}},\ }\bibfield  {title}
  {\bibinfo {title} {Quantum-{Confined} {Stark} {Effect} due to {Piezoelectric}
  {Fields} in {GaInN} {Strained} {Quantum} {Wells}},\ }\href
  {https://doi.org/10.1143/JJAP.36.L382} {\bibfield  {journal} {\bibinfo
  {journal} {Japanese Journal of Applied Physics}\ }\textbf {\bibinfo {volume}
  {36}},\ \bibinfo {pages} {L382} (\bibinfo {year} {1997})}\BibitemShut
  {NoStop}%
\bibitem [{\citenamefont {Gerthsen}\ \emph {et~al.}(2000)\citenamefont
  {Gerthsen}, \citenamefont {Hahn}, \citenamefont {Neubauer}, \citenamefont
  {Rosenauer}, \citenamefont {Schön}, \citenamefont {Heuken},\ and\
  \citenamefont {Rizzi}}]{Gerthsen00}%
  \BibitemOpen
  \bibfield  {author} {\bibinfo {author} {\bibfnamefont {D.}~\bibnamefont
  {Gerthsen}}, \bibinfo {author} {\bibfnamefont {E.}~\bibnamefont {Hahn}},
  \bibinfo {author} {\bibfnamefont {B.}~\bibnamefont {Neubauer}}, \bibinfo
  {author} {\bibfnamefont {A.}~\bibnamefont {Rosenauer}}, \bibinfo {author}
  {\bibfnamefont {O.}~\bibnamefont {Schön}}, \bibinfo {author} {\bibfnamefont
  {M.}~\bibnamefont {Heuken}},\ and\ \bibinfo {author} {\bibfnamefont
  {A.}~\bibnamefont {Rizzi}},\ }\bibfield  {title} {\bibinfo {title}
  {Composition {Fluctuations} in {InGaN} {Analyzed} by {Transmission}
  {Electron} {Microscopy}},\ }\href
  {https://doi.org/10.1002/(SICI)1521-396X(200001)177:1<145::AID-PSSA145>3.0.CO;2-0}
  {\bibfield  {journal} {\bibinfo  {journal} {physica status solidi (a)}\
  }\textbf {\bibinfo {volume} {177}},\ \bibinfo {pages} {145} (\bibinfo {year}
  {2000})}\BibitemShut {NoStop}%
\bibitem [{\citenamefont {Galtrey}\ \emph {et~al.}(2008)\citenamefont
  {Galtrey}, \citenamefont {Oliver}, \citenamefont {Kappers}, \citenamefont
  {Humphreys}, \citenamefont {Clifton}, \citenamefont {Larson}, \citenamefont
  {Saxey},\ and\ \citenamefont {Cerezo}}]{Galtrey08}%
  \BibitemOpen
  \bibfield  {author} {\bibinfo {author} {\bibfnamefont {M.~J.}\ \bibnamefont
  {Galtrey}}, \bibinfo {author} {\bibfnamefont {R.~A.}\ \bibnamefont {Oliver}},
  \bibinfo {author} {\bibfnamefont {M.~J.}\ \bibnamefont {Kappers}}, \bibinfo
  {author} {\bibfnamefont {C.~J.}\ \bibnamefont {Humphreys}}, \bibinfo {author}
  {\bibfnamefont {P.~H.}\ \bibnamefont {Clifton}}, \bibinfo {author}
  {\bibfnamefont {D.}~\bibnamefont {Larson}}, \bibinfo {author} {\bibfnamefont
  {D.~W.}\ \bibnamefont {Saxey}},\ and\ \bibinfo {author} {\bibfnamefont
  {A.}~\bibnamefont {Cerezo}},\ }\bibfield  {title} {\bibinfo {title}
  {Three-dimensional atom probe analysis of green- and blue-emitting
  {InxGa1}-{xN}/{GaN} multiple quantum well structures},\ }\href
  {https://doi.org/10.1063/1.2938081} {\bibfield  {journal} {\bibinfo
  {journal} {Journal of Applied Physics}\ }\textbf {\bibinfo {volume} {104}},\
  \bibinfo {pages} {013524} (\bibinfo {year} {2008})}\BibitemShut {NoStop}%
\bibitem [{\citenamefont {John}\ \emph {et~al.}(1986)\citenamefont {John},
  \citenamefont {Soukoulis}, \citenamefont {Cohen},\ and\ \citenamefont
  {Economou}}]{John86}%
  \BibitemOpen
  \bibfield  {author} {\bibinfo {author} {\bibfnamefont {S.}~\bibnamefont
  {John}}, \bibinfo {author} {\bibfnamefont {C.}~\bibnamefont {Soukoulis}},
  \bibinfo {author} {\bibfnamefont {M.~H.}\ \bibnamefont {Cohen}},\ and\
  \bibinfo {author} {\bibfnamefont {E.~N.}\ \bibnamefont {Economou}},\
  }\bibfield  {title} {\bibinfo {title} {Theory of {Electron} {Band} {Tails}
  and the {Urbach} {Optical}-{Absorption} {Edge}},\ }\href
  {https://doi.org/10.1103/PhysRevLett.57.1777} {\bibfield  {journal} {\bibinfo
   {journal} {Physical Review Letters}\ }\textbf {\bibinfo {volume} {57}},\
  \bibinfo {pages} {1777} (\bibinfo {year} {1986})}\BibitemShut {NoStop}%
\bibitem [{\citenamefont {Watson-Parris}\ \emph {et~al.}(2011)\citenamefont
  {Watson-Parris}, \citenamefont {Godfrey}, \citenamefont {Dawson},
  \citenamefont {Oliver}, \citenamefont {Galtrey}, \citenamefont {Kappers},\
  and\ \citenamefont {Humphreys}}]{Watson-Parris11}%
  \BibitemOpen
  \bibfield  {author} {\bibinfo {author} {\bibfnamefont {D.}~\bibnamefont
  {Watson-Parris}}, \bibinfo {author} {\bibfnamefont {M.~J.}\ \bibnamefont
  {Godfrey}}, \bibinfo {author} {\bibfnamefont {P.}~\bibnamefont {Dawson}},
  \bibinfo {author} {\bibfnamefont {R.~A.}\ \bibnamefont {Oliver}}, \bibinfo
  {author} {\bibfnamefont {M.~J.}\ \bibnamefont {Galtrey}}, \bibinfo {author}
  {\bibfnamefont {M.~J.}\ \bibnamefont {Kappers}},\ and\ \bibinfo {author}
  {\bibfnamefont {C.~J.}\ \bibnamefont {Humphreys}},\ }\bibfield  {title}
  {\bibinfo {title} {Carrier localization mechanisms in {InxGa1}-{xN}/{GaN}
  quantum wells},\ }\href@noop {} {\bibfield  {journal} {\bibinfo  {journal}
  {Physical Review B}\ }\textbf {\bibinfo {volume} {83}},\ \bibinfo {pages}
  {115321} (\bibinfo {year} {2011})}\BibitemShut {NoStop}%
\bibitem [{\citenamefont {Auf~der Maur}\ and\ \citenamefont
  {Galler}(2015)}]{Aufdermaur15}%
  \BibitemOpen
  \bibfield  {author} {\bibinfo {author} {\bibfnamefont {M.}~\bibnamefont
  {Auf~der Maur}}\ and\ \bibinfo {author} {\bibfnamefont {B.}~\bibnamefont
  {Galler}},\ }\bibfield  {title} {\bibinfo {title} {Multiscale approaches for
  the simulation of {InGaN}/{GaN} {LEDs}},\ }\href@noop {} {\bibfield
  {journal} {\bibinfo  {journal} {Journal of Computational Electronics}\
  }\textbf {\bibinfo {volume} {14}},\ \bibinfo {pages} {398} (\bibinfo {year}
  {2015})}\BibitemShut {NoStop}%
\bibitem [{\citenamefont {Schulz}\ \emph {et~al.}(2015)\citenamefont {Schulz},
  \citenamefont {Caro}, \citenamefont {Coughlan},\ and\ \citenamefont
  {O'Reilly}}]{Schulz15}%
  \BibitemOpen
  \bibfield  {author} {\bibinfo {author} {\bibfnamefont {S.}~\bibnamefont
  {Schulz}}, \bibinfo {author} {\bibfnamefont {M.~A.}\ \bibnamefont {Caro}},
  \bibinfo {author} {\bibfnamefont {C.}~\bibnamefont {Coughlan}},\ and\
  \bibinfo {author} {\bibfnamefont {E.~P.}\ \bibnamefont {O'Reilly}},\
  }\bibfield  {title} {\bibinfo {title} {Atomistic analysis of the impact of
  alloy and well-width fluctuations on the electronic and optical properties of
  {InGaN}/{GaN} quantum wells},\ }\href@noop {} {\bibfield  {journal} {\bibinfo
   {journal} {Physical Review B}\ }\textbf {\bibinfo {volume} {91}},\ \bibinfo
  {pages} {035439} (\bibinfo {year} {2015})}\BibitemShut {NoStop}%
\bibitem [{\citenamefont {Tanner}\ \emph {et~al.}(2018)\citenamefont {Tanner},
  \citenamefont {McMahon},\ and\ \citenamefont {Schulz}}]{Tanner18}%
  \BibitemOpen
  \bibfield  {author} {\bibinfo {author} {\bibfnamefont {D.~S.~P.}\
  \bibnamefont {Tanner}}, \bibinfo {author} {\bibfnamefont {J.~M.}\
  \bibnamefont {McMahon}},\ and\ \bibinfo {author} {\bibfnamefont
  {S.}~\bibnamefont {Schulz}},\ }\bibfield  {title} {\bibinfo {title}
  {Interface {Roughness}, {Carrier} {Localization}, and {Wave} {Function}
  {Overlap} in c-{Plane} ({In}, {Ga}){N}/{GaN} {Quantum} {Wells}: {Interplay}
  of {Well} {Width}, {Alloy} {Microstructure}, {Structural} {Inhomogeneities},
  and {Coulomb} {Effects}},\ }\href@noop {} {\bibfield  {journal} {\bibinfo
  {journal} {Physical Review Applied}\ }\textbf {\bibinfo {volume} {10}},\
  \bibinfo {pages} {034027} (\bibinfo {year} {2018})}\BibitemShut {NoStop}%
\bibitem [{\citenamefont {Tanner}\ \emph {et~al.}(2020)\citenamefont {Tanner},
  \citenamefont {Dawson}, \citenamefont {Kappers}, \citenamefont {Oliver},\
  and\ \citenamefont {Schulz}}]{Tanner20b}%
  \BibitemOpen
  \bibfield  {author} {\bibinfo {author} {\bibfnamefont {D.~S.}\ \bibnamefont
  {Tanner}}, \bibinfo {author} {\bibfnamefont {P.}~\bibnamefont {Dawson}},
  \bibinfo {author} {\bibfnamefont {M.~J.}\ \bibnamefont {Kappers}}, \bibinfo
  {author} {\bibfnamefont {R.~A.}\ \bibnamefont {Oliver}},\ and\ \bibinfo
  {author} {\bibfnamefont {S.}~\bibnamefont {Schulz}},\ }\bibfield  {title}
  {\bibinfo {title} {Polar {InGaN}/{GaN} {Quantum} {Wells}: {Revisiting} the
  {Impact} of {Carrier} {Localization} on the ``{Green} {Gap}'' {Problem}},\
  }\href {https://doi.org/10.1103/PhysRevApplied.13.044068} {\bibfield
  {journal} {\bibinfo  {journal} {Physical Review Applied}\ }\textbf {\bibinfo
  {volume} {13}},\ \bibinfo {pages} {044068} (\bibinfo {year}
  {2020})}\BibitemShut {NoStop}%
\bibitem [{\citenamefont {McMahon}\ \emph {et~al.}(2020)\citenamefont
  {McMahon}, \citenamefont {Tanner}, \citenamefont {Kioupakis},\ and\
  \citenamefont {Schulz}}]{McMahon20}%
  \BibitemOpen
  \bibfield  {author} {\bibinfo {author} {\bibfnamefont {J.~M.}\ \bibnamefont
  {McMahon}}, \bibinfo {author} {\bibfnamefont {D.~S.~P.}\ \bibnamefont
  {Tanner}}, \bibinfo {author} {\bibfnamefont {E.}~\bibnamefont {Kioupakis}},\
  and\ \bibinfo {author} {\bibfnamefont {S.}~\bibnamefont {Schulz}},\
  }\bibfield  {title} {\bibinfo {title} {Atomistic analysis of radiative
  recombination rate, {Stokes} shift, and density of states in c-plane
  {InGaN}/{GaN} quantum wells},\ }\href {https://doi.org/10.1063/5.0006128}
  {\bibfield  {journal} {\bibinfo  {journal} {Applied Physics Letters}\
  }\textbf {\bibinfo {volume} {116}},\ \bibinfo {pages} {181104} (\bibinfo
  {year} {2020})}\BibitemShut {NoStop}%
\bibitem [{\citenamefont {Piccardo}\ \emph {et~al.}(2017)\citenamefont
  {Piccardo}, \citenamefont {Li}, \citenamefont {Wu}, \citenamefont {Speck},
  \citenamefont {Bonef}, \citenamefont {Farrell}, \citenamefont {Filoche},
  \citenamefont {Martinelli}, \citenamefont {Peretti},\ and\ \citenamefont
  {Weisbuch}}]{Piccardo17}%
  \BibitemOpen
  \bibfield  {author} {\bibinfo {author} {\bibfnamefont {M.}~\bibnamefont
  {Piccardo}}, \bibinfo {author} {\bibfnamefont {C.-K.}\ \bibnamefont {Li}},
  \bibinfo {author} {\bibfnamefont {Y.-R.}\ \bibnamefont {Wu}}, \bibinfo
  {author} {\bibfnamefont {J.~S.}\ \bibnamefont {Speck}}, \bibinfo {author}
  {\bibfnamefont {B.}~\bibnamefont {Bonef}}, \bibinfo {author} {\bibfnamefont
  {R.~M.}\ \bibnamefont {Farrell}}, \bibinfo {author} {\bibfnamefont
  {M.}~\bibnamefont {Filoche}}, \bibinfo {author} {\bibfnamefont
  {L.}~\bibnamefont {Martinelli}}, \bibinfo {author} {\bibfnamefont
  {J.}~\bibnamefont {Peretti}},\ and\ \bibinfo {author} {\bibfnamefont
  {C.}~\bibnamefont {Weisbuch}},\ }\bibfield  {title} {\bibinfo {title}
  {Localization landscape theory of disorder in semiconductors. {II}. {Urbach}
  tails of disordered quantum well layers},\ }\href@noop {} {\bibfield
  {journal} {\bibinfo  {journal} {Physical Review B}\ }\textbf {\bibinfo
  {volume} {95}},\ \bibinfo {pages} {144205} (\bibinfo {year}
  {2017})}\BibitemShut {NoStop}%
\bibitem [{\citenamefont {Filoche}\ and\ \citenamefont
  {Mayboroda}(2012)}]{Filoche12}%
  \BibitemOpen
  \bibfield  {author} {\bibinfo {author} {\bibfnamefont {M.}~\bibnamefont
  {Filoche}}\ and\ \bibinfo {author} {\bibfnamefont {S.}~\bibnamefont
  {Mayboroda}},\ }\bibfield  {title} {\bibinfo {title} {Universal mechanism for
  {Anderson} and weak localization},\ }\href
  {https://doi.org/10.1073/pnas.1120432109} {\bibfield  {journal} {\bibinfo
  {journal} {Proceedings of the National Academy of Sciences}\ }\textbf
  {\bibinfo {volume} {109}},\ \bibinfo {pages} {14761} (\bibinfo {year}
  {2012})}\BibitemShut {NoStop}%
\bibitem [{\citenamefont {Yang}\ \emph {et~al.}(2014)\citenamefont {Yang},
  \citenamefont {Shivaraman}, \citenamefont {Speck},\ and\ \citenamefont
  {Wu}}]{Yang14}%
  \BibitemOpen
  \bibfield  {author} {\bibinfo {author} {\bibfnamefont {T.-J.}\ \bibnamefont
  {Yang}}, \bibinfo {author} {\bibfnamefont {R.}~\bibnamefont {Shivaraman}},
  \bibinfo {author} {\bibfnamefont {J.~S.}\ \bibnamefont {Speck}},\ and\
  \bibinfo {author} {\bibfnamefont {Y.-R.}\ \bibnamefont {Wu}},\ }\bibfield
  {title} {\bibinfo {title} {The influence of random indium alloy fluctuations
  in indium gallium nitride quantum wells on the device behavior},\ }\href@noop
  {} {\bibfield  {journal} {\bibinfo  {journal} {Journal of Applied Physics}\
  }\textbf {\bibinfo {volume} {116}},\ \bibinfo {pages} {113104} (\bibinfo
  {year} {2014})}\BibitemShut {NoStop}%
\bibitem [{\citenamefont {Li}\ \emph {et~al.}(2017)\citenamefont {Li},
  \citenamefont {Piccardo}, \citenamefont {Lu}, \citenamefont {Mayboroda},
  \citenamefont {Martinelli}, \citenamefont {Peretti}, \citenamefont {Speck},
  \citenamefont {Weisbuch}, \citenamefont {Filoche},\ and\ \citenamefont
  {Wu}}]{Li17}%
  \BibitemOpen
  \bibfield  {author} {\bibinfo {author} {\bibfnamefont {C.-K.}\ \bibnamefont
  {Li}}, \bibinfo {author} {\bibfnamefont {M.}~\bibnamefont {Piccardo}},
  \bibinfo {author} {\bibfnamefont {L.-S.}\ \bibnamefont {Lu}}, \bibinfo
  {author} {\bibfnamefont {S.}~\bibnamefont {Mayboroda}}, \bibinfo {author}
  {\bibfnamefont {L.}~\bibnamefont {Martinelli}}, \bibinfo {author}
  {\bibfnamefont {J.}~\bibnamefont {Peretti}}, \bibinfo {author} {\bibfnamefont
  {J.~S.}\ \bibnamefont {Speck}}, \bibinfo {author} {\bibfnamefont
  {C.}~\bibnamefont {Weisbuch}}, \bibinfo {author} {\bibfnamefont
  {M.}~\bibnamefont {Filoche}},\ and\ \bibinfo {author} {\bibfnamefont {Y.-R.}\
  \bibnamefont {Wu}},\ }\bibfield  {title} {\bibinfo {title} {Localization
  landscape theory of disorder in semiconductors. {III}. {Application} to
  carrier transport and recombination in light emitting diodes},\ }\href@noop
  {} {\bibfield  {journal} {\bibinfo  {journal} {Physical Review B}\ }\textbf
  {\bibinfo {volume} {95}},\ \bibinfo {pages} {144206} (\bibinfo {year}
  {2017})}\BibitemShut {NoStop}%
\bibitem [{\citenamefont {David}\ \emph {et~al.}(2016)\citenamefont {David},
  \citenamefont {Hurni}, \citenamefont {Young},\ and\ \citenamefont
  {Craven}}]{David16b}%
  \BibitemOpen
  \bibfield  {author} {\bibinfo {author} {\bibfnamefont {A.}~\bibnamefont
  {David}}, \bibinfo {author} {\bibfnamefont {C.~A.}\ \bibnamefont {Hurni}},
  \bibinfo {author} {\bibfnamefont {N.~G.}\ \bibnamefont {Young}},\ and\
  \bibinfo {author} {\bibfnamefont {M.~D.}\ \bibnamefont {Craven}},\ }\bibfield
   {title} {\bibinfo {title} {Electrical properties of {III}-{Nitride} {LEDs}:
  {Recombination}-based injection model and theoretical limits to electrical
  efficiency and electroluminescent cooling},\ }\href@noop {} {\bibfield
  {journal} {\bibinfo  {journal} {Applied Physics Letters}\ }\textbf {\bibinfo
  {volume} {109}},\ \bibinfo {pages} {083501} (\bibinfo {year}
  {2016})}\BibitemShut {NoStop}%
\bibitem [{\citenamefont {Pant}\ \emph {et~al.}(2022)\citenamefont {Pant},
  \citenamefont {Li}, \citenamefont {DeJong}, \citenamefont {Feezell},
  \citenamefont {Armitage},\ and\ \citenamefont {Kioupakis}}]{Pant22}%
  \BibitemOpen
  \bibfield  {author} {\bibinfo {author} {\bibfnamefont {N.}~\bibnamefont
  {Pant}}, \bibinfo {author} {\bibfnamefont {X.}~\bibnamefont {Li}}, \bibinfo
  {author} {\bibfnamefont {E.}~\bibnamefont {DeJong}}, \bibinfo {author}
  {\bibfnamefont {D.}~\bibnamefont {Feezell}}, \bibinfo {author} {\bibfnamefont
  {R.}~\bibnamefont {Armitage}},\ and\ \bibinfo {author} {\bibfnamefont
  {E.}~\bibnamefont {Kioupakis}},\ }\bibfield  {title} {\bibinfo {title}
  {Origin of the injection-dependent emission blueshift and linewidth
  broadening of {III}-nitride light-emitting diodes},\ }\href
  {https://doi.org/10.1063/5.0134995} {\bibfield  {journal} {\bibinfo
  {journal} {AIP Advances}\ }\textbf {\bibinfo {volume} {12}},\ \bibinfo
  {pages} {125020} (\bibinfo {year} {2022})}\BibitemShut {NoStop}%
\bibitem [{\citenamefont {Abrahams}\ \emph {et~al.}(1979)\citenamefont
  {Abrahams}, \citenamefont {Anderson}, \citenamefont {Licciardello},\ and\
  \citenamefont {Ramakrishnan}}]{Anderson79}%
  \BibitemOpen
  \bibfield  {author} {\bibinfo {author} {\bibfnamefont {E.}~\bibnamefont
  {Abrahams}}, \bibinfo {author} {\bibfnamefont {P.~W.}\ \bibnamefont
  {Anderson}}, \bibinfo {author} {\bibfnamefont {D.~C.}\ \bibnamefont
  {Licciardello}},\ and\ \bibinfo {author} {\bibfnamefont {T.~V.}\ \bibnamefont
  {Ramakrishnan}},\ }\bibfield  {title} {\bibinfo {title} {Scaling {Theory} of
  {Localization}: {Absence} of {Quantum} {Diffusion} in {Two} {Dimensions}},\
  }\href {https://doi.org/10.1103/PhysRevLett.42.673} {\bibfield  {journal}
  {\bibinfo  {journal} {Physical Review Letters}\ }\textbf {\bibinfo {volume}
  {42}},\ \bibinfo {pages} {673} (\bibinfo {year} {1979})}\BibitemShut
  {NoStop}%
\bibitem [{\citenamefont {Jones}\ \emph {et~al.}(2017)\citenamefont {Jones},
  \citenamefont {Teng}, \citenamefont {Yan}, \citenamefont {Ku},\ and\
  \citenamefont {Kioupakis}}]{Jones17}%
  \BibitemOpen
  \bibfield  {author} {\bibinfo {author} {\bibfnamefont {C.~M.}\ \bibnamefont
  {Jones}}, \bibinfo {author} {\bibfnamefont {C.-H.}\ \bibnamefont {Teng}},
  \bibinfo {author} {\bibfnamefont {Q.}~\bibnamefont {Yan}}, \bibinfo {author}
  {\bibfnamefont {P.-C.}\ \bibnamefont {Ku}},\ and\ \bibinfo {author}
  {\bibfnamefont {E.}~\bibnamefont {Kioupakis}},\ }\bibfield  {title} {\bibinfo
  {title} {Impact of carrier localization on recombination in {InGaN} quantum
  wells and the efficiency of nitride light-emitting diodes: {Insights} from
  theory and numerical simulations},\ }\href
  {https://doi.org/10.1063/1.5002104} {\bibfield  {journal} {\bibinfo
  {journal} {Applied Physics Letters}\ }\textbf {\bibinfo {volume} {111}},\
  \bibinfo {pages} {113501} (\bibinfo {year} {2017})}\BibitemShut {NoStop}%
\bibitem [{\citenamefont {Di~Vito}\ \emph {et~al.}(2020)\citenamefont
  {Di~Vito}, \citenamefont {Pecchia}, \citenamefont {Di~Carlo},\ and\
  \citenamefont {Auf~der Maur}}]{Divito20}%
  \BibitemOpen
  \bibfield  {author} {\bibinfo {author} {\bibfnamefont {A.}~\bibnamefont
  {Di~Vito}}, \bibinfo {author} {\bibfnamefont {A.}~\bibnamefont {Pecchia}},
  \bibinfo {author} {\bibfnamefont {A.}~\bibnamefont {Di~Carlo}},\ and\
  \bibinfo {author} {\bibfnamefont {M.}~\bibnamefont {Auf~der Maur}},\
  }\bibfield  {title} {\bibinfo {title} {Simulating random alloy effects in
  {III}-nitride light emitting diodes},\ }\href
  {https://doi.org/10.1063/5.0005862} {\bibfield  {journal} {\bibinfo
  {journal} {Journal of Applied Physics}\ }\textbf {\bibinfo {volume} {128}},\
  \bibinfo {pages} {041102} (\bibinfo {year} {2020})}\BibitemShut {NoStop}%
\bibitem [{\citenamefont {Banon}\ \emph {et~al.}(2022)\citenamefont {Banon},
  \citenamefont {Pelletier}, \citenamefont {Weisbuch}, \citenamefont
  {Mayboroda},\ and\ \citenamefont {Filoche}}]{Banon22}%
  \BibitemOpen
  \bibfield  {author} {\bibinfo {author} {\bibfnamefont {J.-P.}\ \bibnamefont
  {Banon}}, \bibinfo {author} {\bibfnamefont {P.}~\bibnamefont {Pelletier}},
  \bibinfo {author} {\bibfnamefont {C.}~\bibnamefont {Weisbuch}}, \bibinfo
  {author} {\bibfnamefont {S.}~\bibnamefont {Mayboroda}},\ and\ \bibinfo
  {author} {\bibfnamefont {M.}~\bibnamefont {Filoche}},\ }\bibfield  {title}
  {\bibinfo {title} {Wigner-{Weyl} description of light absorption in
  disordered semiconductor alloys using the localization landscape theory},\
  }\href {https://doi.org/10.1103/PhysRevB.105.125422} {\bibfield  {journal}
  {\bibinfo  {journal} {Physical Review B}\ }\textbf {\bibinfo {volume}
  {105}},\ \bibinfo {pages} {125422} (\bibinfo {year} {2022})}\BibitemShut
  {NoStop}%
\bibitem [{\citenamefont {David}\ \emph
  {et~al.}(2019{\natexlab{a}})\citenamefont {David}, \citenamefont {Young},\
  and\ \citenamefont {Craven}}]{David19b}%
  \BibitemOpen
  \bibfield  {author} {\bibinfo {author} {\bibfnamefont {A.}~\bibnamefont
  {David}}, \bibinfo {author} {\bibfnamefont {N.~G.}\ \bibnamefont {Young}},\
  and\ \bibinfo {author} {\bibfnamefont {M.~D.}\ \bibnamefont {Craven}},\
  }\bibfield  {title} {\bibinfo {title} {Many-body effects in
  strongly-disordered {III}-nitride quantum wells: interplay between carrier
  localization and {Coulomb} interaction},\ }\href@noop {} {\bibfield
  {journal} {\bibinfo  {journal} {Physical Review Applied}\ }\textbf {\bibinfo
  {volume} {12}},\ \bibinfo {pages} {044059} (\bibinfo {year}
  {2019}{\natexlab{a}})}\BibitemShut {NoStop}%
\bibitem [{\citenamefont {Vurgaftman}\ and\ \citenamefont
  {Meyer}(2003)}]{Vurgaftman03}%
  \BibitemOpen
  \bibfield  {author} {\bibinfo {author} {\bibfnamefont {I.}~\bibnamefont
  {Vurgaftman}}\ and\ \bibinfo {author} {\bibfnamefont {J.~R.}\ \bibnamefont
  {Meyer}},\ }\bibfield  {title} {\bibinfo {title} {Band parameters for
  nitrogen-containing semiconductors},\ }\href@noop {} {\bibfield  {journal}
  {\bibinfo  {journal} {Journal of applied physics}\ }\textbf {\bibinfo
  {volume} {94}},\ \bibinfo {pages} {3675} (\bibinfo {year}
  {2003})}\BibitemShut {NoStop}%
\bibitem [{\citenamefont {Chuang}\ and\ \citenamefont
  {Chang}(1996)}]{Chuang96}%
  \BibitemOpen
  \bibfield  {author} {\bibinfo {author} {\bibfnamefont {S.~L.}\ \bibnamefont
  {Chuang}}\ and\ \bibinfo {author} {\bibfnamefont {C.~S.}\ \bibnamefont
  {Chang}},\ }\bibfield  {title} {\bibinfo {title} {k.p method for strained
  wurtzite semiconductors},\ }\href@noop {} {\bibfield  {journal} {\bibinfo
  {journal} {Physical Review B}\ }\textbf {\bibinfo {volume} {54}},\ \bibinfo
  {pages} {2491} (\bibinfo {year} {1996})}\BibitemShut {NoStop}%
\bibitem [{\citenamefont {Christmas}\ \emph {et~al.}(2005)\citenamefont
  {Christmas}, \citenamefont {Andreev},\ and\ \citenamefont
  {Faux}}]{Christmas05}%
  \BibitemOpen
  \bibfield  {author} {\bibinfo {author} {\bibfnamefont {U.~M.~E.}\
  \bibnamefont {Christmas}}, \bibinfo {author} {\bibfnamefont {A.~D.}\
  \bibnamefont {Andreev}},\ and\ \bibinfo {author} {\bibfnamefont {D.~A.}\
  \bibnamefont {Faux}},\ }\bibfield  {title} {\bibinfo {title} {Calculation of
  electric field and optical transitions in {InGaN}/{GaN} quantum wells},\
  }\href@noop {} {\bibfield  {journal} {\bibinfo  {journal} {Journal of applied
  physics}\ }\textbf {\bibinfo {volume} {98}},\ \bibinfo {pages} {073522}
  (\bibinfo {year} {2005})}\BibitemShut {NoStop}%
\bibitem [{\citenamefont {David}\ and\ \citenamefont
  {Weisbuch}(2022)}]{David22}%
  \BibitemOpen
  \bibfield  {author} {\bibinfo {author} {\bibfnamefont {A.}~\bibnamefont
  {David}}\ and\ \bibinfo {author} {\bibfnamefont {C.}~\bibnamefont
  {Weisbuch}},\ }\bibfield  {title} {\bibinfo {title} {Excitons in a disordered
  medium: {A} numerical study in {InGaN} quantum wells},\ }\href
  {https://doi.org/10.1103/PhysRevResearch.4.043004} {\bibfield  {journal}
  {\bibinfo  {journal} {Physical Review Research}\ }\textbf {\bibinfo {volume}
  {4}},\ \bibinfo {pages} {043004} (\bibinfo {year} {2022})}\BibitemShut
  {NoStop}%
\bibitem [{\citenamefont {Römer}\ \emph {et~al.}(2021)\citenamefont {Römer},
  \citenamefont {Guttmann}, \citenamefont {Wernicke}, \citenamefont {Kneissl},\
  and\ \citenamefont {Witzigmann}}]{Romer21}%
  \BibitemOpen
  \bibfield  {author} {\bibinfo {author} {\bibfnamefont {F.}~\bibnamefont
  {Römer}}, \bibinfo {author} {\bibfnamefont {M.}~\bibnamefont {Guttmann}},
  \bibinfo {author} {\bibfnamefont {T.}~\bibnamefont {Wernicke}}, \bibinfo
  {author} {\bibfnamefont {M.}~\bibnamefont {Kneissl}},\ and\ \bibinfo {author}
  {\bibfnamefont {B.}~\bibnamefont {Witzigmann}},\ }\bibfield  {title}
  {\bibinfo {title} {Effect of {Inhomogeneous} {Broadening} in {Ultraviolet}
  {III}-{Nitride} {Light}-{Emitting} {Diodes}},\ }\href
  {https://doi.org/10.3390/ma14247890} {\bibfield  {journal} {\bibinfo
  {journal} {Materials}\ }\textbf {\bibinfo {volume} {14}},\ \bibinfo {pages}
  {7890} (\bibinfo {year} {2021})}\BibitemShut {NoStop}%
\bibitem [{\citenamefont {David}\ \emph {et~al.}(2017)\citenamefont {David},
  \citenamefont {Young}, \citenamefont {Hurni},\ and\ \citenamefont
  {Craven}}]{David17a}%
  \BibitemOpen
  \bibfield  {author} {\bibinfo {author} {\bibfnamefont {A.}~\bibnamefont
  {David}}, \bibinfo {author} {\bibfnamefont {N.~G.}\ \bibnamefont {Young}},
  \bibinfo {author} {\bibfnamefont {C.~A.}\ \bibnamefont {Hurni}},\ and\
  \bibinfo {author} {\bibfnamefont {M.~D.}\ \bibnamefont {Craven}},\ }\bibfield
   {title} {\bibinfo {title} {All-optical measurements of carrier dynamics in
  bulk-{GaN} {LEDs}: {Beyond} the {ABC} approximation},\ }\href@noop {}
  {\bibfield  {journal} {\bibinfo  {journal} {Applied Physics Letters}\
  }\textbf {\bibinfo {volume} {110}},\ \bibinfo {pages} {253504} (\bibinfo
  {year} {2017})}\BibitemShut {NoStop}%
\bibitem [{\citenamefont {David}\ \emph
  {et~al.}(2019{\natexlab{b}})\citenamefont {David}, \citenamefont {Young},
  \citenamefont {Lund},\ and\ \citenamefont {Craven}}]{David20}%
  \BibitemOpen
  \bibfield  {author} {\bibinfo {author} {\bibfnamefont {A.}~\bibnamefont
  {David}}, \bibinfo {author} {\bibfnamefont {N.~G.}\ \bibnamefont {Young}},
  \bibinfo {author} {\bibfnamefont {C.}~\bibnamefont {Lund}},\ and\ \bibinfo
  {author} {\bibfnamefont {M.~D.}\ \bibnamefont {Craven}},\ }\bibfield  {title}
  {\bibinfo {title} {Review—{The} {Physics} of {Recombinations} in
  {III}-{Nitride} {Emitters}},\ }\href {https://doi.org/10.1149/2.0372001JSS}
  {\bibfield  {journal} {\bibinfo  {journal} {ECS Journal of Solid State
  Science and Technology}\ }\textbf {\bibinfo {volume} {9}},\ \bibinfo {pages}
  {016021} (\bibinfo {year} {2019}{\natexlab{b}})}\BibitemShut {NoStop}%
\bibitem [{\citenamefont {David}(2021)}]{David21}%
  \BibitemOpen
  \bibfield  {author} {\bibinfo {author} {\bibfnamefont {A.}~\bibnamefont
  {David}},\ }\bibfield  {title} {\bibinfo {title} {Long-{Range} {Carrier}
  {Diffusion} in ({In},{Ga}){N} {Quantum} {Wells} and {Implications} from
  {Fundamentals} to {Devices}},\ }\href
  {https://doi.org/10.1103/PhysRevApplied.15.054015} {\bibfield  {journal}
  {\bibinfo  {journal} {Physical Review Applied}\ }\textbf {\bibinfo {volume}
  {15}},\ \bibinfo {pages} {054015} (\bibinfo {year} {2021})}\BibitemShut
  {NoStop}%
\bibitem [{\citenamefont {Nakamura}(1998)}]{Nakamura98}%
  \BibitemOpen
  \bibfield  {author} {\bibinfo {author} {\bibfnamefont {S.}~\bibnamefont
  {Nakamura}},\ }\bibfield  {title} {\bibinfo {title} {The {Roles} of
  {Structural} {Imperfections} in {InGaN}-{Based} {Blue} {Light}-{Emitting}
  {Diodes} and {Laser} {Diodes}},\ }\href
  {https://doi.org/10.1126/science.281.5379.956} {\bibfield  {journal}
  {\bibinfo  {journal} {Science}\ }\textbf {\bibinfo {volume} {281}},\ \bibinfo
  {pages} {956} (\bibinfo {year} {1998})}\BibitemShut {NoStop}%
\bibitem [{\citenamefont {O'Donnell}\ \emph {et~al.}(1999)\citenamefont
  {O'Donnell}, \citenamefont {Martin},\ and\ \citenamefont
  {Middleton}}]{Odonnell99}%
  \BibitemOpen
  \bibfield  {author} {\bibinfo {author} {\bibfnamefont {K.~P.}\ \bibnamefont
  {O'Donnell}}, \bibinfo {author} {\bibfnamefont {R.~W.}\ \bibnamefont
  {Martin}},\ and\ \bibinfo {author} {\bibfnamefont {P.~G.}\ \bibnamefont
  {Middleton}},\ }\bibfield  {title} {\bibinfo {title} {Origin of
  {Luminescence} from {InGaN} {Diodes}},\ }\href
  {https://doi.org/10.1103/PhysRevLett.82.237} {\bibfield  {journal} {\bibinfo
  {journal} {Physical Review Letters}\ }\textbf {\bibinfo {volume} {82}},\
  \bibinfo {pages} {237} (\bibinfo {year} {1999})}\BibitemShut {NoStop}%
\bibitem [{\citenamefont {Chichibu}\ \emph {et~al.}(2006)\citenamefont
  {Chichibu}, \citenamefont {Uedono}, \citenamefont {Onuma}, \citenamefont
  {Haskell}, \citenamefont {Chakraborty}, \citenamefont {Koyama}, \citenamefont
  {Fini}, \citenamefont {Keller}, \citenamefont {Denbaars}, \citenamefont
  {Speck}, \citenamefont {Mishra}, \citenamefont {Nakamura}, \citenamefont
  {Yamaguchi}, \citenamefont {Kamiyama}, \citenamefont {Amano}, \citenamefont
  {Akasaki}, \citenamefont {Han},\ and\ \citenamefont {Sota}}]{Chichibu06}%
  \BibitemOpen
  \bibfield  {author} {\bibinfo {author} {\bibfnamefont {S.~F.}\ \bibnamefont
  {Chichibu}}, \bibinfo {author} {\bibfnamefont {A.}~\bibnamefont {Uedono}},
  \bibinfo {author} {\bibfnamefont {T.}~\bibnamefont {Onuma}}, \bibinfo
  {author} {\bibfnamefont {B.~A.}\ \bibnamefont {Haskell}}, \bibinfo {author}
  {\bibfnamefont {A.}~\bibnamefont {Chakraborty}}, \bibinfo {author}
  {\bibfnamefont {T.}~\bibnamefont {Koyama}}, \bibinfo {author} {\bibfnamefont
  {P.~T.}\ \bibnamefont {Fini}}, \bibinfo {author} {\bibfnamefont
  {S.}~\bibnamefont {Keller}}, \bibinfo {author} {\bibfnamefont {S.~P.}\
  \bibnamefont {Denbaars}}, \bibinfo {author} {\bibfnamefont {J.~S.}\
  \bibnamefont {Speck}}, \bibinfo {author} {\bibfnamefont {U.~K.}\ \bibnamefont
  {Mishra}}, \bibinfo {author} {\bibfnamefont {S.}~\bibnamefont {Nakamura}},
  \bibinfo {author} {\bibfnamefont {S.}~\bibnamefont {Yamaguchi}}, \bibinfo
  {author} {\bibfnamefont {S.}~\bibnamefont {Kamiyama}}, \bibinfo {author}
  {\bibfnamefont {H.}~\bibnamefont {Amano}}, \bibinfo {author} {\bibfnamefont
  {I.}~\bibnamefont {Akasaki}}, \bibinfo {author} {\bibfnamefont
  {J.}~\bibnamefont {Han}},\ and\ \bibinfo {author} {\bibfnamefont
  {T.}~\bibnamefont {Sota}},\ }\bibfield  {title} {\bibinfo {title} {Origin of
  defect-insensitive emission probability in {In}-containing
  ({Al},{In},{Ga}){N} alloy semiconductors},\ }\href@noop {} {\bibfield
  {journal} {\bibinfo  {journal} {Nature Materials}\ }\textbf {\bibinfo
  {volume} {5}},\ \bibinfo {pages} {810} (\bibinfo {year} {2006})}\BibitemShut
  {NoStop}%
\bibitem [{\citenamefont {Hammersley}\ \emph {et~al.}(2012)\citenamefont
  {Hammersley}, \citenamefont {Watson-Parris}, \citenamefont {Dawson},
  \citenamefont {Godfrey}, \citenamefont {Badcock}, \citenamefont {Kappers},
  \citenamefont {McAleese}, \citenamefont {Oliver},\ and\ \citenamefont
  {Humphreys}}]{Hammersley12}%
  \BibitemOpen
  \bibfield  {author} {\bibinfo {author} {\bibfnamefont {S.}~\bibnamefont
  {Hammersley}}, \bibinfo {author} {\bibfnamefont {D.}~\bibnamefont
  {Watson-Parris}}, \bibinfo {author} {\bibfnamefont {P.}~\bibnamefont
  {Dawson}}, \bibinfo {author} {\bibfnamefont {M.~J.}\ \bibnamefont {Godfrey}},
  \bibinfo {author} {\bibfnamefont {T.~J.}\ \bibnamefont {Badcock}}, \bibinfo
  {author} {\bibfnamefont {M.~J.}\ \bibnamefont {Kappers}}, \bibinfo {author}
  {\bibfnamefont {C.}~\bibnamefont {McAleese}}, \bibinfo {author}
  {\bibfnamefont {R.~A.}\ \bibnamefont {Oliver}},\ and\ \bibinfo {author}
  {\bibfnamefont {C.~J.}\ \bibnamefont {Humphreys}},\ }\bibfield  {title}
  {\bibinfo {title} {The consequences of high injected carrier densities on
  carrier localization and efficiency droop in {InGaN}/{GaN} quantum well
  structures},\ }\href@noop {} {\bibfield  {journal} {\bibinfo  {journal}
  {Journal of Applied Physics}\ }\textbf {\bibinfo {volume} {111}},\ \bibinfo
  {pages} {083512} (\bibinfo {year} {2012})}\BibitemShut {NoStop}%
\bibitem [{\citenamefont {Becht}\ \emph {et~al.}(2023)\citenamefont {Becht},
  \citenamefont {Schwarz}, \citenamefont {Binder},\ and\ \citenamefont
  {Galler}}]{Becht23}%
  \BibitemOpen
  \bibfield  {author} {\bibinfo {author} {\bibfnamefont {C.}~\bibnamefont
  {Becht}}, \bibinfo {author} {\bibfnamefont {U.~T.}\ \bibnamefont {Schwarz}},
  \bibinfo {author} {\bibfnamefont {M.}~\bibnamefont {Binder}},\ and\ \bibinfo
  {author} {\bibfnamefont {B.}~\bibnamefont {Galler}},\ }\bibfield  {title}
  {\bibinfo {title} {Diffusion {Analysis} of {Charge} {Carriers} in
  {InGaN}/{GaN} {Heterostructures} by {Microphotoluminescence}},\ }\href
  {https://doi.org/10.1002/pssb.202200565} {\bibfield  {journal} {\bibinfo
  {journal} {physica status solidi (b)}\ }\textbf {\bibinfo {volume} {260}},\
  \bibinfo {pages} {2200565} (\bibinfo {year} {2023})}\BibitemShut {NoStop}%
\bibitem [{\citenamefont {Slawinska}\ \emph {et~al.}(2025)\citenamefont
  {Slawinska}, \citenamefont {Muziol}, \citenamefont {Kafar},\ and\
  \citenamefont {Skierbiszewski}}]{Slawinska25}%
  \BibitemOpen
  \bibfield  {author} {\bibinfo {author} {\bibfnamefont {J.}~\bibnamefont
  {Slawinska}}, \bibinfo {author} {\bibfnamefont {G.}~\bibnamefont {Muziol}},
  \bibinfo {author} {\bibfnamefont {A.}~\bibnamefont {Kafar}},\ and\ \bibinfo
  {author} {\bibfnamefont {C.}~\bibnamefont {Skierbiszewski}},\ }\bibfield
  {title} {\bibinfo {title} {Lateral {Carrier} {Diffusion} in {Ion}-{Implanted}
  {Ultra}-{Small} {Blue} {III}-{Nitride} {MicroLEDs}},\ }\href
  {https://doi.org/10.1021/acsami.4c14784} {\bibfield  {journal} {\bibinfo
  {journal} {ACS Applied Materials \& Interfaces}\ }\textbf {\bibinfo {volume}
  {17}},\ \bibinfo {pages} {6473} (\bibinfo {year} {2025})}\BibitemShut
  {NoStop}%
\bibitem [{\citenamefont {Schomig}\ \emph {et~al.}(2004)\citenamefont
  {Schomig}, \citenamefont {Halm}, \citenamefont {Forchel}, \citenamefont
  {Bacher}, \citenamefont {Off},\ and\ \citenamefont {Scholz}}]{Schomig04}%
  \BibitemOpen
  \bibfield  {author} {\bibinfo {author} {\bibfnamefont {H.}~\bibnamefont
  {Schomig}}, \bibinfo {author} {\bibfnamefont {S.}~\bibnamefont {Halm}},
  \bibinfo {author} {\bibfnamefont {A.}~\bibnamefont {Forchel}}, \bibinfo
  {author} {\bibfnamefont {G.}~\bibnamefont {Bacher}}, \bibinfo {author}
  {\bibfnamefont {J.}~\bibnamefont {Off}},\ and\ \bibinfo {author}
  {\bibfnamefont {F.}~\bibnamefont {Scholz}},\ }\bibfield  {title} {\bibinfo
  {title} {Probing individual localization centers in an {InGaN}/{GaN} quantum
  well},\ }\href@noop {} {\bibfield  {journal} {\bibinfo  {journal} {Physical
  Review Letters}\ }\textbf {\bibinfo {volume} {92}},\ \bibinfo {pages}
  {106802} (\bibinfo {year} {2004})}\BibitemShut {NoStop}%
\bibitem [{\citenamefont {Sauty}\ \emph {et~al.}(2022)\citenamefont {Sauty},
  \citenamefont {Lopes}, \citenamefont {Banon}, \citenamefont {Lassailly},
  \citenamefont {Martinelli}, \citenamefont {Alhassan}, \citenamefont {Chow},
  \citenamefont {Nakamura}, \citenamefont {Speck}, \citenamefont {Weisbuch},\
  and\ \citenamefont {Peretti}}]{Sauty22}%
  \BibitemOpen
  \bibfield  {author} {\bibinfo {author} {\bibfnamefont {M.}~\bibnamefont
  {Sauty}}, \bibinfo {author} {\bibfnamefont {N.~M.}\ \bibnamefont {Lopes}},
  \bibinfo {author} {\bibfnamefont {J.-P.}\ \bibnamefont {Banon}}, \bibinfo
  {author} {\bibfnamefont {Y.}~\bibnamefont {Lassailly}}, \bibinfo {author}
  {\bibfnamefont {L.}~\bibnamefont {Martinelli}}, \bibinfo {author}
  {\bibfnamefont {A.}~\bibnamefont {Alhassan}}, \bibinfo {author}
  {\bibfnamefont {Y.~C.}\ \bibnamefont {Chow}}, \bibinfo {author}
  {\bibfnamefont {S.}~\bibnamefont {Nakamura}}, \bibinfo {author}
  {\bibfnamefont {J.~S.}\ \bibnamefont {Speck}}, \bibinfo {author}
  {\bibfnamefont {C.}~\bibnamefont {Weisbuch}},\ and\ \bibinfo {author}
  {\bibfnamefont {J.}~\bibnamefont {Peretti}},\ }\bibfield  {title} {\bibinfo
  {title} {Localization {Effect} in {Photoelectron} {Transport} {Induced} by
  {Alloy} {Disorder} in {Nitride} {Semiconductor} {Compounds}},\ }\href
  {https://doi.org/10.1103/PhysRevLett.129.216602} {\bibfield  {journal}
  {\bibinfo  {journal} {Physical Review Letters}\ }\textbf {\bibinfo {volume}
  {129}},\ \bibinfo {pages} {216602} (\bibinfo {year} {2022})}\BibitemShut
  {NoStop}%
\bibitem [{\citenamefont {Paskov}\ \emph {et~al.}(2002)\citenamefont {Paskov},
  \citenamefont {Holtz}, \citenamefont {Monemar}, \citenamefont {Kamiyama},
  \citenamefont {Iwaya}, \citenamefont {Amano},\ and\ \citenamefont
  {Akasaki}}]{Paskov02}%
  \BibitemOpen
  \bibfield  {author} {\bibinfo {author} {\bibfnamefont {P.}~\bibnamefont
  {Paskov}}, \bibinfo {author} {\bibfnamefont {P.}~\bibnamefont {Holtz}},
  \bibinfo {author} {\bibfnamefont {B.}~\bibnamefont {Monemar}}, \bibinfo
  {author} {\bibfnamefont {S.}~\bibnamefont {Kamiyama}}, \bibinfo {author}
  {\bibfnamefont {M.}~\bibnamefont {Iwaya}}, \bibinfo {author} {\bibfnamefont
  {H.}~\bibnamefont {Amano}},\ and\ \bibinfo {author} {\bibfnamefont
  {I.}~\bibnamefont {Akasaki}},\ }\bibfield  {title} {\bibinfo {title}
  {Phonon-{Assisted} {Photoluminescence} in {InGaN}/{GaN} {Multiple} {Quantum}
  {Wells}},\ }\href
  {https://doi.org/10.1002/1521-3951(200212)234:3<755::AID-PSSB755>3.0.CO;2-0}
  {\bibfield  {journal} {\bibinfo  {journal} {physica status solidi (b)}\
  }\textbf {\bibinfo {volume} {234}},\ \bibinfo {pages} {755} (\bibinfo {year}
  {2002})}\BibitemShut {NoStop}%
\bibitem [{\citenamefont {Tan}\ \emph {et~al.}(2006)\citenamefont {Tan},
  \citenamefont {Martin}, \citenamefont {O’Donnell},\ and\ \citenamefont
  {Watson}}]{Tan06}%
  \BibitemOpen
  \bibfield  {author} {\bibinfo {author} {\bibfnamefont {L.~T.}\ \bibnamefont
  {Tan}}, \bibinfo {author} {\bibfnamefont {R.~W.}\ \bibnamefont {Martin}},
  \bibinfo {author} {\bibfnamefont {K.~P.}\ \bibnamefont {O’Donnell}},\ and\
  \bibinfo {author} {\bibfnamefont {I.~M.}\ \bibnamefont {Watson}},\ }\bibfield
   {title} {\bibinfo {title} {Photoluminescence and phonon satellites of single
  {InGaN}/{GaN} quantum wells with varying {GaN} cap thickness},\ }\href
  {https://doi.org/10.1063/1.2345246} {\bibfield  {journal} {\bibinfo
  {journal} {Applied Physics Letters}\ }\textbf {\bibinfo {volume} {89}},\
  \bibinfo {pages} {101910} (\bibinfo {year} {2006})}\BibitemShut {NoStop}%
\bibitem [{\citenamefont {Kalliakos}\ \emph {et~al.}(2002)\citenamefont
  {Kalliakos}, \citenamefont {Zhang}, \citenamefont {Taliercio}, \citenamefont
  {Lefebvre}, \citenamefont {Gil}, \citenamefont {Grandjean}, \citenamefont
  {Damilano},\ and\ \citenamefont {Massies}}]{Kalliakos02}%
  \BibitemOpen
  \bibfield  {author} {\bibinfo {author} {\bibfnamefont {S.}~\bibnamefont
  {Kalliakos}}, \bibinfo {author} {\bibfnamefont {X.~B.}\ \bibnamefont
  {Zhang}}, \bibinfo {author} {\bibfnamefont {T.}~\bibnamefont {Taliercio}},
  \bibinfo {author} {\bibfnamefont {P.}~\bibnamefont {Lefebvre}}, \bibinfo
  {author} {\bibfnamefont {B.}~\bibnamefont {Gil}}, \bibinfo {author}
  {\bibfnamefont {N.}~\bibnamefont {Grandjean}}, \bibinfo {author}
  {\bibfnamefont {B.}~\bibnamefont {Damilano}},\ and\ \bibinfo {author}
  {\bibfnamefont {J.}~\bibnamefont {Massies}},\ }\bibfield  {title} {\bibinfo
  {title} {Large size dependence of exciton-longitudinal-optical-phonon
  coupling in nitride-based quantum wells and quantum boxes},\ }\href
  {https://doi.org/10.1063/1.1433165} {\bibfield  {journal} {\bibinfo
  {journal} {Applied Physics Letters}\ }\textbf {\bibinfo {volume} {80}},\
  \bibinfo {pages} {428} (\bibinfo {year} {2002})}\BibitemShut {NoStop}%
\bibitem [{\citenamefont {Graham}\ \emph {et~al.}(2005)\citenamefont {Graham},
  \citenamefont {Soltani-Vala}, \citenamefont {Dawson}, \citenamefont
  {Godfrey}, \citenamefont {Smeeton}, \citenamefont {Barnard}, \citenamefont
  {Kappers}, \citenamefont {Humphreys},\ and\ \citenamefont
  {Thrush}}]{Graham05}%
  \BibitemOpen
  \bibfield  {author} {\bibinfo {author} {\bibfnamefont {D.~M.}\ \bibnamefont
  {Graham}}, \bibinfo {author} {\bibfnamefont {A.}~\bibnamefont
  {Soltani-Vala}}, \bibinfo {author} {\bibfnamefont {P.}~\bibnamefont
  {Dawson}}, \bibinfo {author} {\bibfnamefont {M.~J.}\ \bibnamefont {Godfrey}},
  \bibinfo {author} {\bibfnamefont {T.~M.}\ \bibnamefont {Smeeton}}, \bibinfo
  {author} {\bibfnamefont {J.~S.}\ \bibnamefont {Barnard}}, \bibinfo {author}
  {\bibfnamefont {M.~J.}\ \bibnamefont {Kappers}}, \bibinfo {author}
  {\bibfnamefont {C.~J.}\ \bibnamefont {Humphreys}},\ and\ \bibinfo {author}
  {\bibfnamefont {E.~J.}\ \bibnamefont {Thrush}},\ }\bibfield  {title}
  {\bibinfo {title} {Optical and microstructural studies of {InGaN}/{GaN}
  single-quantum-well structures},\ }\href@noop {} {\bibfield  {journal}
  {\bibinfo  {journal} {Journal of applied physics}\ }\textbf {\bibinfo
  {volume} {97}},\ \bibinfo {pages} {103508} (\bibinfo {year}
  {2005})}\BibitemShut {NoStop}%
\bibitem [{\citenamefont {Coldren}\ and\ \citenamefont
  {Corzine}(1995)}]{Coldren95b}%
  \BibitemOpen
  \bibfield  {author} {\bibinfo {author} {\bibfnamefont {L.}~\bibnamefont
  {Coldren}}\ and\ \bibinfo {author} {\bibfnamefont {S.~W.}\ \bibnamefont
  {Corzine}},\ }\href@noop {} {\emph {\bibinfo {title} {Diode lasers and
  photonic integrated circuits}}},\ edited by\ \bibinfo {editor} {\bibnamefont
  {Wiley}}\ (\bibinfo  {publisher} {Wiley},\ \bibinfo {address} {New York},\
  \bibinfo {year} {1995})\BibitemShut {NoStop}%
\bibitem [{\citenamefont {Nagai}\ \emph {et~al.}(2004)\citenamefont {Nagai},
  \citenamefont {Inagaki},\ and\ \citenamefont {Kanemitsu}}]{Nagai04}%
  \BibitemOpen
  \bibfield  {author} {\bibinfo {author} {\bibfnamefont {T.}~\bibnamefont
  {Nagai}}, \bibinfo {author} {\bibfnamefont {T.~J.}\ \bibnamefont {Inagaki}},\
  and\ \bibinfo {author} {\bibfnamefont {Y.}~\bibnamefont {Kanemitsu}},\
  }\bibfield  {title} {\bibinfo {title} {Band-gap renormalization in highly
  excited {GaN}},\ }\href {https://doi.org/10.1063/1.1650552} {\bibfield
  {journal} {\bibinfo  {journal} {Applied Physics Letters}\ }\textbf {\bibinfo
  {volume} {84}},\ \bibinfo {pages} {1284} (\bibinfo {year}
  {2004})}\BibitemShut {NoStop}%
\bibitem [{\citenamefont {Cho}\ \emph {et~al.}(1998)\citenamefont {Cho},
  \citenamefont {Gainer}, \citenamefont {Fischer}, \citenamefont {Song},
  \citenamefont {Keller}, \citenamefont {Mishra},\ and\ \citenamefont
  {DenBaars}}]{Cho98}%
  \BibitemOpen
  \bibfield  {author} {\bibinfo {author} {\bibfnamefont {Y.-H.}\ \bibnamefont
  {Cho}}, \bibinfo {author} {\bibfnamefont {G.~H.}\ \bibnamefont {Gainer}},
  \bibinfo {author} {\bibfnamefont {A.~J.}\ \bibnamefont {Fischer}}, \bibinfo
  {author} {\bibfnamefont {J.~J.}\ \bibnamefont {Song}}, \bibinfo {author}
  {\bibfnamefont {S.}~\bibnamefont {Keller}}, \bibinfo {author} {\bibfnamefont
  {U.~K.}\ \bibnamefont {Mishra}},\ and\ \bibinfo {author} {\bibfnamefont
  {S.~P.}\ \bibnamefont {DenBaars}},\ }\bibfield  {title} {\bibinfo {title}
  {“{S}-shaped” temperature-dependent emission shift and carrier dynamics
  in {InGaN}/{GaN} multiple quantum wells},\ }\href
  {https://doi.org/10.1063/1.122164} {\bibfield  {journal} {\bibinfo  {journal}
  {Applied Physics Letters}\ }\textbf {\bibinfo {volume} {73}},\ \bibinfo
  {pages} {1370} (\bibinfo {year} {1998})}\BibitemShut {NoStop}%
\bibitem [{\citenamefont {Di~Vito}\ \emph {et~al.}(2019)\citenamefont
  {Di~Vito}, \citenamefont {Pecchia}, \citenamefont {Di~Carlo},\ and\
  \citenamefont {Auf~der Maur}}]{DiVito19}%
  \BibitemOpen
  \bibfield  {author} {\bibinfo {author} {\bibfnamefont {A.}~\bibnamefont
  {Di~Vito}}, \bibinfo {author} {\bibfnamefont {A.}~\bibnamefont {Pecchia}},
  \bibinfo {author} {\bibfnamefont {A.}~\bibnamefont {Di~Carlo}},\ and\
  \bibinfo {author} {\bibfnamefont {M.}~\bibnamefont {Auf~der Maur}},\
  }\bibfield  {title} {\bibinfo {title} {Impact of {Compositional}
  {Nonuniformity} in ({In},{Ga}){N}-{Based} {Light}-{Emitting} {Diodes}},\
  }\href {https://doi.org/10.1103/PhysRevApplied.12.014055} {\bibfield
  {journal} {\bibinfo  {journal} {Physical Review Applied}\ }\textbf {\bibinfo
  {volume} {12}},\ \bibinfo {pages} {014055} (\bibinfo {year}
  {2019})}\BibitemShut {NoStop}%
\bibitem [{\citenamefont {Auf~der Maur}\ \emph {et~al.}(2016)\citenamefont
  {Auf~der Maur}, \citenamefont {Pecchia}, \citenamefont {Penazzi},
  \citenamefont {Rodrigues},\ and\ \citenamefont {Di~Carlo}}]{Aufdermaur16}%
  \BibitemOpen
  \bibfield  {author} {\bibinfo {author} {\bibfnamefont {M.}~\bibnamefont
  {Auf~der Maur}}, \bibinfo {author} {\bibfnamefont {A.}~\bibnamefont
  {Pecchia}}, \bibinfo {author} {\bibfnamefont {G.}~\bibnamefont {Penazzi}},
  \bibinfo {author} {\bibfnamefont {W.}~\bibnamefont {Rodrigues}},\ and\
  \bibinfo {author} {\bibfnamefont {A.}~\bibnamefont {Di~Carlo}},\ }\bibfield
  {title} {\bibinfo {title} {Efficiency {Drop} in {Green} {InGaN}/{GaN} {Light}
  {Emitting} {Diodes}: {The} {Role} of {Random} {Alloy} {Fluctuations}},\
  }\href@noop {} {\bibfield  {journal} {\bibinfo  {journal} {Physical Review
  Letters}\ }\textbf {\bibinfo {volume} {116}},\ \bibinfo {pages} {027401}
  (\bibinfo {year} {2016})}\BibitemShut {NoStop}%
\bibitem [{\citenamefont {Pant}\ and\ \citenamefont
  {Kioupakis}(2023)}]{Pant23}%
  \BibitemOpen
  \bibfield  {author} {\bibinfo {author} {\bibfnamefont {N.}~\bibnamefont
  {Pant}}\ and\ \bibinfo {author} {\bibfnamefont {E.}~\bibnamefont
  {Kioupakis}},\ }\bibfield  {title} {\bibinfo {title} {Increased
  light-emission efficiency in disordered ({In},{Ga}){N} through the correlated
  reduction of recombination rates},\ }\href
  {https://doi.org/10.1103/PhysRevApplied.20.064049} {\bibfield  {journal}
  {\bibinfo  {journal} {Physical Review Applied}\ }\textbf {\bibinfo {volume}
  {20}},\ \bibinfo {pages} {064049} (\bibinfo {year} {2023})}\BibitemShut
  {NoStop}%
\bibitem [{Note1()}]{Note1}%
  \BibitemOpen
  \bibinfo {note} {More precisely, the sum is computed over all electron states
  $q$ that are in the ground state along the $z$ direction, i.e. whose
  wavefunction has only one lobe along $z$. If higher-order states along $z$
  were included in the sum, $O_p$ could exceed unity.}\BibitemShut {Stop}%
\bibitem [{\citenamefont {Galler}\ \emph {et~al.}(2012)\citenamefont {Galler},
  \citenamefont {Drechsel}, \citenamefont {Monnard}, \citenamefont {Rode},
  \citenamefont {Stauss}, \citenamefont {Froehlich}, \citenamefont {Bergbauer},
  \citenamefont {Binder}, \citenamefont {Sabathil},\ and\ \citenamefont
  {Hahn}}]{Galler12}%
  \BibitemOpen
  \bibfield  {author} {\bibinfo {author} {\bibfnamefont {B.}~\bibnamefont
  {Galler}}, \bibinfo {author} {\bibfnamefont {P.}~\bibnamefont {Drechsel}},
  \bibinfo {author} {\bibfnamefont {R.}~\bibnamefont {Monnard}}, \bibinfo
  {author} {\bibfnamefont {P.}~\bibnamefont {Rode}}, \bibinfo {author}
  {\bibfnamefont {P.}~\bibnamefont {Stauss}}, \bibinfo {author} {\bibfnamefont
  {S.}~\bibnamefont {Froehlich}}, \bibinfo {author} {\bibfnamefont
  {W.}~\bibnamefont {Bergbauer}}, \bibinfo {author} {\bibfnamefont
  {M.}~\bibnamefont {Binder}}, \bibinfo {author} {\bibfnamefont
  {M.}~\bibnamefont {Sabathil}},\ and\ \bibinfo {author} {\bibfnamefont
  {B.}~\bibnamefont {Hahn}},\ }\bibfield  {title} {\bibinfo {title} {Influence
  of indium content and temperature on {Auger}-like recombination in {InGaN}
  quantum wells grown on (111) silicon substrates},\ }\href@noop {} {\bibfield
  {journal} {\bibinfo  {journal} {Applied Physics Letters}\ }\textbf {\bibinfo
  {volume} {101}},\ \bibinfo {pages} {131111} (\bibinfo {year}
  {2012})}\BibitemShut {NoStop}%
\bibitem [{\citenamefont {Tian}\ \emph {et~al.}(2014)\citenamefont {Tian},
  \citenamefont {McKendry}, \citenamefont {Herrnsdorf}, \citenamefont {Watson},
  \citenamefont {Ferreira}, \citenamefont {Watson}, \citenamefont {Gu},
  \citenamefont {Kelly},\ and\ \citenamefont {Dawson}}]{Tian14}%
  \BibitemOpen
  \bibfield  {author} {\bibinfo {author} {\bibfnamefont {P.}~\bibnamefont
  {Tian}}, \bibinfo {author} {\bibfnamefont {J.~J.~D.}\ \bibnamefont
  {McKendry}}, \bibinfo {author} {\bibfnamefont {J.}~\bibnamefont
  {Herrnsdorf}}, \bibinfo {author} {\bibfnamefont {S.}~\bibnamefont {Watson}},
  \bibinfo {author} {\bibfnamefont {R.}~\bibnamefont {Ferreira}}, \bibinfo
  {author} {\bibfnamefont {I.~M.}\ \bibnamefont {Watson}}, \bibinfo {author}
  {\bibfnamefont {E.}~\bibnamefont {Gu}}, \bibinfo {author} {\bibfnamefont
  {A.~E.}\ \bibnamefont {Kelly}},\ and\ \bibinfo {author} {\bibfnamefont
  {M.~D.}\ \bibnamefont {Dawson}},\ }\bibfield  {title} {\bibinfo {title}
  {Temperature-dependent efficiency droop of blue {InGaN} micro-light emitting
  diodes},\ }\href {https://doi.org/10.1063/1.4900865} {\bibfield  {journal}
  {\bibinfo  {journal} {Applied Physics Letters}\ }\textbf {\bibinfo {volume}
  {105}},\ \bibinfo {pages} {171107} (\bibinfo {year} {2014})}\BibitemShut
  {NoStop}%
\bibitem [{\citenamefont {David}\ \emph
  {et~al.}(2019{\natexlab{c}})\citenamefont {David}, \citenamefont {Young},
  \citenamefont {Lund},\ and\ \citenamefont {Craven}}]{David19_HC}%
  \BibitemOpen
  \bibfield  {author} {\bibinfo {author} {\bibfnamefont {A.}~\bibnamefont
  {David}}, \bibinfo {author} {\bibfnamefont {N.}~\bibnamefont {Young}},
  \bibinfo {author} {\bibfnamefont {C.}~\bibnamefont {Lund}},\ and\ \bibinfo
  {author} {\bibfnamefont {M.}~\bibnamefont {Craven}},\ }\bibfield  {title}
  {\bibinfo {title} {Thermal droop in high-quality {InGaN} {LEDs}},\
  }\href@noop {} {\bibfield  {journal} {\bibinfo  {journal} {Applied Physics
  Letters}\ }\textbf {\bibinfo {volume} {115}},\ \bibinfo {pages} {223502}
  (\bibinfo {year} {2019}{\natexlab{c}})}\BibitemShut {NoStop}%
\bibitem [{\citenamefont {David}\ \emph
  {et~al.}(2019{\natexlab{d}})\citenamefont {David}, \citenamefont {Young},
  \citenamefont {Lund},\ and\ \citenamefont {Craven}}]{David19_BC}%
  \BibitemOpen
  \bibfield  {author} {\bibinfo {author} {\bibfnamefont {A.}~\bibnamefont
  {David}}, \bibinfo {author} {\bibfnamefont {N.}~\bibnamefont {Young}},
  \bibinfo {author} {\bibfnamefont {C.}~\bibnamefont {Lund}},\ and\ \bibinfo
  {author} {\bibfnamefont {M.}~\bibnamefont {Craven}},\ }\bibfield  {title}
  {\bibinfo {title} {Compensation between radiative and {Auger} recombinations
  in {III}-{Nitrides}: the scaling law of separated-wavefunction
  recombinations},\ }\href@noop {} {\bibfield  {journal} {\bibinfo  {journal}
  {Applied Physics Letters}\ }\textbf {\bibinfo {volume} {115}},\ \bibinfo
  {pages} {193502} (\bibinfo {year} {2019}{\natexlab{d}})}\BibitemShut
  {NoStop}%
\bibitem [{\citenamefont {Kudo}\ \emph {et~al.}(2002)\citenamefont {Kudo},
  \citenamefont {Murakami}, \citenamefont {Zheng}, \citenamefont {Yamada},
  \citenamefont {Taguchi}, \citenamefont {Tadatomo}, \citenamefont {Okagawa},
  \citenamefont {Ohuchi}, \citenamefont {Tsunekawa}, \citenamefont {Imada},\
  and\ \citenamefont {Kato}}]{Kudo02}%
  \BibitemOpen
  \bibfield  {author} {\bibinfo {author} {\bibfnamefont {H.}~\bibnamefont
  {Kudo}}, \bibinfo {author} {\bibfnamefont {K.}~\bibnamefont {Murakami}},
  \bibinfo {author} {\bibfnamefont {R.}~\bibnamefont {Zheng}}, \bibinfo
  {author} {\bibfnamefont {Y.}~\bibnamefont {Yamada}}, \bibinfo {author}
  {\bibfnamefont {T.}~\bibnamefont {Taguchi}}, \bibinfo {author} {\bibfnamefont
  {K.}~\bibnamefont {Tadatomo}}, \bibinfo {author} {\bibfnamefont
  {H.}~\bibnamefont {Okagawa}}, \bibinfo {author} {\bibfnamefont
  {Y.}~\bibnamefont {Ohuchi}}, \bibinfo {author} {\bibfnamefont
  {T.}~\bibnamefont {Tsunekawa}}, \bibinfo {author} {\bibfnamefont
  {Y.}~\bibnamefont {Imada}},\ and\ \bibinfo {author} {\bibfnamefont
  {M.}~\bibnamefont {Kato}},\ }\bibfield  {title} {\bibinfo {title} {Intense
  {Ultraviolet} {Electroluminescence} {Properties} of the {High}-{Power}
  {InGaN}-{Based} {Light}-{Emitting} {Diodes} {Fabricated} on {Patterned}
  {Sapphire} {Substrates}},\ }\href {https://doi.org/10.1143/JJAP.41.2484}
  {\bibfield  {journal} {\bibinfo  {journal} {Japanese Journal of Applied
  Physics}\ }\textbf {\bibinfo {volume} {41}},\ \bibinfo {pages} {2484}
  (\bibinfo {year} {2002})}\BibitemShut {NoStop}%
\bibitem [{\citenamefont {Shen}\ \emph {et~al.}(2007)\citenamefont {Shen},
  \citenamefont {Mueller}, \citenamefont {Watanabe}, \citenamefont {Gardner},
  \citenamefont {Munkholm},\ and\ \citenamefont {Krames}}]{Shen07}%
  \BibitemOpen
  \bibfield  {author} {\bibinfo {author} {\bibfnamefont {Y.~C.}\ \bibnamefont
  {Shen}}, \bibinfo {author} {\bibfnamefont {G.~O.}\ \bibnamefont {Mueller}},
  \bibinfo {author} {\bibfnamefont {S.}~\bibnamefont {Watanabe}}, \bibinfo
  {author} {\bibfnamefont {N.~F.}\ \bibnamefont {Gardner}}, \bibinfo {author}
  {\bibfnamefont {A.}~\bibnamefont {Munkholm}},\ and\ \bibinfo {author}
  {\bibfnamefont {M.~R.}\ \bibnamefont {Krames}},\ }\bibfield  {title}
  {\bibinfo {title} {Auger recombination in {InGaN} measured by
  photoluminescence},\ }\href@noop {} {\bibfield  {journal} {\bibinfo
  {journal} {Applied Physics Letters}\ }\textbf {\bibinfo {volume} {91}},\
  \bibinfo {pages} {141101} (\bibinfo {year} {2007})}\BibitemShut {NoStop}%
\bibitem [{\citenamefont {Keppens}\ \emph {et~al.}(2010)\citenamefont
  {Keppens}, \citenamefont {Ryckaert}, \citenamefont {Deconinck},\ and\
  \citenamefont {Hanselaer}}]{Keppens10}%
  \BibitemOpen
  \bibfield  {author} {\bibinfo {author} {\bibfnamefont {A.}~\bibnamefont
  {Keppens}}, \bibinfo {author} {\bibfnamefont {W.~R.}\ \bibnamefont
  {Ryckaert}}, \bibinfo {author} {\bibfnamefont {G.}~\bibnamefont
  {Deconinck}},\ and\ \bibinfo {author} {\bibfnamefont {P.}~\bibnamefont
  {Hanselaer}},\ }\bibfield  {title} {\bibinfo {title} {Modeling high power
  light-emitting diode spectra and their variation with junction temperature},\
  }\href {https://doi.org/10.1063/1.3463411} {\bibfield  {journal} {\bibinfo
  {journal} {Journal of Applied Physics}\ }\textbf {\bibinfo {volume} {108}},\
  \bibinfo {pages} {043104} (\bibinfo {year} {2010})}\BibitemShut {NoStop}%
\bibitem [{\citenamefont {Han}\ \emph {et~al.}(2020)\citenamefont {Han},
  \citenamefont {Lee}, \citenamefont {Min}, \citenamefont {Shin}, \citenamefont
  {Shim}, \citenamefont {Iwaya}, \citenamefont {Takeuchi}, \citenamefont
  {Kamiyama},\ and\ \citenamefont {Akasaki}}]{Han20}%
  \BibitemOpen
  \bibfield  {author} {\bibinfo {author} {\bibfnamefont {D.-P.}\ \bibnamefont
  {Han}}, \bibinfo {author} {\bibfnamefont {G.~W.}\ \bibnamefont {Lee}},
  \bibinfo {author} {\bibfnamefont {S.}~\bibnamefont {Min}}, \bibinfo {author}
  {\bibfnamefont {D.-S.}\ \bibnamefont {Shin}}, \bibinfo {author}
  {\bibfnamefont {J.-I.}\ \bibnamefont {Shim}}, \bibinfo {author}
  {\bibfnamefont {M.}~\bibnamefont {Iwaya}}, \bibinfo {author} {\bibfnamefont
  {T.}~\bibnamefont {Takeuchi}}, \bibinfo {author} {\bibfnamefont
  {S.}~\bibnamefont {Kamiyama}},\ and\ \bibinfo {author} {\bibfnamefont
  {I.}~\bibnamefont {Akasaki}},\ }\bibfield  {title} {\bibinfo {title}
  {Identifying the cause of thermal droop in {GaInN}-based {LEDs} by carrier-
  and thermo-dynamics analysis},\ }\href
  {https://doi.org/10.1038/s41598-020-74585-w} {\bibfield  {journal} {\bibinfo
  {journal} {Scientific Reports}\ }\textbf {\bibinfo {volume} {10}},\ \bibinfo
  {pages} {17433} (\bibinfo {year} {2020})}\BibitemShut {NoStop}%
\bibitem [{\citenamefont {Sun}\ \emph {et~al.}(1997)\citenamefont {Sun},
  \citenamefont {Vallee}, \citenamefont {Keller}, \citenamefont {Bowers},\ and\
  \citenamefont {DenBaars}}]{Sun97}%
  \BibitemOpen
  \bibfield  {author} {\bibinfo {author} {\bibfnamefont {C.~K.}\ \bibnamefont
  {Sun}}, \bibinfo {author} {\bibfnamefont {F.}~\bibnamefont {Vallee}},
  \bibinfo {author} {\bibfnamefont {S.}~\bibnamefont {Keller}}, \bibinfo
  {author} {\bibfnamefont {J.~E.}\ \bibnamefont {Bowers}},\ and\ \bibinfo
  {author} {\bibfnamefont {S.~P.}\ \bibnamefont {DenBaars}},\ }\bibfield
  {title} {\bibinfo {title} {Femtosecond studies of carrier dynamics in
  {InGaN}},\ }\href@noop {} {\bibfield  {journal} {\bibinfo  {journal} {Applied
  Physics Letters}\ }\textbf {\bibinfo {volume} {70}},\ \bibinfo {pages} {2004}
  (\bibinfo {year} {1997})}\BibitemShut {NoStop}%
\bibitem [{\citenamefont {Binder}\ \emph {et~al.}(2013)\citenamefont {Binder},
  \citenamefont {Korona}, \citenamefont {Wysmolek}, \citenamefont {Kaminska},
  \citenamefont {Köhler}, \citenamefont {Kirste}, \citenamefont {Ambacher},
  \citenamefont {Zajac},\ and\ \citenamefont {Dwilinski}}]{Binder13b}%
  \BibitemOpen
  \bibfield  {author} {\bibinfo {author} {\bibfnamefont {J.}~\bibnamefont
  {Binder}}, \bibinfo {author} {\bibfnamefont {K.~P.}\ \bibnamefont {Korona}},
  \bibinfo {author} {\bibfnamefont {A.}~\bibnamefont {Wysmolek}}, \bibinfo
  {author} {\bibfnamefont {M.}~\bibnamefont {Kaminska}}, \bibinfo {author}
  {\bibfnamefont {K.}~\bibnamefont {Köhler}}, \bibinfo {author} {\bibfnamefont
  {L.}~\bibnamefont {Kirste}}, \bibinfo {author} {\bibfnamefont
  {O.}~\bibnamefont {Ambacher}}, \bibinfo {author} {\bibfnamefont
  {M.}~\bibnamefont {Zajac}},\ and\ \bibinfo {author} {\bibfnamefont
  {R.}~\bibnamefont {Dwilinski}},\ }\bibfield  {title} {\bibinfo {title}
  {Dynamics of thermalization in {GaInN}/{GaN} quantum wells grown on
  ammonothermal {GaN}},\ }\href {https://doi.org/10.1063/1.4845715} {\bibfield
  {journal} {\bibinfo  {journal} {Journal of Applied Physics}\ }\textbf
  {\bibinfo {volume} {114}},\ \bibinfo {pages} {223504} (\bibinfo {year}
  {2013})}\BibitemShut {NoStop}%
\bibitem [{\citenamefont {Martin}\ \emph {et~al.}(1999)\citenamefont {Martin},
  \citenamefont {Middleton}, \citenamefont {O’Donnell},\ and\ \citenamefont
  {Van Der~Stricht}}]{Martin99}%
  \BibitemOpen
  \bibfield  {author} {\bibinfo {author} {\bibfnamefont {R.~W.}\ \bibnamefont
  {Martin}}, \bibinfo {author} {\bibfnamefont {P.~G.}\ \bibnamefont
  {Middleton}}, \bibinfo {author} {\bibfnamefont {K.~P.}\ \bibnamefont
  {O’Donnell}},\ and\ \bibinfo {author} {\bibfnamefont {W.}~\bibnamefont {Van
  Der~Stricht}},\ }\bibfield  {title} {\bibinfo {title} {Exciton localization
  and the {Stokes}’ shift in {InGaN} epilayers},\ }\href
  {https://doi.org/10.1063/1.123275} {\bibfield  {journal} {\bibinfo  {journal}
  {Applied Physics Letters}\ }\textbf {\bibinfo {volume} {74}},\ \bibinfo
  {pages} {263} (\bibinfo {year} {1999})}\BibitemShut {NoStop}%
\bibitem [{\citenamefont {Chichibu}\ \emph {et~al.}(1999)\citenamefont
  {Chichibu}, \citenamefont {Abare}, \citenamefont {Mack}, \citenamefont
  {Minsky}, \citenamefont {Deguchi}, \citenamefont {Cohen}, \citenamefont
  {Kozodoy}, \citenamefont {Fleischer}, \citenamefont {Keller}, \citenamefont
  {Speck}, \citenamefont {Bowers}, \citenamefont {Hu}, \citenamefont {Mishra},
  \citenamefont {Coldren}, \citenamefont {DenBaars}, \citenamefont {Wada},
  \citenamefont {Sota},\ and\ \citenamefont {Nakamura}}]{Chichibu99}%
  \BibitemOpen
  \bibfield  {author} {\bibinfo {author} {\bibfnamefont {S.~F.}\ \bibnamefont
  {Chichibu}}, \bibinfo {author} {\bibfnamefont {A.~C.}\ \bibnamefont {Abare}},
  \bibinfo {author} {\bibfnamefont {M.~P.}\ \bibnamefont {Mack}}, \bibinfo
  {author} {\bibfnamefont {M.~S.}\ \bibnamefont {Minsky}}, \bibinfo {author}
  {\bibfnamefont {T.}~\bibnamefont {Deguchi}}, \bibinfo {author} {\bibfnamefont
  {D.}~\bibnamefont {Cohen}}, \bibinfo {author} {\bibfnamefont
  {P.}~\bibnamefont {Kozodoy}}, \bibinfo {author} {\bibfnamefont {S.~B.}\
  \bibnamefont {Fleischer}}, \bibinfo {author} {\bibfnamefont {S.}~\bibnamefont
  {Keller}}, \bibinfo {author} {\bibfnamefont {J.~S.}\ \bibnamefont {Speck}},
  \bibinfo {author} {\bibfnamefont {J.~E.}\ \bibnamefont {Bowers}}, \bibinfo
  {author} {\bibfnamefont {E.}~\bibnamefont {Hu}}, \bibinfo {author}
  {\bibfnamefont {U.~K.}\ \bibnamefont {Mishra}}, \bibinfo {author}
  {\bibfnamefont {L.~A.}\ \bibnamefont {Coldren}}, \bibinfo {author}
  {\bibfnamefont {S.~P.}\ \bibnamefont {DenBaars}}, \bibinfo {author}
  {\bibfnamefont {K.}~\bibnamefont {Wada}}, \bibinfo {author} {\bibfnamefont
  {T.}~\bibnamefont {Sota}},\ and\ \bibinfo {author} {\bibfnamefont
  {S.}~\bibnamefont {Nakamura}},\ }\bibfield  {title} {\bibinfo {title}
  {Optical properties of {InGaN} quantum wells},\ }\href
  {https://doi.org/10.1016/S0921-5107(98)00359-6} {\bibfield  {journal}
  {\bibinfo  {journal} {Materials Science and Engineering: B}\ }\textbf
  {\bibinfo {volume} {59}},\ \bibinfo {pages} {298} (\bibinfo {year}
  {1999})}\BibitemShut {NoStop}%
\bibitem [{\citenamefont {Shan}\ \emph {et~al.}(1998)\citenamefont {Shan},
  \citenamefont {Walukiewicz}, \citenamefont {Haller}, \citenamefont {Little},
  \citenamefont {Song}, \citenamefont {McCluskey}, \citenamefont {Johnson},
  \citenamefont {Feng}, \citenamefont {Schurman},\ and\ \citenamefont
  {Stall}}]{Shan98}%
  \BibitemOpen
  \bibfield  {author} {\bibinfo {author} {\bibfnamefont {W.}~\bibnamefont
  {Shan}}, \bibinfo {author} {\bibfnamefont {W.}~\bibnamefont {Walukiewicz}},
  \bibinfo {author} {\bibfnamefont {E.~E.}\ \bibnamefont {Haller}}, \bibinfo
  {author} {\bibfnamefont {B.~D.}\ \bibnamefont {Little}}, \bibinfo {author}
  {\bibfnamefont {J.~J.}\ \bibnamefont {Song}}, \bibinfo {author}
  {\bibfnamefont {M.~D.}\ \bibnamefont {McCluskey}}, \bibinfo {author}
  {\bibfnamefont {N.~M.}\ \bibnamefont {Johnson}}, \bibinfo {author}
  {\bibfnamefont {Z.~C.}\ \bibnamefont {Feng}}, \bibinfo {author}
  {\bibfnamefont {M.}~\bibnamefont {Schurman}},\ and\ \bibinfo {author}
  {\bibfnamefont {R.~A.}\ \bibnamefont {Stall}},\ }\bibfield  {title} {\bibinfo
  {title} {Optical properties of {InxGa1}-{xN} alloys grown by metalorganic
  chemical vapor deposition},\ }\href {https://doi.org/10.1063/1.368669}
  {\bibfield  {journal} {\bibinfo  {journal} {Journal of Applied Physics}\
  }\textbf {\bibinfo {volume} {84}},\ \bibinfo {pages} {4452} (\bibinfo {year}
  {1998})}\BibitemShut {NoStop}%
\bibitem [{\citenamefont {Berkowicz}\ \emph {et~al.}(2000)\citenamefont
  {Berkowicz}, \citenamefont {Gershoni}, \citenamefont {Bahir}, \citenamefont
  {Lakin}, \citenamefont {Shilo}, \citenamefont {Zolotoyabko}, \citenamefont
  {Abare}, \citenamefont {Denbaars},\ and\ \citenamefont
  {Coldren}}]{Berkowicz00}%
  \BibitemOpen
  \bibfield  {author} {\bibinfo {author} {\bibfnamefont {E.}~\bibnamefont
  {Berkowicz}}, \bibinfo {author} {\bibfnamefont {D.}~\bibnamefont {Gershoni}},
  \bibinfo {author} {\bibfnamefont {G.}~\bibnamefont {Bahir}}, \bibinfo
  {author} {\bibfnamefont {E.}~\bibnamefont {Lakin}}, \bibinfo {author}
  {\bibfnamefont {D.}~\bibnamefont {Shilo}}, \bibinfo {author} {\bibfnamefont
  {E.}~\bibnamefont {Zolotoyabko}}, \bibinfo {author} {\bibfnamefont {A.~C.}\
  \bibnamefont {Abare}}, \bibinfo {author} {\bibfnamefont {S.~P.}\ \bibnamefont
  {Denbaars}},\ and\ \bibinfo {author} {\bibfnamefont {L.~A.}\ \bibnamefont
  {Coldren}},\ }\bibfield  {title} {\bibinfo {title} {Measured and calculated
  radiative lifetime and optical absorption of
  {InxGa}(1-x){N}\vphantom{\{}\}/{GaN} quantum structures},\ }\href
  {https://doi.org/10.1103/PhysRevB.61.10994} {\bibfield  {journal} {\bibinfo
  {journal} {Physical Review B}\ }\textbf {\bibinfo {volume} {61}},\ \bibinfo
  {pages} {10994} (\bibinfo {year} {2000})}\BibitemShut {NoStop}%
\bibitem [{\citenamefont {Weisbuch}\ \emph {et~al.}(2021)\citenamefont
  {Weisbuch}, \citenamefont {Nakamura}, \citenamefont {Wu},\ and\ \citenamefont
  {Speck}}]{Weisbuch21}%
  \BibitemOpen
  \bibfield  {author} {\bibinfo {author} {\bibfnamefont {C.}~\bibnamefont
  {Weisbuch}}, \bibinfo {author} {\bibfnamefont {S.}~\bibnamefont {Nakamura}},
  \bibinfo {author} {\bibfnamefont {Y.-R.}\ \bibnamefont {Wu}},\ and\ \bibinfo
  {author} {\bibfnamefont {J.~S.}\ \bibnamefont {Speck}},\ }\bibfield  {title}
  {\bibinfo {title} {Disorder effects in nitride semiconductors: impact on
  fundamental and device properties},\ }\href
  {https://doi.org/10.1515/nanoph-2020-0590} {\bibfield  {journal} {\bibinfo
  {journal} {Nanophotonics}\ }\textbf {\bibinfo {volume} {10}},\ \bibinfo
  {pages} {3} (\bibinfo {year} {2021})}\BibitemShut {NoStop}%
\bibitem [{\citenamefont {Nippert}\ \emph {et~al.}(2016)\citenamefont
  {Nippert}, \citenamefont {Karpov}, \citenamefont {Callsen}, \citenamefont
  {Galler}, \citenamefont {Kure}, \citenamefont {Nenstiel}, \citenamefont
  {Wagner}, \citenamefont {Strassburg}, \citenamefont {Lugauer},\ and\
  \citenamefont {Hoffmann}}]{Nippert16b}%
  \BibitemOpen
  \bibfield  {author} {\bibinfo {author} {\bibfnamefont {F.}~\bibnamefont
  {Nippert}}, \bibinfo {author} {\bibfnamefont {S.~Y.}\ \bibnamefont {Karpov}},
  \bibinfo {author} {\bibfnamefont {G.}~\bibnamefont {Callsen}}, \bibinfo
  {author} {\bibfnamefont {B.}~\bibnamefont {Galler}}, \bibinfo {author}
  {\bibfnamefont {T.}~\bibnamefont {Kure}}, \bibinfo {author} {\bibfnamefont
  {C.}~\bibnamefont {Nenstiel}}, \bibinfo {author} {\bibfnamefont {M.~R.}\
  \bibnamefont {Wagner}}, \bibinfo {author} {\bibfnamefont {M.}~\bibnamefont
  {Strassburg}}, \bibinfo {author} {\bibfnamefont {H.-J.}\ \bibnamefont
  {Lugauer}},\ and\ \bibinfo {author} {\bibfnamefont {A.}~\bibnamefont
  {Hoffmann}},\ }\bibfield  {title} {\bibinfo {title} {Temperature-dependent
  recombination coefficients in {InGaN} light-emitting diodes: {Hole}
  localization, {Auger} processes, and the green gap},\ }\href@noop {}
  {\bibfield  {journal} {\bibinfo  {journal} {Applied Physics Letters}\
  }\textbf {\bibinfo {volume} {109}},\ \bibinfo {pages} {161103} (\bibinfo
  {year} {2016})}\BibitemShut {NoStop}%
\bibitem [{\citenamefont {Karpov}(2018)}]{Karpov18}%
  \BibitemOpen
  \bibfield  {author} {\bibinfo {author} {\bibfnamefont {S.~Y.}\ \bibnamefont
  {Karpov}},\ }\bibfield  {title} {\bibinfo {title} {Effect of {Carrier}
  {Localization} on {Recombination} {Processes} and {Efficiency} of
  {InGaN}-{Based} {LEDs} {Operating} in the “{Green} {Gap}”},\ }\href
  {https://doi.org/10.3390/app8050818} {\bibfield  {journal} {\bibinfo
  {journal} {Applied Sciences}\ }\textbf {\bibinfo {volume} {8}},\ \bibinfo
  {pages} {818} (\bibinfo {year} {2018})}\BibitemShut {NoStop}%
\bibitem [{\citenamefont {David}\ and\ \citenamefont
  {Grundmann}(2010)}]{David10b}%
  \BibitemOpen
  \bibfield  {author} {\bibinfo {author} {\bibfnamefont {A.}~\bibnamefont
  {David}}\ and\ \bibinfo {author} {\bibfnamefont {M.~J.}\ \bibnamefont
  {Grundmann}},\ }\bibfield  {title} {\bibinfo {title} {Influence of
  polarization fields on carrier lifetime and recombination rates in
  {InGaN}-based light-emitting diodes},\ }\href@noop {} {\bibfield  {journal}
  {\bibinfo  {journal} {Applied Physics Letters}\ }\textbf {\bibinfo {volume}
  {97}},\ \bibinfo {pages} {033501} (\bibinfo {year} {2010})}\BibitemShut
  {NoStop}%
\bibitem [{Lum()}]{LumiledsLuxeonC}%
  \BibitemOpen
  \href@noop {} {\bibinfo {title} {Lumileds {Luxeon} {C} {Color} datasheet,
  retrieved from
  https://lumileds.com/products/color-leds/luxeon-c-colors/}}\BibitemShut
  {NoStop}%
\bibitem [{\citenamefont {Zhang}\ \emph {et~al.}(2020)\citenamefont {Zhang},
  \citenamefont {Zhang}, \citenamefont {Gao}, \citenamefont {Wang},
  \citenamefont {Zheng}, \citenamefont {Zhang}, \citenamefont {Wu},
  \citenamefont {Xu}, \citenamefont {Ding}, \citenamefont {Quan},\ and\
  \citenamefont {Jiang}}]{Zhang20}%
  \BibitemOpen
  \bibfield  {author} {\bibinfo {author} {\bibfnamefont {S.}~\bibnamefont
  {Zhang}}, \bibinfo {author} {\bibfnamefont {J.}~\bibnamefont {Zhang}},
  \bibinfo {author} {\bibfnamefont {J.}~\bibnamefont {Gao}}, \bibinfo {author}
  {\bibfnamefont {X.}~\bibnamefont {Wang}}, \bibinfo {author} {\bibfnamefont
  {C.}~\bibnamefont {Zheng}}, \bibinfo {author} {\bibfnamefont
  {M.}~\bibnamefont {Zhang}}, \bibinfo {author} {\bibfnamefont
  {X.}~\bibnamefont {Wu}}, \bibinfo {author} {\bibfnamefont {L.}~\bibnamefont
  {Xu}}, \bibinfo {author} {\bibfnamefont {J.}~\bibnamefont {Ding}}, \bibinfo
  {author} {\bibfnamefont {Z.}~\bibnamefont {Quan}},\ and\ \bibinfo {author}
  {\bibfnamefont {F.}~\bibnamefont {Jiang}},\ }\bibfield  {title} {\bibinfo
  {title} {Efficient emission of {InGaN}-based light-emitting diodes: toward
  orange and red},\ }\href {https://doi.org/10.1364/PRJ.402555} {\bibfield
  {journal} {\bibinfo  {journal} {Photonics Research}\ }\textbf {\bibinfo
  {volume} {8}},\ \bibinfo {pages} {1671} (\bibinfo {year} {2020})}\BibitemShut
  {NoStop}%
\bibitem [{\citenamefont {Armitage}\ \emph {et~al.}(2024)\citenamefont
  {Armitage}, \citenamefont {Ren}, \citenamefont {Holmes},\ and\ \citenamefont
  {Flemish}}]{Armitage24}%
  \BibitemOpen
  \bibfield  {author} {\bibinfo {author} {\bibfnamefont {R.}~\bibnamefont
  {Armitage}}, \bibinfo {author} {\bibfnamefont {Z.}~\bibnamefont {Ren}},
  \bibinfo {author} {\bibfnamefont {M.}~\bibnamefont {Holmes}},\ and\ \bibinfo
  {author} {\bibfnamefont {J.}~\bibnamefont {Flemish}},\ }\bibfield  {title}
  {\bibinfo {title} {True-{Red} {InGaN} {Light}-{Emitting} {Diodes} for
  {Display} {Applications}},\ }\href {https://doi.org/10.1002/pssr.202400012}
  {\bibfield  {journal} {\bibinfo  {journal} {physica status solidi (RRL) –
  Rapid Research Letters}\ }\textbf {\bibinfo {volume} {18}},\ \bibinfo {pages}
  {2400012} (\bibinfo {year} {2024})}\BibitemShut {NoStop}%
\bibitem [{\citenamefont {Iida}\ \emph {et~al.}(2022)\citenamefont {Iida},
  \citenamefont {Kirilenko}, \citenamefont {Velazquez-Rizo}, \citenamefont
  {Zhuang}, \citenamefont {Najmi},\ and\ \citenamefont {Ohkawa}}]{Iida22}%
  \BibitemOpen
  \bibfield  {author} {\bibinfo {author} {\bibfnamefont {D.}~\bibnamefont
  {Iida}}, \bibinfo {author} {\bibfnamefont {P.}~\bibnamefont {Kirilenko}},
  \bibinfo {author} {\bibfnamefont {M.}~\bibnamefont {Velazquez-Rizo}},
  \bibinfo {author} {\bibfnamefont {Z.}~\bibnamefont {Zhuang}}, \bibinfo
  {author} {\bibfnamefont {M.~A.}\ \bibnamefont {Najmi}},\ and\ \bibinfo
  {author} {\bibfnamefont {K.}~\bibnamefont {Ohkawa}},\ }\bibfield  {title}
  {\bibinfo {title} {Demonstration of 621-nm-wavelength {InGaN}-based
  single-quantum-well {LEDs} with an external quantum efficiency of 4.3\% at
  10.1 {A}/cm2},\ }\href {https://doi.org/10.1063/5.0097761} {\bibfield
  {journal} {\bibinfo  {journal} {AIP Advances}\ }\textbf {\bibinfo {volume}
  {12}},\ \bibinfo {pages} {065125} (\bibinfo {year} {2022})}\BibitemShut
  {NoStop}%
\bibitem [{\citenamefont {Hwang}\ \emph {et~al.}(2014)\citenamefont {Hwang},
  \citenamefont {Hashimoto}, \citenamefont {Saito},\ and\ \citenamefont
  {Nunoue}}]{Hwang14}%
  \BibitemOpen
  \bibfield  {author} {\bibinfo {author} {\bibfnamefont {J.-I.}\ \bibnamefont
  {Hwang}}, \bibinfo {author} {\bibfnamefont {R.}~\bibnamefont {Hashimoto}},
  \bibinfo {author} {\bibfnamefont {S.}~\bibnamefont {Saito}},\ and\ \bibinfo
  {author} {\bibfnamefont {S.}~\bibnamefont {Nunoue}},\ }\bibfield  {title}
  {\bibinfo {title} {Development of {InGaN}-based red {LED} grown on (0001)
  polar surface},\ }\href {https://doi.org/10.7567/APEX.7.071003} {\bibfield
  {journal} {\bibinfo  {journal} {Applied Physics Express}\ }\textbf {\bibinfo
  {volume} {7}},\ \bibinfo {pages} {071003} (\bibinfo {year}
  {2014})}\BibitemShut {NoStop}%
\bibitem [{Note2()}]{Note2}%
  \BibitemOpen
  \bibinfo {note} {Specifically, the plotted current density is the radiative
  current density $J_{rad}$, calculated from $J_rad=J \times IQE$.}\BibitemShut
  {Stop}%
\bibitem [{\citenamefont {Funato}\ \emph {et~al.}(2006)\citenamefont {Funato},
  \citenamefont {Ueda}, \citenamefont {Kawakami}, \citenamefont {Narukawa},
  \citenamefont {Kosugi}, \citenamefont {Takahashi},\ and\ \citenamefont
  {Mukai}}]{Funato06}%
  \BibitemOpen
  \bibfield  {author} {\bibinfo {author} {\bibfnamefont {M.}~\bibnamefont
  {Funato}}, \bibinfo {author} {\bibfnamefont {M.}~\bibnamefont {Ueda}},
  \bibinfo {author} {\bibfnamefont {Y.}~\bibnamefont {Kawakami}}, \bibinfo
  {author} {\bibfnamefont {Y.}~\bibnamefont {Narukawa}}, \bibinfo {author}
  {\bibfnamefont {T.}~\bibnamefont {Kosugi}}, \bibinfo {author} {\bibfnamefont
  {M.}~\bibnamefont {Takahashi}},\ and\ \bibinfo {author} {\bibfnamefont
  {T.}~\bibnamefont {Mukai}},\ }\bibfield  {title} {\bibinfo {title} {Blue,
  {Green}, and {Amber} {InGaN}/{GaN} {Light}-{Emitting} {Diodes} on {Semipolar}
  \{11-22\} {GaN} {Bulk} {Substrates}},\ }\href
  {https://doi.org/10.1143/JJAP.45.L659} {\bibfield  {journal} {\bibinfo
  {journal} {Japanese Journal of Applied Physics}\ }\textbf {\bibinfo {volume}
  {45}},\ \bibinfo {pages} {L659} (\bibinfo {year} {2006})}\BibitemShut
  {NoStop}%
\bibitem [{\citenamefont {Muyeed}\ \emph {et~al.}(2021)\citenamefont {Muyeed},
  \citenamefont {Borovac}, \citenamefont {Xue}, \citenamefont {Wei},
  \citenamefont {Song}, \citenamefont {Tansu},\ and\ \citenamefont
  {Wierer}}]{Muyeed21}%
  \BibitemOpen
  \bibfield  {author} {\bibinfo {author} {\bibfnamefont {S.~A.~A.}\
  \bibnamefont {Muyeed}}, \bibinfo {author} {\bibfnamefont {D.}~\bibnamefont
  {Borovac}}, \bibinfo {author} {\bibfnamefont {H.}~\bibnamefont {Xue}},
  \bibinfo {author} {\bibfnamefont {X.}~\bibnamefont {Wei}}, \bibinfo {author}
  {\bibfnamefont {R.}~\bibnamefont {Song}}, \bibinfo {author} {\bibfnamefont
  {N.}~\bibnamefont {Tansu}},\ and\ \bibinfo {author} {\bibfnamefont {J.~J.}\
  \bibnamefont {Wierer}},\ }\bibfield  {title} {\bibinfo {title} {Recombination
  {Rates} of {InxGa1}-{xN}/{AlyGa1}-{yN}/{GaN} {Multiple} {Quantum} {Wells}
  {Emitting} {From} 640 to 565 nm},\ }\href
  {https://doi.org/10.1109/JQE.2021.3111402} {\bibfield  {journal} {\bibinfo
  {journal} {IEEE Journal of Quantum Electronics}\ }\textbf {\bibinfo {volume}
  {57}},\ \bibinfo {pages} {1} (\bibinfo {year} {2021})}\BibitemShut {NoStop}%
\bibitem [{\citenamefont {Lv}\ \emph {et~al.}(2019)\citenamefont {Lv},
  \citenamefont {Liu}, \citenamefont {Mo}, \citenamefont {Zhang}, \citenamefont
  {Wu}, \citenamefont {Wu},\ and\ \citenamefont {Jiang}}]{Lv18}%
  \BibitemOpen
  \bibfield  {author} {\bibinfo {author} {\bibfnamefont {Q.}~\bibnamefont
  {Lv}}, \bibinfo {author} {\bibfnamefont {J.}~\bibnamefont {Liu}}, \bibinfo
  {author} {\bibfnamefont {C.}~\bibnamefont {Mo}}, \bibinfo {author}
  {\bibfnamefont {J.}~\bibnamefont {Zhang}}, \bibinfo {author} {\bibfnamefont
  {X.}~\bibnamefont {Wu}}, \bibinfo {author} {\bibfnamefont {Q.}~\bibnamefont
  {Wu}},\ and\ \bibinfo {author} {\bibfnamefont {F.}~\bibnamefont {Jiang}},\
  }\bibfield  {title} {\bibinfo {title} {Realization of {Highly} {Efficient}
  {InGaN} {Green} {LEDs} with {Sandwich}-like {Multiple} {Quantum} {Well}
  {Structure}: {Role} of {Enhanced} {Interwell} {Carrier} {Transport}},\ }\href
  {https://doi.org/10.1021/acsphotonics.8b01040} {\bibfield  {journal}
  {\bibinfo  {journal} {ACS Photonics}\ }\textbf {\bibinfo {volume} {6}},\
  \bibinfo {pages} {130} (\bibinfo {year} {2019})}\BibitemShut {NoStop}%
\bibitem [{\citenamefont {Hsiao}\ \emph {et~al.}(2023)\citenamefont {Hsiao},
  \citenamefont {Lee}, \citenamefont {Miao}, \citenamefont {Pai}, \citenamefont
  {Iida}, \citenamefont {Lin}, \citenamefont {Chen}, \citenamefont {Chow},
  \citenamefont {Lin}, \citenamefont {Horng}, \citenamefont {He}, \citenamefont
  {Ohkawa}, \citenamefont {Hong}, \citenamefont {Chang},\ and\ \citenamefont
  {Kuo}}]{Hsiao23}%
  \BibitemOpen
  \bibfield  {author} {\bibinfo {author} {\bibfnamefont {F.-H.}\ \bibnamefont
  {Hsiao}}, \bibinfo {author} {\bibfnamefont {T.-Y.}\ \bibnamefont {Lee}},
  \bibinfo {author} {\bibfnamefont {W.-C.}\ \bibnamefont {Miao}}, \bibinfo
  {author} {\bibfnamefont {Y.-H.}\ \bibnamefont {Pai}}, \bibinfo {author}
  {\bibfnamefont {D.}~\bibnamefont {Iida}}, \bibinfo {author} {\bibfnamefont
  {C.-L.}\ \bibnamefont {Lin}}, \bibinfo {author} {\bibfnamefont {F.-C.}\
  \bibnamefont {Chen}}, \bibinfo {author} {\bibfnamefont {C.-W.}\ \bibnamefont
  {Chow}}, \bibinfo {author} {\bibfnamefont {C.-C.}\ \bibnamefont {Lin}},
  \bibinfo {author} {\bibfnamefont {R.-H.}\ \bibnamefont {Horng}}, \bibinfo
  {author} {\bibfnamefont {J.-H.}\ \bibnamefont {He}}, \bibinfo {author}
  {\bibfnamefont {K.}~\bibnamefont {Ohkawa}}, \bibinfo {author} {\bibfnamefont
  {Y.-H.}\ \bibnamefont {Hong}}, \bibinfo {author} {\bibfnamefont {C.-Y.}\
  \bibnamefont {Chang}},\ and\ \bibinfo {author} {\bibfnamefont {H.-C.}\
  \bibnamefont {Kuo}},\ }\bibfield  {title} {\bibinfo {title} {Investigations
  on the high performance of {InGaN} red micro-{LEDs} with single quantum well
  for visible light communication applications},\ }\href
  {https://doi.org/10.1186/s11671-023-03871-z} {\bibfield  {journal} {\bibinfo
  {journal} {Discover Nano}\ }\textbf {\bibinfo {volume} {18}},\ \bibinfo
  {pages} {95} (\bibinfo {year} {2023})}\BibitemShut {NoStop}%
\bibitem [{\citenamefont {Xue}\ \emph {et~al.}(2023)\citenamefont {Xue},
  \citenamefont {Muyeed}, \citenamefont {Palmese}, \citenamefont {Rogers},
  \citenamefont {Song}, \citenamefont {Tansu},\ and\ \citenamefont
  {Wierer}}]{Muyeed23}%
  \BibitemOpen
  \bibfield  {author} {\bibinfo {author} {\bibfnamefont {H.}~\bibnamefont
  {Xue}}, \bibinfo {author} {\bibfnamefont {S.~A.~A.}\ \bibnamefont {Muyeed}},
  \bibinfo {author} {\bibfnamefont {E.}~\bibnamefont {Palmese}}, \bibinfo
  {author} {\bibfnamefont {D.}~\bibnamefont {Rogers}}, \bibinfo {author}
  {\bibfnamefont {R.}~\bibnamefont {Song}}, \bibinfo {author} {\bibfnamefont
  {N.}~\bibnamefont {Tansu}},\ and\ \bibinfo {author} {\bibfnamefont {J.~J.}\
  \bibnamefont {Wierer}},\ }\bibfield  {title} {\bibinfo {title} {Recombination
  {Rate} {Analysis} of {InGaN}-{Based} {Red}-{Emitting} {Light}-{Emitting}
  {Diodes}},\ }\href {https://doi.org/10.1109/JQE.2023.3246981} {\bibfield
  {journal} {\bibinfo  {journal} {IEEE Journal of Quantum Electronics}\
  }\textbf {\bibinfo {volume} {59}},\ \bibinfo {pages} {1} (\bibinfo {year}
  {2023})}\BibitemShut {NoStop}%
\bibitem [{\citenamefont {Kawakami}(2025)}]{Kawakami25}%
  \BibitemOpen
  \bibfield  {author} {\bibinfo {author} {\bibfnamefont {Y.}~\bibnamefont
  {Kawakami}},\ }\bibfield  {title} {\bibinfo {title} {Conference presentation:
  {Radiative} and non-radiative recombination mechanisms in red-emitting
  {InGaN} quantum wells}\ }(\bibinfo {address} {ICNS-15, Malmo, Sweden},\
  \bibinfo {year} {2025})\BibitemShut {NoStop}%
\bibitem [{\citenamefont {Dreyer}\ \emph {et~al.}(2016)\citenamefont {Dreyer},
  \citenamefont {Janotti}, \citenamefont {Van~de Walle},\ and\ \citenamefont
  {Vanderbilt}}]{Dreyer16}%
  \BibitemOpen
  \bibfield  {author} {\bibinfo {author} {\bibfnamefont {C.~E.}\ \bibnamefont
  {Dreyer}}, \bibinfo {author} {\bibfnamefont {A.}~\bibnamefont {Janotti}},
  \bibinfo {author} {\bibfnamefont {C.~G.}\ \bibnamefont {Van~de Walle}},\ and\
  \bibinfo {author} {\bibfnamefont {D.}~\bibnamefont {Vanderbilt}},\ }\bibfield
   {title} {\bibinfo {title} {Correct {Implementation} of {Polarization}
  {Constants} in {Wurtzite} {Materials} and {Impact} on {III}-{Nitrides}},\
  }\href {https://doi.org/10.1103/PhysRevX.6.021038} {\bibfield  {journal}
  {\bibinfo  {journal} {Physical Review X}\ }\textbf {\bibinfo {volume} {6}},\
  \bibinfo {pages} {021038} (\bibinfo {year} {2016})}\BibitemShut {NoStop}%
\bibitem [{\citenamefont {Liu}\ \emph {et~al.}(2016)\citenamefont {Liu},
  \citenamefont {Butte}, \citenamefont {Dussaigne}, \citenamefont {Grandjean},
  \citenamefont {Deveaud},\ and\ \citenamefont {Jacopin}}]{Liu16}%
  \BibitemOpen
  \bibfield  {author} {\bibinfo {author} {\bibfnamefont {W.}~\bibnamefont
  {Liu}}, \bibinfo {author} {\bibfnamefont {R.}~\bibnamefont {Butte}}, \bibinfo
  {author} {\bibfnamefont {A.}~\bibnamefont {Dussaigne}}, \bibinfo {author}
  {\bibfnamefont {N.}~\bibnamefont {Grandjean}}, \bibinfo {author}
  {\bibfnamefont {B.}~\bibnamefont {Deveaud}},\ and\ \bibinfo {author}
  {\bibfnamefont {G.}~\bibnamefont {Jacopin}},\ }\bibfield  {title} {\bibinfo
  {title} {Carrier-density-dependent recombination dynamics of excitons and
  electron-hole plasma in m-plane {InGaN}/{GaN} quantum wells},\ }\href@noop {}
  {\bibfield  {journal} {\bibinfo  {journal} {Physical Review B}\ }\textbf
  {\bibinfo {volume} {94}},\ \bibinfo {pages} {195411} (\bibinfo {year}
  {2016})}\BibitemShut {NoStop}%
\bibitem [{\citenamefont {Snoke}(2008)}]{Snoke08}%
  \BibitemOpen
  \bibfield  {author} {\bibinfo {author} {\bibfnamefont {D.}~\bibnamefont
  {Snoke}},\ }\bibfield  {title} {\bibinfo {title} {Predicting the ionization
  threshold for carriers in excited semiconductors},\ }\href@noop {} {\bibfield
   {journal} {\bibinfo  {journal} {Solid State Communications}\ }\textbf
  {\bibinfo {volume} {146}},\ \bibinfo {pages} {73} (\bibinfo {year}
  {2008})}\BibitemShut {NoStop}%
\bibitem [{\citenamefont {Koch}\ \emph {et~al.}(2006)\citenamefont {Koch},
  \citenamefont {Kira}, \citenamefont {Khitrova},\ and\ \citenamefont
  {Gibbs}}]{Koch06}%
  \BibitemOpen
  \bibfield  {author} {\bibinfo {author} {\bibfnamefont {S.~W.}\ \bibnamefont
  {Koch}}, \bibinfo {author} {\bibfnamefont {M.}~\bibnamefont {Kira}}, \bibinfo
  {author} {\bibfnamefont {G.}~\bibnamefont {Khitrova}},\ and\ \bibinfo
  {author} {\bibfnamefont {H.~M.}\ \bibnamefont {Gibbs}},\ }\bibfield  {title}
  {\bibinfo {title} {Semiconductor excitons in new light},\ }\href@noop {}
  {\bibfield  {journal} {\bibinfo  {journal} {Nature Materials}\ }\textbf
  {\bibinfo {volume} {5}},\ \bibinfo {pages} {523} (\bibinfo {year}
  {2006})}\BibitemShut {NoStop}%
\bibitem [{\citenamefont {Miller}\ and\ \citenamefont
  {Abrahams}(1960)}]{Miller60}%
  \BibitemOpen
  \bibfield  {author} {\bibinfo {author} {\bibfnamefont {A.}~\bibnamefont
  {Miller}}\ and\ \bibinfo {author} {\bibfnamefont {E.}~\bibnamefont
  {Abrahams}},\ }\bibfield  {title} {\bibinfo {title} {Impurity {Conduction} at
  {Low} {Concentrations}},\ }\href {https://doi.org/10.1103/PhysRev.120.745}
  {\bibfield  {journal} {\bibinfo  {journal} {Physical Review}\ }\textbf
  {\bibinfo {volume} {120}},\ \bibinfo {pages} {745} (\bibinfo {year}
  {1960})}\BibitemShut {NoStop}%
\bibitem [{\citenamefont {Mott}(1969)}]{Mott68}%
  \BibitemOpen
  \bibfield  {author} {\bibinfo {author} {\bibfnamefont {N.~F.}\ \bibnamefont
  {Mott}},\ }\bibfield  {title} {\bibinfo {title} {Conduction in
  non-crystalline materials: {III}. {Localized} states in a pseudogap and near
  extremities of conduction and valence bands},\ }\href
  {https://doi.org/10.1080/14786436908216338} {\bibfield  {journal} {\bibinfo
  {journal} {The Philosophical Magazine: A Journal of Theoretical Experimental
  and Applied Physics}\ }\textbf {\bibinfo {volume} {19}},\ \bibinfo {pages}
  {835} (\bibinfo {year} {1969})}\BibitemShut {NoStop}%
\bibitem [{\citenamefont {Hamaguchi}(2021)}]{Hamaguchi21}%
  \BibitemOpen
  \bibfield  {author} {\bibinfo {author} {\bibfnamefont {C.}~\bibnamefont
  {Hamaguchi}},\ }\bibfield  {title} {\bibinfo {title} {Electron and hole
  mobilities of {GaN} with bulk, quantum well, and {HEMT} structures},\ }\href
  {https://doi.org/10.1063/5.0060630} {\bibfield  {journal} {\bibinfo
  {journal} {Journal of Applied Physics}\ }\textbf {\bibinfo {volume} {130}},\
  \bibinfo {pages} {125701} (\bibinfo {year} {2021})}\BibitemShut {NoStop}%
\end{thebibliography}

%

\end{document}